\documentclass[aps,twocolumn,amsmath,amssymb,showpacs,prb,floatfix]{revtex4-1}
\usepackage[normalem]{ulem}
\usepackage{amsmath}
\usepackage{color}
\usepackage{graphicx}
\usepackage{epsfig}
\usepackage{SIunits}
\usepackage{mathtools}

\def\ltdash{\raise-1.8pt\hbox{$\scriptscriptstyle |$}}
\newlength{\upit}\upit=0.1truein

\newcommand{\ltappr}{{{\lower4pt\hbox{$<$} } \atop \widetilde{ \ \ \
}}}
\newcommand{\gtappr}{{{\lower4pt\hbox{$>$} } \atop \widetilde{ \ \ \ }}}
\newlength{\bxwidth}\bxwidth=1.5 truein

\newcommand{\dg}{^{\dagger }}

\newcommand{\rarrow}{\rightarrow}

\newcommand{\B}{\Bigg}

\newlength{\figwidth}
\newlength{\shift}
\newlength{\fight}
\newcommand{\fg}[3]
{
\begin{figure}[ht]

\vspace*{-0cm}
\[
\includegraphics[width=\fight]{#1}
\]
\vskip -0.2cm
\caption{\label{#2}
\small #3
}
\end{figure}}

\newcommand \bea {\begin{eqnarray} }
\newcommand \eea {\end{eqnarray}}
\newcommand{\bk}{{\bf{k}}}

\newcommand{\bQ}{{\bf{Q}}}

\newcommand{\bR}{{\bf{R}}}
\newcommand{\urs}{URu$_{2}$Si$_{2}$\ }

\begin{document}
\title{Cubic hastatic order in the two-channel Kondo-Heisenberg model}
\author{Guanghua Zhang,$^1$ John Van Dyke,$^{1}$ and Rebecca Flint$^1$}
\affiliation{$^1$ 
Department of Physics and Astronomy, Iowa State University, Ames, Iowa
50011, USA}
\date{\today}
\begin{abstract}

Materials with non-Kramers doublet ground states naturally manifest the two-channel Kondo effect, as the valence fluctuations are from a non-Kramers doublet ground state to an excited Kramers doublet. Here, the development of a heavy Fermi liquid requires a channel symmetry breaking spinorial hybridization that breaks both single and double time-reversal symmetry, and is known as hastatic order. Motivated by cubic Pr-based materials with $\Gamma_3$ non-Kramers ground state doublets, this paper provides a survey of cubic hastatic order using the simple two-channel Kondo-Heisenberg model. Hastatic order necessarily breaks time-reversal symmetry, but the spatial arrangement of the hybridization spinor can be either uniform (ferrohastatic) or break additional lattice symmetries (antiferrohastatic). The experimental signatures of both orders are presented in detail, and include tiny conduction electron magnetic moments. Interestingly, there can be several distinct antiferrohastatic orders with the same moment pattern that break different lattice symmetries, revealing a potential experimental route to detect the spinorial nature of the hybridization. We employ an SU(N) fermionic mean-field treatment on square and simple cubic lattices, and examine how the nature and stability of hastatic order varies as we vary the Heisenberg coupling, conduction electron density, band degeneracies, and apply both channel and spin symmetry breaking fields. We find that both ferrohastatic and several types of antiferrohastatic orders are stabilized in different regions of the mean-field phase diagram, and evolve differently in strain and magnetic fields. 
\end{abstract}
\maketitle

\section{Introduction}

Kondo physics in heavy fermion materials yields the particularly rich \emph{Doniach phase diagram}\cite{doniach77}, where the competition between heavy Fermi liquid formation and magnetism leads to quantum criticality\cite{coleman05,gegenwart08} and unconventional superconductivity\cite{stewart17}, as well as topological Kondo insulators\cite{dzero10} and exotic magnetism\cite{qphase,pr2271,pr2272,cepdal}. However, this single-channel Kondo physics applies only to Kramers ions, those with an odd number of $f$-electrons, such as Ce and Yb. Non-Kramers ions, with an even number of $f$-electrons like U, Pr and Tb can have \emph{non-Kramers doublet} ground states\cite{zawadowski}. These non-Kramers doublets always manifest the two-channel Kondo effect, since virtual valence fluctuations must involve an excited Kramers doublet\cite{blandin}. This two-channel Kondo physics was originally and extensively explored by Daniel Cox\cite{cox87,cox88,cox96,jarrell96,jarrell97,zawadowski} as a potential origin of unconventional superconductivity in UBe$_{13}$\cite{ott83}. Recently, interest in this physics has been revived, due to new Pr-based materials with non-Kramers doublets, signs of Kondo physics \cite{sakai11,tokunaga13} and quantum criticality \cite{onimaru10,onimaru11,sato12,onimaru12,nagasawa12,iwasa13,tsujimoto14}, and the proposal that the hidden order in \urs might be a type of spinorial hybridization, \emph{hastatic order}, originating from two-channel Kondo physics in tetragonal symmetry\cite{chandra13}.  

These non-Kramers doublets require a new \emph{non-Kramers Doniach phase diagram}, with novel Kondo phases. As the two-channel Kondo impurity is quantum critical, with a $\frac{1}{2} R \ln 2$ zero point entropy\cite{andrei84,emery92}, no conventional heavy Fermi liquids can emerge from a non-Kramers doublet ground state. Instead, the usual heavy Fermi liquid is replaced by a channel symmetry breaking heavy Fermi liquid, where the hybridization between conduction electrons and local moments acquires a spinorial nature, called hastatic order\cite{chandra13,chandra15} or also known as diagonal composite order\cite{hoshino11}. This spinorial hybridization can lead to a number of exotic effects, including nematicity and subtle time-reversal symmetry breaking. Of course, non-Kramers doublet materials can also simply order magnetically or via a cooperative Jahn-Teller distortion, depending on the type of doublet, and so the non-Kramers Doniach phase diagram will also manifest the competition between heavy Fermi liquid formation and magnetism, now with the twist that the heavy Fermi liquid must break channel symmetry. The goal of this paper is to explore the generic features of this hastatic order in a simple Kondo-Heisenberg model.

\fight=8.5cm
\fg{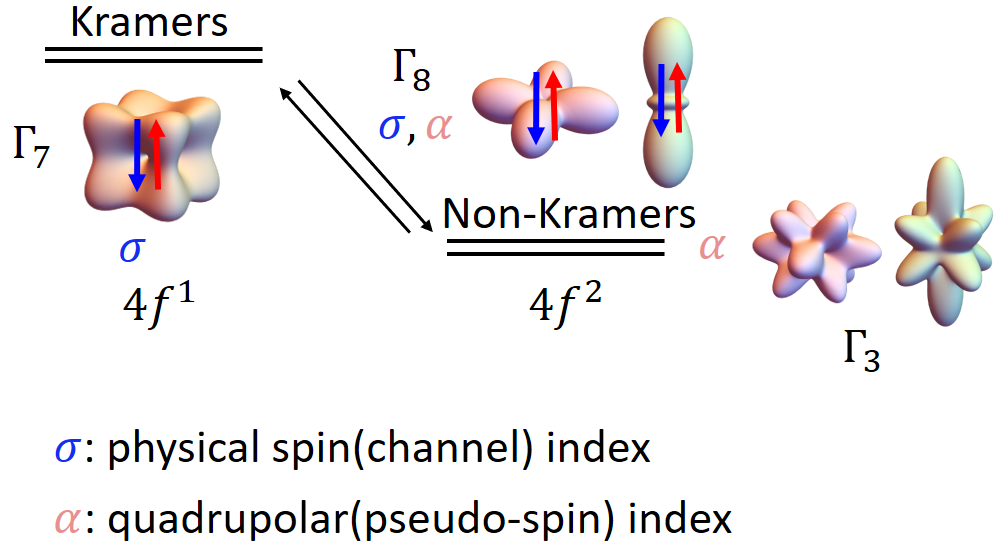}{atomic_levels}{Atomic levels of Pr and the two-channel quadrupolar Kondo effect. In Pr$^{3+}$, valence fluctuations from a $4f^2$ $\Gamma_3$ non-Kramers doublet ground state into a $4f^1$ $\Gamma_7$ Kramers doublet excited state via $\Gamma_8$ conduction electrons generate a two-channel Kondo effect. In this figure, $\sigma$ (red and blue arrows) is the physical spin (channel) index and $\alpha$(light red and light green) is the quadrupolar (pseudospin) index. The charge densities of the $\Gamma_3$ (red/green), $\Gamma_8$ (red/green) and $\Gamma_7$ (golden) orbitals are also depicted. }

Non-Kramers materials with cubic symmetry provide the most straightforward realization of this physics, as these can have a non-magnetic doublet ground state, $\Gamma_3$ with quadrupolar degrees of freedom. In a metallic material, these doublets realize the \emph{quadrupolar Kondo effect}, where the conduction electrons' quadrupolar moments screen the local $\Gamma_3$ quadrupolar moment in two different spin channels\cite{cox87,zawadowski}. The pseudospin and channel degrees of freedom are described by two independent $SU(2)$ symmetries, in contrast to the tetragonal non-Kramers doublet, $\Gamma_5$, where these are entangled \cite{chandra13}. In this paper, we explore the generic realizations of hastatic order in cubic systems via a simple two-channel Kondo-Heisenberg model whose symmetry properties are derived from the $\Gamma_3$ doublet. We study both ferro- and antiferrohastatic phases, finding multiple antiferrohastatic phases with the same pattern of magnetic moments that break double-time-reversal symmetry in different ways. In this simplified model, we explore the global phase diagram as the relative strength of Kondo and quadrupolar couplings are varied, as well as the conduction electron density, magnetic (channel symmetry breaking) and strain (pseudospin symmetry breaking) fields. We also discuss the experimental signatures of hastatic order and the potential relevance to the Pr ``1-2-20'' materials.

The structure of this paper is as follows. In the rest of this section, we give a brief introduction to non-Kramers doublets and the relevant Pr-based materials. In Sec. \ref{sec:model}, we describe our simple two-channel Kondo-Heisenberg model, the effect of magnetic field on realistic systems, and the symmetries of the model. We motivate our choice of mean-field ansatzes with a strong coupling analysis in Sec. \ref{strong}, and discuss the definitions and bandstructures of the ansatzes in Sec. \ref{sec:mfansatzes}. In Sec. \ref{sec:multipolarmoments}, we discuss the symmetry-breaking moments and susceptibilities. Next, we present the phase diagram at zero temperature, finite temperature, and in applied magnetic field and strain in Sec. \ref{sec:PD} to Sec. \ref{sec:strain field}. Finally, we discuss experimental signatures of hastatic order (Sec. \ref{sec:experiments}), the connection to previous theoretical results (Sec. \ref{sec:theory}), qualitatively suggest a generic non-Kramers Doniach phase diagram (Sec. \ref{sec:Doniach}), and summarize our conclusions in Sec. \ref{sec:conclusions}.

\subsection{Introduction to the $\Gamma_3$ non-Kramers doublet}

Rare earth and actinide ions have extremely strong spin-orbit coupling, making the total angular momentum, $J = L+S$ the relevant quantum number; this $2J+1$ degeneracy is then split by the crystalline electric fields into crystal field multiplets. Ions with odd and even numbers of $f$ electons therefore have half-integer and integer $J$, respectively. These two classes behave quite differently under the time-reversal operation $\theta$, as integer $J$ states are left invariant under double-time-reversal symmetry, $\theta^2 = +1$, while half-integer $J$ states invert, $\theta^2 = -1$. This difference manifests most clearly in Kramers theorem, which guarantees that half-integer $J$ states split at most to doublets under any time-reversal symmetry-preserving perturbation: such ions are called Kramers ions and their states Kramers doublets \cite{sakurai}. Integer $J$ states, however, may be split down to time-reversal invariant singlets, and these ions are called non-Kramers ions. If the crystal symmetry is sufficiently high, their states may form doublets and triplets. Non-Kramers doublets can be split by lowering the point group symmetry.

There are two types of non-Kramers doublets: Ising doublets that are magnetic along the local $\hat z$ axis and non-magnetic in the basal plane (tetragonal, hexagonal or trigonal symmetries); and essentially non-magnetic doublets (cubic symmetry). Here, we focus on the cubic case. The cubic $\Gamma_3$ doublet for $J=4$, which is relevant for Pr$^{3+}$ and U$^{4+}$, can be written as \cite{lealeaskwolf}:
\bea
|\Gamma_3 +\rangle & = & \sqrt{\frac{7}{24}} (|4\rangle +|-4\rangle)-\sqrt{\frac{5}{12}}|0\rangle\cr
|\Gamma_3 -\rangle & = & \sqrt{\frac{1}{2}} (|2\rangle + |-2\rangle).
\eea
in terms of the $|J_z\rangle$ eigenstates. This doublet is non-magnetic, with $\langle \vec{J} \rangle = 0$, but has a pseudospin $\frac{1}{2}$ degree of freedom that we describe with the Pauli matrices, $\vec{\alpha}$. $\alpha_1$ and $\alpha_3$, respectively correspond to the quadrupolar moments, $Q_{x^2-y^2} \propto \langle J_x^2-J_y^2\rangle$, and $Q_{3z^2-r^2} \propto \langle 3J_z^2-J(J+1)\rangle$, while $\alpha_2$ corresponds to the octupolar moment, $T_{xyz} \propto \langle \overline{J_x J_y J_z} \rangle$; the overline indicates symmetric permutation of indices. $\alpha_1$ and $\alpha_3$ couple to strains with the same symmetry, and their quadrupolar ordering would be a cooperative Jahn-Teller distortion that lowers the point group symmetry. $\alpha_2$ couples to a linear combination of strain and magnetic field both along the $[111]$ direction\cite{zawadowski}; these octupolar moments can also order, as proposed for PrV$_2$Al$_{20}$\cite{freyer17}.

Pr$^{3+}$ ions can fluctuate from $4f^2$ to either $4f^1$ or $4f^3$, both of which are Kramers configurations with only doublet and quartet states. Here, for simplicity we take the $4f^1$ $\Gamma_7$ excited doublet to be the relevant excited state,
\bea
|\Gamma_7 \pm\rangle = \sqrt{\frac{1}{6}}|\pm 5/2\rangle -\sqrt{\frac{5}{6}}|\mp 3/2\rangle,
\eea
although the $4f^3$ excited $\Gamma_6$ is perhaps more likely \cite{matsunami11}; the physics is the same.
These valence fluctuations involve conduction electrons in the $\Gamma_8$ symmetry, due to group-theoretic selection rules \cite{zawadowski, cox93}. $\Gamma_8$ is a quartet with both quadrupolar ($\Gamma_3$) and dipolar ($\Gamma_7$) degrees of freedom,
\bea
|\Gamma_8 a \pm \rangle & = & \sqrt{\frac{5}{6}} |\pm 5/2\rangle + \sqrt{\frac{1}{6}}|\mp 3/2\rangle; \cr
|\Gamma_8 b \pm\rangle & = & |\pm 1/2\rangle.
\eea
This atomic level diagram is shown in Fig. \ref{atomic_levels}.

Cubic symmetry renders the valence fluctuation Hamiltonian particularly simple \cite{zawadowski}:
\bea
H_{VF}(j) = V \sum_{\bk \alpha \mu}  \left[ \tilde{\mu}|\Gamma_3 \alpha\rangle \langle \Gamma_7 -\mu|\psi_{j 8\alpha \mu} + H.c.\right],
\eea
where $\mu$ and $\alpha$ label the magnetic and quadrupolar indices, respectively. The factor $\tilde{\mu} = \rm{sgn}(\mu)$ ensures that the $\Gamma_3$ doublet hybridizes with the two-particle states comprised of a conduction and $\Gamma_7$ $f$-electron; the latter two states form a singlet in magnetic ($\mu$) space and a doublet in quadrupolar ($\alpha$) space. 

The conduction electrons that directly hybridize with the Pr$^{3+}$ ion are $\Gamma_8$ Wannier functions, $\psi_{j 8 \alpha \mu}$, which possess the symmetries of a $\Gamma_8$ $f$-electron on the $f$ site. These may be constructed from any type of conduction electron that overlaps with the $f$-electron site, including simple plane waves. For simplicity, we consider a quartet of conduction electrons with $\Gamma_8$ symmetry. These could be a quartet of $e_g \otimes \frac{1}{2}$ $d$-electrons, which have $\Gamma_8$ symmetry. These have even parity in contrast to the odd parity $f$-electrons, and so must be overlapping from neighboring sites; see extensive recent work on this model for SmB$_6$, which has this conduction electron bandstructure \cite{alexandrov13,Vojta2014}. In this paper, we neglect the details of the overlap, which will generically be a complicated momentum dependent, spin-orbit coupled matrix, and consider only an onsite hybridization that leads to a momentum independent Kondo coupling.

A Schrieffer-Wolff transformation takes the valence fluctuation term, along with appropriate atomic and conduction terms, into a two-channel Kondo model\cite{zawadowski},
\bea
\label{2CKM}
H = \sum_{\bk \alpha \sigma} \epsilon_{\bk \alpha} c\dg_{\bk \alpha \sigma} c_{\bk \alpha \sigma} + J_K \sum_{j \sigma \alpha \beta} \psi\dg_{j \alpha \sigma} \vec{\alpha}_{\alpha \beta} \psi_{j \beta \sigma} \cdot \vec{\alpha}_{fj},
\eea
where $\sigma$ represents the $e_g$ conduction electron spin. As the Kondo couplings obey $J_{K\sigma} = J_K$, this is a completely degenerate two-channel Kondo lattice model. If the conduction bands are not degenerate everywhere in momentum space, the quadrupolar Kondo couplings, $J_{K}^{x}$ and $J_{K}^{z}$ may differ from the octupolar Kondo coupling, $J_{K}^{y}$; note that this anisotropy does not break cubic symmetry. The anisotropy is irrelevant, in the renormalization group sense, for the two channel Kondo impurity \cite{pang91}, and so we choose to neglect it here. The two-channel Kondo model will give rise to RKKY coupling between the $f$-electron quadrupole and octupole moments, also generically with $J_{RKKY}^{y} \neq J_{RKKY}^{x} = J_{RKKY}^{z}$ \cite{freyer17}. Again, we neglect this potential anisotropy.



\subsection{Relevant Pr-based materials}


Praseodymium is the simplest non-Kramers ion, as its $4f^2$ configuration has the lowest allowed $J = 4$, and in cubic symmetry, the $\Gamma_3$ doublet is the ground state doublet in about half of parameter space\cite{lealeaskwolf}. There are several Pr-based intermetallic materials where the ground state has been identified as $\Gamma_3$ by inelastic neutron scattering. The most promising are the ``1-2-20'' cage compounds Pr$T_2X_{20}$, where $T$ is a transition metal and $X$ = Al or Zn; these cubic ($Fd\bar{3}m$) materials have particularly strong Kondo coupling, as the Pr sit within Frank-Kasper cages of 16 Al or Zn atoms, allowing for strong $c$--$f$ hybridization \cite{sakai11,tokunaga13}. The Pr ions are then arranged on a diamond lattice. Considerable evidence exists for Kondo physics in these materials. At high temperatures, there is only partial quenching of the R $\ln 2$ entropy \cite{sakai11}, logarithmic scattering in the resistivity \cite{tsujimoto14}, relatively large hyperfine coupling \cite{tokunaga13}, enhanced effective masses \cite{shimura15}, and a Kondo resonance in photoemission \cite{matsunami11}. At low temperatures, most of these materials order, and then become superconducting at even lower temperatures. PrTi$_2$Al$_{20}$ and PrIr$_2$Zn$_{20}$ order ferro- and antiferro-quadrupolarly at $T_Q = 2$K\cite{sakai11, ito11,sato12, taniguchi16} and $0.11$K\cite{onimaru11, iwasa17}, respectively, while the ordering in PrV$_2$Al$_{20}$ \cite{sakai11, ito11} and PrRh$_2$Zn$_{20}$ \cite{onimaru12} is still undetermined. PrNb$_2$Al$_{20}$ does not order to the lowest temperatures, instead exhibiting non-Fermi liquid behavior\cite{higashinaka11,kubo15}. The quadrupolar order can be suppressed both with pressure (PrTi$_2$Al$_{20}$) \cite{matsubayashi12} and magnetic field [Pr(Ir,Rh)$_2$Zn$_{20}$\cite{onimaru11, onimaru12} and PrV$_2$Al$_{20}$] \cite{sakai11}, leading to extended non-Fermi liquid regions. Pressure enhances the superconductivity\cite{matsubayashi12}, which is almost certainly unconventional. The in-field phase diagrams are even more interesting, as there is an intermediate heavy Fermi liquid region in all three materials, sandwiched between the zero-field order and a fully polarized high field state where all Kondo physics is lost\cite{onimaru16b, yoshida17}.

PrPb$_3$ is another $\Gamma_3$ material with quadrupolar density wave ordering ($T_c = 0.35$K) that shows signs of heavy fermion behavior within the ordered phase at high fields, making it a candidate for hastatic order\cite{morin82,onimaru05,kawae06,sato10}.
The $\Gamma_3$ Heusler materials PrInAg$_2$\cite{yatskar96} and PrMg$_3$\cite{tanida06} exhibit non-Fermi liquid behavior, with extremely large Sommerfeld coefficients, but no clear phase transitions.

\section{A simple model for hastatic order}\label{sec:model}

While we are motivated by the rich physics of the $\Gamma_3$ doublet, in this paper, we consider a simpler model that captures much of the same physics. This simpler model allows us to fully explore the fundamental properties of hastatic order before looking at more complicated, realistic models in the future.

We begin with the two-channel Kondo model, \eqref{2CKM}, and add a nearest-neighbor Heisenberg term for the local moments in order to treat both magnetism and Kondo physics at the mean-field level \cite{andrei89,senthil03},
\bea
\label{2CKHM}
H = && \sum_{\bk \alpha \sigma} \epsilon_{\bk \alpha} c\dg_{\bk \alpha \sigma} c_{\bk \alpha \sigma} + J_K \sum_{j \sigma\alpha \beta} c\dg_{j \alpha \sigma} \vec{\alpha}_{\alpha \beta} c_{j \beta \sigma} \cdot \vec{\alpha}_{fj}\cr 
& & + J_H \sum_{\langle ij \rangle} \vec{\alpha}_{fi} \cdot \vec{\alpha}_{fj}.
\eea
This Kondo model is valid in any dimension, but it is only connected to the $\Gamma_3$ Anderson lattice model in three dimensions (3D). Nevertheless, the physics is often more transparent in the two-dimensional (2D) model, and so we will treat both 2D and 3D. While the 2D system can not order at any finite temperature, as both hastatic and quadrupolar orders break continuous symmetries, our mean-field picture neglects those fluctuations, and the main difference between our 2D and 3D models is the conduction electron density of states, and the complexity of the calculations. We present both results, but focus on the simpler 2D case.

\subsection{Conduction electron Hamiltonian}

While in realistic materials the c- and $f$-electrons are often on distinct sites, yielding a momentum dependent hybridization, here we assume that the c-electrons are $s$-electrons hybridizing with local moments at the same site, $c_{j\alpha \sigma} = \sum_{\bk} c_{\bk \alpha \sigma} \mathrm{e}^{-i \bk \cdot \bR_j}$. The conduction electron Hamiltonian is generically a matrix in channel ($\sigma$) and pseudospin ($\alpha$) space, spanned by the Pauli matrices, $\vec{\sigma}$ and $\vec{\alpha}$, respectively. Previous two-channel Kondo calculations have taken exactly degenerate conduction bands \cite{ zawadowski,jarrell96,tsvelik00,schauerte05,hoshino11}, making this matrix proportional to $\alpha_0\sigma_0$. This degeneracy is not required, nor particularly likely in real materials. We partially relax this condition to consider conduction electrons coming from two bands that are locally spin degenerate, but are not degenerate everywhere in $k$-space. In a 2D model with square symmetry, we take $p_x$ and $p_y$ orbitals, which generically have different hopping parallel and perpendicular to the orbital orientation. The resulting conduction electron dispersion is,
\bea
\epsilon^{(2D)}_{\bk} =-t\left[(1+\eta)(c_x +c_y)\alpha_0+(1-\eta)(c_x-c_y)\alpha_3 \right],
\eea
with $c_{x,y,z} = \cos k_{x,y,z} a$, and $a$ is the lattice constant. $\eta = 1$ recovers fully degenerate conduction electron bands. For a 3D model, we consider the $e_g$ doublet, $d_{x^2-y^2}, d_{z^2}$; here the cubic symmetry of $e_g$ is a more natural match for the $\Gamma_3$ doublet. For nearest neighbor hopping, we consider hopping between different orbitals on different sites, and obtain the dispersion\cite{Vojta2014}
\bea
\!\!\!\epsilon^{(3D)}_{\bk}\! &  =& \!\!\!-t[(1+\eta)(\!c_x\!+c_y\!+c_z\!)\alpha_0\cr
& &\!\!\!\!\!\!\!\!\!\!\! + \frac{\sqrt{3}(\eta -\! 1)}{2}(\!c_x\!-c_y\!)\alpha_1\!+\!\frac{\eta-\!1}{2}(\!c_x\!+c_y\!-2 c_z\!)\alpha_3].
\eea
For $\eta =1$, again we recover fully degenerate conduction electron bands that are diagonal in this basis.  

The full conduction electron band structure is then $\epsilon^{(2D,3D)}_\bk \sigma_0-\mu \alpha_0 \sigma_0$. We work in the canonical ensemble, where $\mu$ is adjusted to keep the total number of conduction electrons fixed,
\begin{equation}
n_c = \sum_{\alpha \sigma}\int d^dk f(\epsilon_{\bk\alpha}-\mu).
\end{equation}
Here $f(x)$ is the Fermi function.

Our conduction electrons couple both to channel symmetry breaking magnetic fields ($\sigma$), $H_c-g\mu_B \vec{B}\cdot \vec{\sigma} \alpha_0$, and pseudospin symmetry breaking strain fields ($\alpha$), $H_c-\kappa \vec{\epsilon}\cdot \vec{\alpha} \sigma_0$, where $\vec{\epsilon}$ is a vector of strains with the appropriate symmetries and $\kappa$ is the materials dependent coupling coefficient. If desired, the orbital degeneracy of the conduction electron bands can be broken by shifting the two bands by different chemical potentials, $\Delta \mu \alpha_3$, which effectively acts as a conduction electron strain term. This splitting will eventually destroy the quadrupolar Kondo effect, just as magnetic field destroys the usual Kondo effect. In a more realistic model, the Wannier functions screening the local moments are constructed out of partial wave expansions of both conduction electron orbitals and both spins at other sites, and so full screening can still occur even with a single conduction electron band\cite{chandra13}.

\subsection{Effect of magnetic field on realistic systems}
\label{section:mag_field}

An isolated $\Gamma_3$ doublet does not couple to magnetic field, however virtual fluctuations to excited crystal field states induce a $B^2$ coupling. As the crystal field splitting is typically on the order of 50K, relatively small magnetic fields will already mix in excited states, and for any realistic model we must consider their effect. Here, we take the excited state to be the $\Gamma_4$ triplet at energy $\Delta$, as in PrTi$_2$Al$_{20}$ \cite{sato12}. For simplicity, we neglect higher excited states and keep $\mu_B B < \Delta$. Including all excited states yields similar effects. The $\Gamma_4$ triplet for $J=4$ is
\bea
|\Gamma_4, a/b\rangle & = & \sqrt{\frac{7}{8}} |\pm 1\rangle + \sqrt{\frac{1}{8}} |\mp 3\rangle\cr
|\Gamma_4, c\rangle & = & \frac{1}{\sqrt{2}}\left(|4\rangle-|-4\rangle\right),
\eea 
and so mixes with the $\Gamma_3$ doublet in fields both along and perpendicular to the quantization axis.

With these crystal fields the $\Gamma_3$ doublet is split approximately quadratically in parallel magnetic field,
\begin{equation}
\Delta_3 = 6 \frac{(\mu_B B)^2}{\Delta} + O(B^4/\Delta^3),
\end{equation}
where $\Delta$ is in units of energy and $B || [001]$, see Fig. \ref{momentsinfield}(a). For fields along $[110]$ and $[111]$, the splitting is two and ten times smaller, respectively. This splitting competes with the Kondo effect and eventually destroys hastatic order. Here, $|\Gamma_3+\rangle$ mixes with the excited $\Gamma_4$ triplet, while $|\Gamma_3 -\rangle$ mixes only with the excited $\Gamma_5$ triplet. Therefore, $|\Gamma_3+\rangle$ is repelled by the excited states, and $|\Gamma_3 -\rangle$ remains at zero. Similarly, $|\Gamma_3+\rangle$ acquires a magnetic dipolar component along the field direction, while $|\Gamma_3-\rangle$ remains non-magnetic. While the $B=0$ doublet has two nonzero quadrupolar moments, $O_{3z^2-r^2}$ and $O_{x^2-y^2}$, and one nonzero octupolar moment, $T_{xyz}$, the $B > 0$ pseudo-doublet, for $B || [001]$, acquires $J_z$ dipolar and $O_{xy}$ quadrupole moments that grow linearly in field, for small $B/\Delta$. The pseudospin moments then correspond to: $\alpha_3\sim O_{3z^2-r^2} + J_z$, $\alpha_1 \sim O_{x^2-y^2}$ and $\alpha_2 \sim T_{xyz} + O_{xy}$. The in-field behavior of the dipolar and quadrupolar moments is shown in Figure \ref{momentsinfield} (b); the octupolar moment does not vary with field. Note that we plot the $\langle J_z \rangle$ associated with $|\Gamma_3\pm\rangle$ independently. More realistic crystal field schemes give slightly different coefficients, but the same nonzero quantities and functional dependencies. These field-induced dipolar moments are already well known, as they can be measured via neutron scattering to resolve quadrupolar order \cite{sato12,iwasa17}. Indeed, the magnetic field considered here could be external, or the internal exchange field; either one induces dipolar moments parallel to the local field.

\fight=8.5cm
\fg{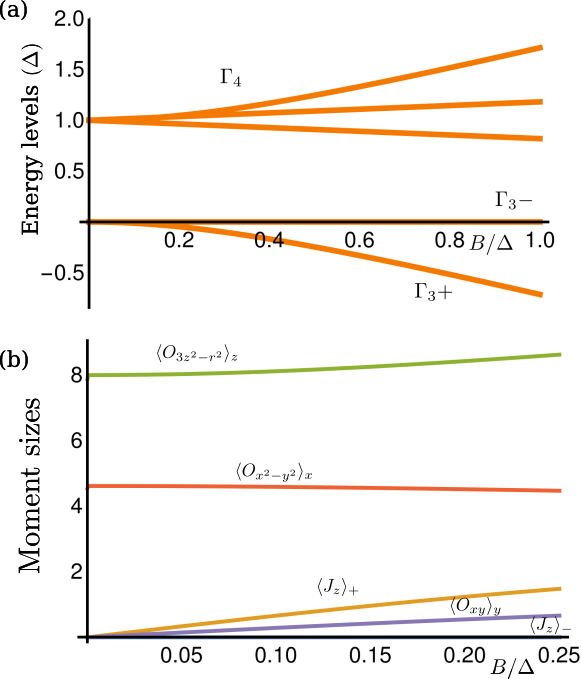}{momentsinfield}{(a) Splitting of the $\Gamma_3$ doublet in magnetic field ($B_z$), and its mixing with the excited $\Gamma_4$ triplet. (b) Single ion $\Gamma_3$ moments in field. This plot shows the magnitude of the $\Gamma_3$ moments as functions of the magnetic field along the z-direction. Note that these moments are the expectation values of the given multipolar operator within the appropriate components of the doublet. For example, $\langle O_{3z^2-r^2} \rangle_z = \frac{1}{2}(\langle \Gamma_3+|3J_z^2-J(J+1)|\Gamma_3 +\rangle-\langle \Gamma_3-|3J_z^2-J(J+1)|\Gamma_3-\rangle)$, where $|\Gamma_3 \pm\rangle$ are the new ground (+) and first excited (-) singlet states. Aside from $T_{xyz}$, which is constant in field, these are the only nonzero moments. Here, $\langle J_z \rangle_\pm = \langle \Gamma_3\pm|J_z|\Gamma_3\pm\rangle$, where $\langle J_z\rangle_- = 0$ due to the excited $\Gamma_5$ triplet being absent.}


\subsection{Large $N$ mean-field treatment}

In order to solve this model in a controlled mean-field theory, we introduce a fermionic representation for the pseudospins, $\vec{\alpha}_j =\frac{1}{2}\sum_{\alpha\beta}f\dg_{j\alpha} \vec{\alpha}_{\alpha \beta} f_{j\beta}$. $\vec{\alpha}$ also represents the $SU(2)$ pseudospin of the $\Gamma_3$ doublet, as it obeys the same symmetries as the conduction electron $\vec{\alpha}$. In this representation, both Kondo and Heisenberg terms become four fermion interactions. As these $f$-``electrons'' are really neutral spinons representing the local moments, we must also implement the constraint that each site is half-filled, $n_{fj} = 1$. We next take the $SU(N)$ limit, where the ground state multiplet has $N$ components, $\alpha = \pm \frac{1}{2},\ldots \pm \frac{N}{2}$, but remains half filled \cite{Coleman83}. In this limit, 
\begin{align}
H= & \sum_{\bk}\epsilon_{\bk \alpha} c\dg_{\bk \alpha \sigma} c_{\bk \alpha \sigma} - \frac{J_K}{N} \sum_{j}(f\dg_{j\beta}c_{j\beta \sigma} )(c\dg_{j\alpha\sigma} f_{j\alpha})\cr
&- \frac{J_H}{N} \sum_{\langle ij\rangle}(f\dg_{j\beta}f_{i\beta})(f\dg_{i\alpha}f_{j\alpha}) +\sum_j \lambda_j(f\dg_{j\alpha}f_{j\alpha} - N)\cr &-\mu \sum_\bk \left( c\dg_{\bk \alpha \sigma} c_{\bk \alpha \sigma}-\frac{N}{2} n_c\right).
\end{align}
We have introduced Einstein summation notation for $\sigma$ and $\alpha$ and rescaled $J_K$ and $J_H$ such that the entire Hamiltonian scales as $N$. The first line reproduces the two-channel Coqblin-Schreiffer model \cite{Coqblin69}, while first term on the second line gives the usual $SU(N)$ fermionic representation of an antiferromagnetic interaction \cite{Arovas88}. The second term on the second line is the half-filling constraint for the $f$'s, which must be enforced locally on each site. The final line implements the global fixing of the conduction electron density, $n_c$. Note that this particular large-$N$ theory does not capture superconductivity, either composite pair \cite{abrahams,catk,anders02,flint08, flint10,hoshino142} or quadrupolarly-mediated \cite{mathur, miyake, scalapino, bourbonnais}. Superconductivity is always a potential coexisting or competing ground state that we neglect here in order to focus on the stability and nature of hastatic order. A more complicated symplectic-$N$ large-$N$ calculation would incorporate both types of superconductivity\cite{flint08, flint10}, and will be considered in the future.

We next decouple the quartic terms with Hubbard-Stratonovich fields and take the saddle-point approximation in real space,
\begin{align}
V_{j\sigma} & =\frac{J_K}{N}\Big\langle f\dg_{j\alpha}c_{j\alpha \sigma} \Big\rangle; \cr
\chi_{Hij} &= \frac{J_H}{N}\Big\langle f\dg_{i\alpha}f_{j\alpha}\Big\rangle.
\end{align}
$V_{j\sigma}$ describes the local hybridization between conduction electrons and local moments at site $j$ in channel $\sigma$. $\chi_{Hij}$ describes ``antiferromagnetic'' correlations between local moment sites; for $\Gamma_3$, these are actually antiferroquadrupolar correlations, but we loosely use the term ``magnetic'' to generally represent the local moment multipolar order here. Note that the choice of fermionic spin representation means that we cannot capture long range magnetic or quadrupolar order in the large-$N$ limit. Instead, in the absence of hybridization, $\chi_{Hij}$ describes a spin, or really quadrupolar, liquid with $f$-spinons hopping from site to site with amplitude and phase given by $\chi_{Hij}$. In the $N=2$ limit, we expect that this quadrupolar liquid is unstable to quadrupolar order at lower temperatures, and take the quadrupole liquid as a proxy for the quadrupolar order that we cannot capture. At high temperatures above the development of $V_{j\sigma}$, this spinon hopping term describes $f$-electron hopping generated by hybridization fluctuations that otherwise would be beyond our mean-field picture.

The resulting mean field Hamiltonian is,
\begin{eqnarray}\label{H_final}
H  & = & \sum_{\bk}\epsilon_{\bk \alpha} c\dg_{\bk \alpha \sigma} c_{\bk \alpha \sigma} 
+ \sum_{j}[ V_{j\sigma}c\dg_{j\alpha\sigma} f_{j\alpha}+V^\ast_{j\sigma}f\dg_{j\alpha}c_{j\alpha \sigma}]  \cr
& & + \sum_{j}\lambda_{j} (f\dg_{j\alpha}f_{j\alpha}-N) -\!\! \sum_{\langle ij\rangle}[\chi_{Hij}f\dg_{i\alpha}f_{j\alpha} +\chi^\ast_{Hij}f\dg_{j\alpha}f_{i\alpha}]  \cr
& & + \sum_{j\sigma} \frac{N　|V_{j\sigma}|^2}{J_K}　+ \sum_{\langle ij\rangle} \frac{N　|\chi_{Hij}|^2}{J_H} + \sum_j \frac{N}{2}\mu n_c.
\end{eqnarray}
The mean-field solution is given by the saddle point values of all of the $V_{j\sigma}$, $\chi_{Hij}$, $\lambda_j$, and $\mu$; in principle, this problem is arbitrarily complicated. We simplify the problem by considering a set of possible mean-field ansatzes motivated by the strong coupling analysis in section \ref{strong}. In general, we assume that $\chi_{Hij} = \chi_H$ takes real, uniform values on nearest-neighbor bonds, and similarly that $\lambda_j = \lambda$ is uniform and real. All of our hybridization ansatzes have a uniform amplitude $\sum_\sigma |V_{j\sigma}|^2 = |V|^2$. We consider both uniform, $\langle V_{j\sigma}\rangle = V_\sigma$ and various N\'{e}el-type staggered, $\langle V\dg_{j}\rangle \vec{\sigma}\langle V_{j}\rangle = (-1)^{j_x + j_y} |V|^2$ hybridization ansatzes; any other spatial arrangements are less likely to occur on the hypercubic lattices we consider.

\fight=9cm
\fg{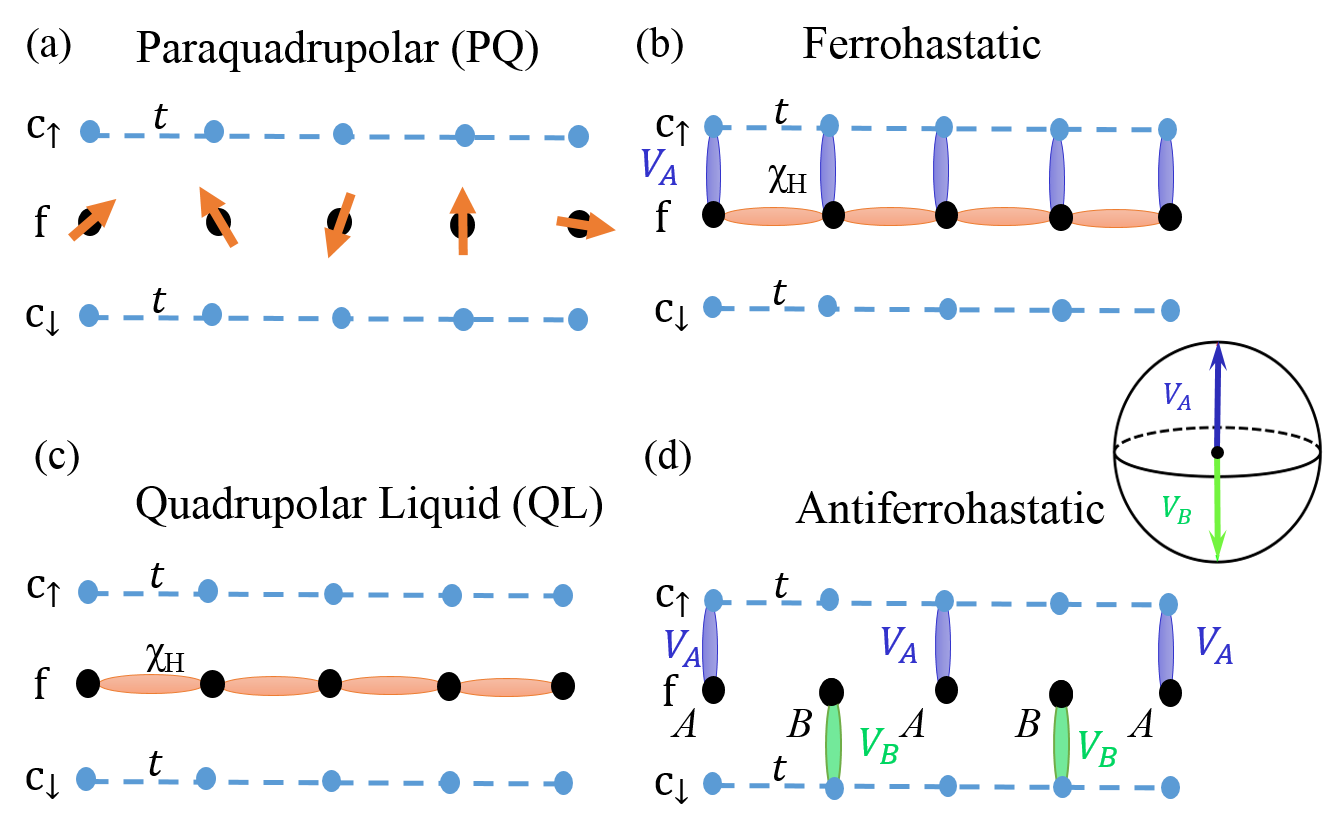}{cartoons}{One-dimensional cartoons of the mean-field ansatzes considered for this model. The upper and lower lines represent the spin-up and spin-down conduction electrons, while the middle line represents the local moments. The arrows represent the free quadrupolar moments, while the orange ovals represent quadrupolar valence bonds between local moments. There are four classes of states: (a) A completely disordered paraquadrupolar state; (b) a quadrupolar liquid state, with $f$-electron hopping between nearest neighbors; (c) a ferrohastatic order in which $f$ moments only hybridize with spin-up conduction electrons - this hybridization is represented by the blue ovals; (d) an antiferrohastatic order, in which the hybridization between $f$ moments and conduction electrons on different sublattices (A/B = blue/green ovals) are related by time reversal, i.e. $V_B = \theta V_A$.}

\subsection{Symmetries of the model}

After the Hubbard-Stratonovich transformation, but prior to the saddle-point approximation, our model has a number of symmetries that may be broken in any particular mean-field ansatz:
\begin{itemize}
\item{\emph{Translation and other lattice symmetries} for the square or cubic lattice. Any non-uniform hybridization ansatz will break some of these symmetries.}
\item{\emph{Particle-hole symmetry}, as we consider nearest-neighbor hopping on a hypercubic lattice; this symmetry will be broken by further neighbor hopping terms. Particle-hole symmetry implies that the physics is invariant under $n_c \rarrow 4 - n_c$, or $\mu \rarrow -\mu$.}
\item{\emph{$SU(2)$ pseudospin symmetry ($\vec{\alpha}$)}, which protects the non-Kramers doublet degeneracy. Physically, this symmetry is the cubic crystal symmetry, and can be broken by coupling to stresses or external fields, which will eventually kill the Kondo effect.}
\item{\emph{$SU(2)$ channel symmetry ($\vec{\sigma}$)}, which protects the degeneracy of the conduction electron bands. Physically, spin is the channel index, so this is the spin rotational symmetry. The hybridization, $V\dg_{j\sigma} = (V^\ast_{j\uparrow}, V^\ast_{j\downarrow})$ is an $SU(2)$ spinor. Condensing this spinor into a mean-field ansatz automatically breaks this $SU(2)$ symmetry.}
\item{\emph{Time-reversal symmetry}, which affects the conduction electrons and $f$-spinons differently. Our $f$-spinons here are spinless fermions from the point of view of time-reversal $\theta$, transforming as $f_\alpha \rarrow f\dg_\alpha$, with $\theta^2 = 1$. By contrast, our conduction electrons are Kramers degenerate, and transform as $c_{j\alpha} \rarrow i \sigma_2 c_{j\alpha}\dg$, with $\theta^2 =-1$. As the hybridization, $V_{j\sigma}$ connects non-Kramers f-spinons and Kramers c-electrons, it is itself Kramers-like, and transforms as $V_{j\sigma} \rarrow -\mathrm{sgn}(\sigma)V_{j,-\sigma}\dg$; with $\theta^2 =-1$. The resulting composite fermions, $\tilde{f}_{\sigma \alpha} \sim V_{\sigma}f_\sigma$ now behave like Kramers electrons. However, once we condense $V_{j\sigma}$, they are no longer operators, and instead transform as complex numbers, $V_{j\sigma} \rarrow V^\ast_{j\sigma}$, due to the complex conjugation in the definition of time reversal. Therefore, any mean-field ansatz for $V_{j\sigma}$ breaks time-reversal symmetry, although time-reversal plus a lattice symmetry may restore it, as in traditional antiferromagnets.}
\item{\emph{Gauge symmetries,} of which there are two in the problem: the original electromagnetic gauge symmetry, $c_j \rarrow c_j \mathrm{e}^{i\phi_j}$, and an emergent gauge symmetry, $V_j \rarrow V_j \mathrm{e}^{i\beta_j}$, $f_j\rarrow f_j \mathrm{e}^{-i\beta_j}$ and $\chi_{Hij} \rarrow \chi_{Hij} \mathrm{e}^{i(\beta_i - \beta_j)}$. The development of hybridization locks together the two gauge fields, which couples the neutral $f$-spinons to the external field and thus turns them into charge-$e$ heavy electrons \cite{Coleman2005}. For the rest of the paper, we will call these spinons $f$-electrons, in anticipation of this gauge field locking.}
\end{itemize}

Any mean-field ansatz with nonzero hybridization necessarily breaks some of the above symmetries. The channel symmetry is always broken, one way or another, which reflects the essential nature of hastatic order as a channel symmetry breaking heavy Fermi liquid. The two types of mean-field ansatzes with zero hybridization, the quadrupolar liquid ($\chi_H \neq 0$) and paramagnetic ($\chi_H = V_\sigma = 0$) phases break no symmetries.

\subsection{Moments and coupling to external fields}
\label{sec:moments}

Both the conduction and $f$-electrons can develop moments corresponding to certain broken symmetries. The conduction electrons have both spin ($\vec{\sigma}$) and quadrupolar moments ($\vec{\alpha}$), and in fact form a $\Gamma_8$ quartet. The generic conduction electron moment is
\begin{equation}
m_{c,j,a,s} = \langle c\dg_{j}\alpha_a \sigma_s c_j\rangle,
\end{equation}
where there are fifteen total moments: three dipoles, five quadrupoles, and seven octupoles \cite{Shiina97}. The irreducible representations and conjugate fields of each of the dipolar and quadrupolar moments are listed in Table I.
\begin{table}[h]
\label{cmoments}
\centering
\begin{tabular}{ cccc } 
 \hline
Operator & Moment & Conjugate field & Symmetry \\ \hline 
$\alpha_0 \sigma_1$ & $S_x$ & $B_x$ & $\Gamma_{4u} = T_{1u}$\\
$\alpha_0 \sigma_2$ & $S_y$ & $B_y$ & \\
$\alpha_0 \sigma_3$ & $S_z$ & $B_z$ & \\
\hline
$\alpha_1 \sigma_0$ & $O_{x^2-y^2}$ & $\epsilon_{xx-yy}$ & $\Gamma_{3g} = E_g$ \\
$\alpha_3 \sigma_0$ & $O_{3z^2-r^2}$ & $\epsilon_{zz}$ & \\
\hline
$\alpha_2 \sigma_1$ & $O_{yz}$ & $\epsilon_{yz}$ & $\Gamma_{5g} = T_{2g}$\\
$\alpha_2 \sigma_2$ & $O_{xz}$ & $\epsilon_{xz}$ & \\
$\alpha_2 \sigma_3$ & $O_{xy}$ & $\epsilon_{xy}$ & \\
\hline
$\alpha_2 \sigma_0$ & $T_{xyz}$ & $B_{[111]} \epsilon_{[111]}$ & $\Gamma_{2u} = A_{2u}$\\
\hline
$\alpha_3 \sigma_3$ & $T_z^\alpha$ & $\epsilon_{zz} B_z$ & $\Gamma_{4u}$\\
\hline
\end{tabular}
\caption{Table of conduction electron dipole and quadrupole moments, as well as the two octupoles relevant to our discussions. Here, the symmetries and physical conjugate fields of each moment are also given, where $u/g$ refers to odd/even under time-reversal symmetry, not the usual parity.}
\end{table}

The $f$-electron has three possible moments, $\vec{m}_{f,j} = \langle f_j\dg \vec{\alpha} f_j\rangle$, which we take to be the quadrupolar and octupolar moments of the $\Gamma_3$ doublet. These moments couple linearly to the appropriate strains, $\epsilon_{xx-yy}$ to $m_{f,1}$ and $\epsilon_{zz}$ to $m_{f,3}$, while the octupolar moment, $m_{f,2}$ couples to the product of strain and magnetic field along $[111]$ \cite{zawadowski}. If we include excited crystal field levels, magnetic fields along the $z$-axis couple as $-\sum_j 6 (\mu_B B_z)^2/\Delta \; f\dg_{j+}f_{j+}$. For the induced moments, see section \ref{section:mag_field}.

\section{Strong coupling limit of the two-channel Kondo model}
\label{strong}

Before going in depth into the mean-field analysis, let us motivate our different hastatic orders by reexamining the strong-coupling limit of the two-channel Kondo lattice model\cite{schauerte05,zawadowski}. In this limit, we drop the Heisenberg term, as it is a small perturbation. As $J_K/t \rarrow \infty$, the Kondo singlet becomes completely local, and is essentially an on-site valence bond between the local moment and a conduction electron on site. If we start from the $n_c = 0$ limit, each conduction electron we add immediately forms a Kondo singlet, until we reach quarter-filling ($n_c = 1$), where every local moment is bound up into a singlet. Below quarter-filling, we have excess local moments, while above quarter-filling we have excess conduction electrons on the background of a lattice of spin-ful Kondo singlets. Below quarter-filling, the local moment behavior is largely the same as in the single-channel Kondo lattice \cite{sigrist91}. The phase diagram will be symmetric above and below half-filling due to the particle-hole symmetry.

First, we consider the relative stability of hastatic order and quadrupolar order in this strong coupling limit. The local (single-site) energy difference is sufficient: the Kondo singlet is essentially a valence bond between local moment and conduction electron, $\frac{1}{\sqrt{2}}\left[\vert c_{\sigma +}\dg f_-\dg\rangle - \vert c_{\sigma -}\dg f_+\dg\rangle\right]$, with energy $-J_KS(S+1) = -3J_K/4$. Here, $\pm$ represent the pseudospin ($\alpha$) degrees of freedom. The local quadrupolar state consists of the local moment antiparallel to any conduction electrons on site; importantly, unlike the Kondo singlet, the local moment is \emph{frozen}. The lowest energy occurs when there are two conduction electrons on site, both anti-parallel to the local moment, $\vert c_{\uparrow -}\dg f_+\dg c_{\downarrow -}\dg\rangle$, with energy $-2J_K S^2 = -J_K/2$. Thus, hastatic order is always favored for sufficiently strong coupling.

\begin{figure}[ht]\centering
	\includegraphics[width=.9\columnwidth]{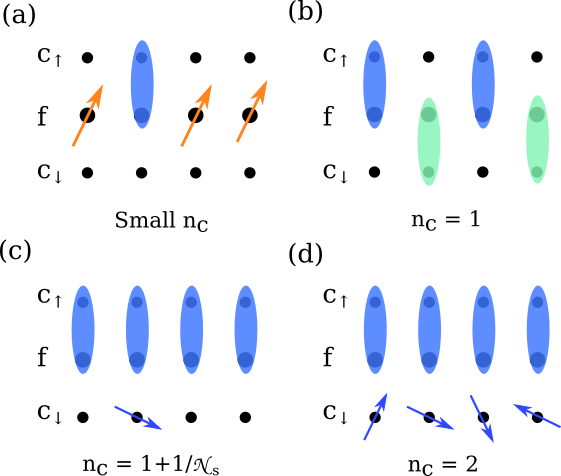}
\caption{One dimensional cartoons of the strong coupling limit at several values of the conduction electron density. The spins here are the quadrupolar moments of the local moments (orange) and conduction electrons (blue). Blue (green) ovals represent Kondo singlets that carry channel $\sigma = \uparrow$ ($\sigma = \downarrow$). (a) At small $n_c$, kinetic energy favors ferroquadrupolar order of the unbound local moments. (b) At $n_c = 1$, superexchange between the Kondo singlets leads to antiferrohastatic order. (c) Just above $n_c = 1$, adding a single conduction electron makes the Kondo singlets ferrohastatic to maximize the kinetic energy. (d) At half-filling, again ferrohastatic order maximizes the kinetic energy. }
\label{fig:strong_coupling}
\end{figure}

Now we turn to the nature of the hastatic order. A few limits of the lattice behavior are well-understood \cite{schauerte05,zawadowski}, as shown in Fig. \ref{fig:strong_coupling},
\begin{itemize}
\item{\emph{Small $n_c$:} For $n_c \ll 1$, the Kondo singlets form a dilute gas of spin-ful bosons. The remaining local moments order ferroquadrupolarly to maximize the kinetic energy of the bosons; this behavior is identical to the single-channel Kondo model \cite{sigrist91,schauerte05}. Two neighboring Kondo singlets gain superexchange energy, $O(t^2/J_K)$ if they are antiparallel, so this region is likely to be antiferrohastatic, in addition to the ferroquadrupolar order of the unscreened local moments. Note that this competing state is absent from our mean-field treatment.}
\item{\emph{Quarter-filling:} With a Kondo singlet at each site, this state is a Kondo insulator, with a remaining channel degree of freedom. As in the infinite $U$ Hubbard model, the $2^{\mathcal{N}_s}$ degeneracy is broken by channel superexchange $O(t^2/J_K)$, leading to a channel Heisenberg model. For our hypercubic lattices, the ground state will be a N\'{e}el type antiferrohastatic ground state.}
\item{\emph{Near quarter-filling:} Adding a single conduction electron to the quarter-filled state immediately turns it ferrohastatic in order to maximize the kinetic energy of the electron, as a variant of the Nagaoka ferromagnetism in the Hubbard model \cite{Nagaoka66}. As $t$ increases, we expect the antiferrohastatic state to extend for $n_c > 1$, by analogy with the Hubbard model. However, the behavior here is not symmetric about quarter-filling. Removing a single conduction electron leaves a single unbound local moment. This local moment moves by conduction electron hopping that moves the Kondo singlets; this process is not affected by the nature of the hastatic order, and superexchange will continue to favor antiferrohastatic order.}
\item{\emph{Half-filling:} Exactly at half-filling, we have a full complement of Kondo singlets, and exactly half a band of conduction electrons. While superexchange ($\sim t^2/J_K$) favors the antiferrohastatic state, the kinetic energy ($\sim t$) will be maximized in the fully decoupled ferrohastatic state, and so we expect ferrohastatic order here.}
\end{itemize}

In the end, we can assemble a simple picture of the hastatic behavior motivated by these limits. In this paper, we neglect non-hastatic behavior, like the small $n_c$ ferroquadrupolar order and potential superconductivity at intermediate coupling. We expect a N\'{e}el-like antiferrohastatic phase below quarter-filling, and extending above it for a finite range, followed by a transition to ferrohastatic order, which is stable out to half-filling. In the hypercubic models studied here, these are likely to occupy most of the phase space. One could study more complicated orders by adding further neighbor hoppings, or by studying frustrated lattices like the triangular lattice. We focus on the ferrohastatic and N\'{e}el-like antiferrohastatic orders in this paper, and indeed the above picture mostly agrees with our mean-field phase diagrams, with small differences at low filling.

\section{Mean-field Ansatzes}\label{sec:mfansatzes}
Here we describe several simple mean-field ansatzes for hastatic order, leaving the detailed description of their physical properties for later sections.

\subsection{Ferrohastatic Order}

The most straightforward ansatz is to assume that the hybridization is uniform, $V_{j\sigma} = V_\sigma$. The hybridization does not break any lattice symmetries, but does break both single and double time-reversal symmetries, as well as the $SU(2)$ channel symmetry (spin-rotational symmetry), as it couples $f$-electrons with conduction electrons of only one spin polarization. If this spin polarization is ``up'', only the spin up conduction electrons hybridize, and we obtain two bands of heavy up electrons and one band of light down electrons.

\begin{figure}[ht]\centering
	\includegraphics[width=.9\columnwidth]{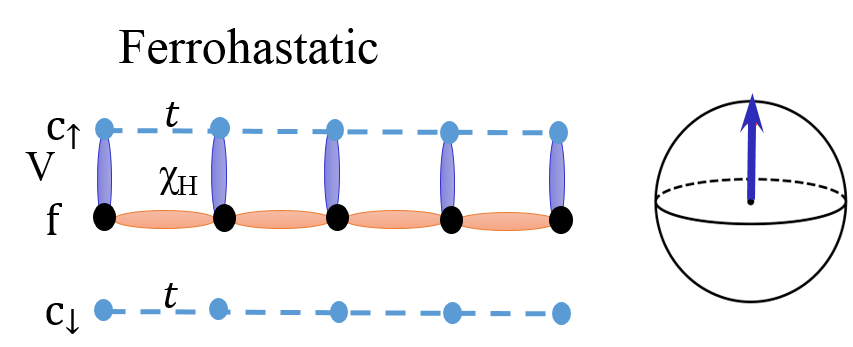}
\caption{\emph{Left}: A simple one dimensional cartoon of ferrohastatic order, where the top and bottom rows represent spin up and spin down conduction electrons, and the middle row represents the quadrupolar local moments. In ferrohastatic order, only one spin species of conduction electrons hybridize (blue ovals), while both the c and $f$-electrons can disperse within their row, with the $f$-electron dispersion generated by the Heisenberg coupling (orange ovals). \emph{Right:} The hybridization is a spinor that can point anywhere in $SU(2)$ space. For this ansatz, it points to the north pole of the Bloch sphere.}
\label{fig:uniform_cartoon}
\end{figure}

In this ansatz, the Hamiltonian in eq. \eqref{H_final} becomes,
\begin{align}
H  & = \frac{1}{\mathcal{N}_s}\sum_\bk (c\dg_{\bk \alpha \uparrow}, c\dg_{\bk \alpha \downarrow}, f\dg_{\bk\alpha} ) 
\left(\begin{array}{ccc}
\epsilon_{\bk\alpha} & 0 & V_\uparrow \\
0 & \epsilon_{\bk\alpha} & V_\downarrow \\
V^\ast_\uparrow & V^\ast_\downarrow & \epsilon_{f\bk} 
\end{array}\right) 
\left(\begin{array}{c}
c_{\bk \alpha \uparrow} \\
c_{\bk \alpha \downarrow} \\
f_{\bk\alpha}
\end{array}\right) \cr
& +  \frac{N}{J_K} \sum_\sigma|V_{\sigma}|^2+  \frac{z N}{2J_H} |\chi_{H}|^2 -\lambda N + \frac{N}{2} \mu n_c,
\label{eq:Huniform3by3}\end{align} 
where we have divided the Hamiltonian by the total number of sites, $\mathcal{N}_s$, $\frac{1}{\mathcal{N}_s}\sum_\bk = \int \frac{d^d k}{(2\pi)^d}$, and $z$ is the coordination number of the lattice: $z = 4,6$ in 2D and 3D, respectively. The ``bare'' $f$-electron dispersion is $\epsilon_{f \bk}\equiv \lambda - 2 \chi_H \sum_\eta \cos(\bk\cdot\vec{\eta})$, where $\vec{\eta}$ are the $z/2$ nearest-neighbor locations with positive coordinates. In the 2D model, the two $\alpha$ states do not mix and the Hamiltonian matrix is block diagonal, allowing for the representation in equation \ref{eq:Huniform3by3}. In 3D, with non-degenerate conduction electron bands ($\eta \neq 1$), the Hamiltonian is slightly more complicated, but the physics is the same. This Hamiltonian can be diagonalized to give the one light and two heavy doubly-degenerate bands\cite{tsvelik00},
\begin{align}\label{uniform bands}
\omega_{\bk\alpha}\!= \!\epsilon_{\bk \alpha}, \frac{\epsilon_{\bk\alpha}+\epsilon_{f\bk}}{2}\! \pm\! \sqrt{ \Big(\frac{\epsilon_{\bk\alpha}-\epsilon_{f\bk}}{2} \Big)^2\!+\!\sum_\sigma |V_\sigma|^2 }.
\end{align}

\begin{figure}[h]\centering
   \begin{minipage}{0.7\columnwidth}
     \includegraphics[width=\columnwidth]{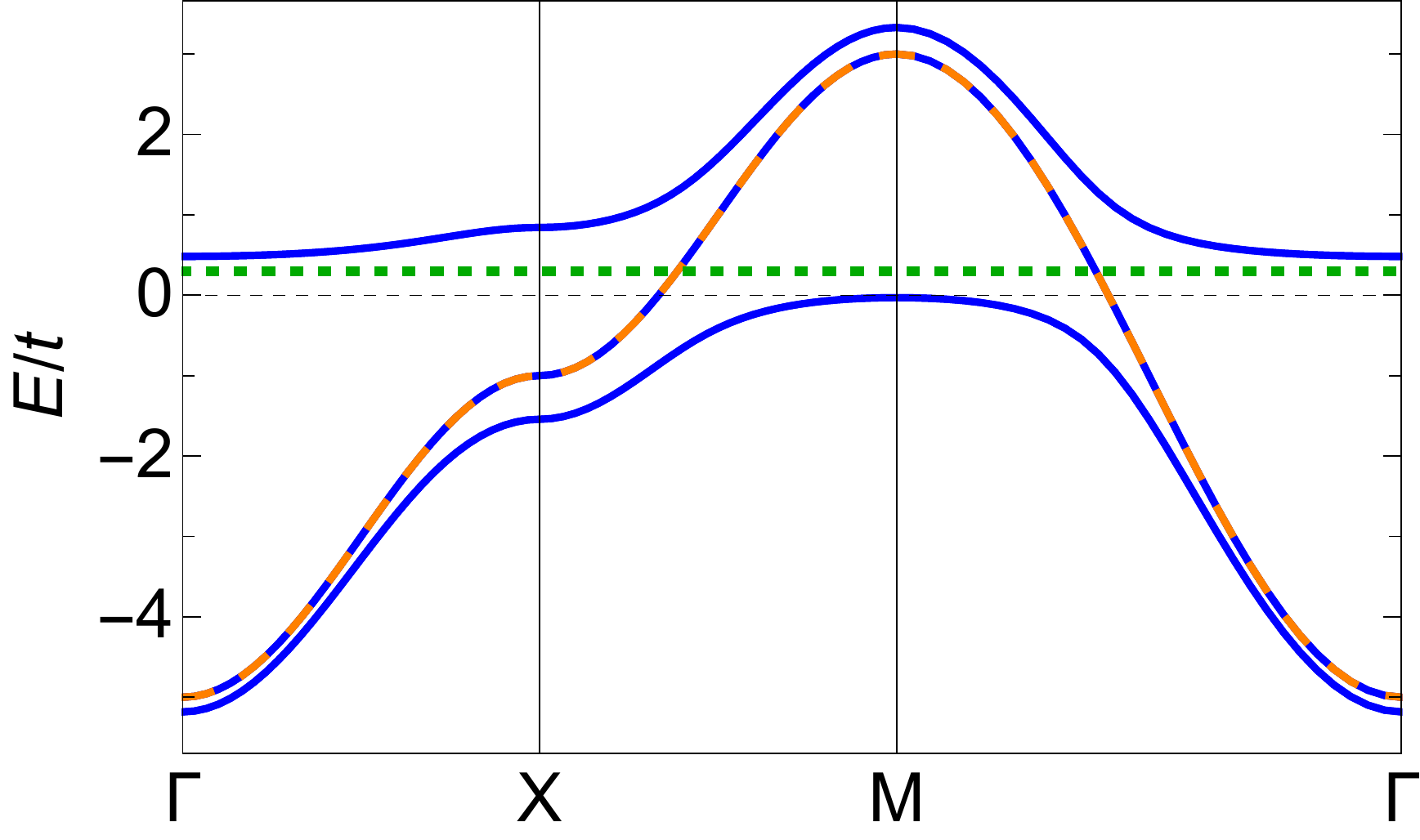}
   \end{minipage}
   \begin {minipage}{0.25\columnwidth}
     \includegraphics[width=\columnwidth]{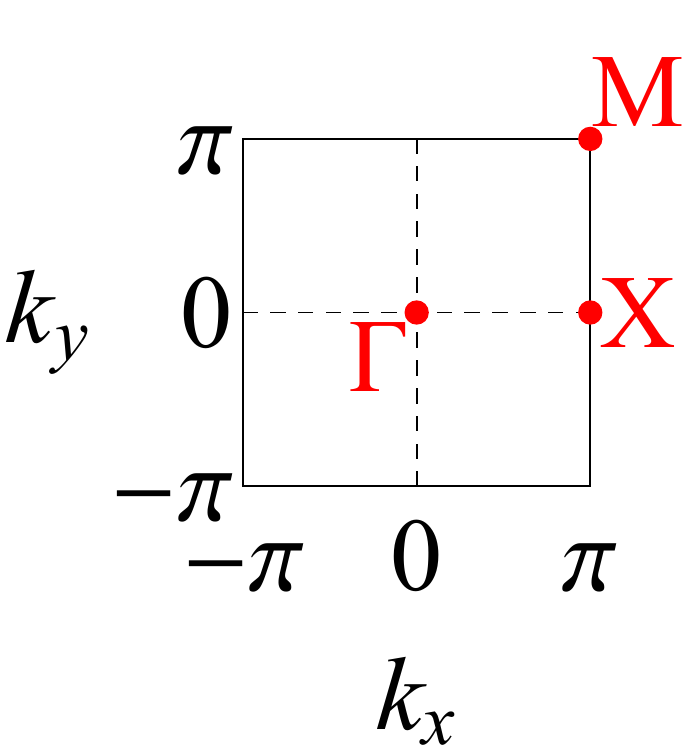}
   \end{minipage}
\caption{Bandstructure along high symmetry lines in ferrohastatic order. Before hybridization, the four bare conduction electron bands(orange dashed lines) have two-fold spin degeneracy and two-fold pseudospin degeneracy, while the bare $f$-electron bands (green dotted line) have only two-fold pseudospin degeneracy. After hybridization, there are six bands (blue) with two unhybridized. This plot is for $V_\uparrow=1, V_\downarrow=0, \lambda=0.3,\chi=0,\mu=1, \eta = 1$. The right figure shows the first Brillouin zone and high symmetry points.}
\label{fig:uniform}
\end{figure}

The band structure is $SU(2)$ invariant and thus independent of the direction of $V_\sigma$, while the eigenvectors, which capture the spin structure of the bands clearly depend on $V_\sigma$. As one conduction band always remains unhybridized, if the original conduction electron bandstructure is metallic, ferrohastatic order will be too. An example bandstructure is shown in Fig. \ref{fig:uniform}.

Aside from the breaking of channel symmetry, ferrohastatic order behaves identically to the usual Kondo effect, and will have similar signatures. In particular, the interaction between the Kondo effect and quadrupolarly mediated superconductivity should be identical. In section \ref{sec:multipolarmoments}, we discuss the moments and susceptibilities associated with the broken channel symmetry, while section \ref{sec:experiments} summarizes the experimental signatures.

\subsection{Antiferrohastatic Order}

While the ferrohastatic ansatz breaks time-reversal, but no lattice symmetries, we also want to consider hybridization ansatzes that break lattice symmetries. In particular, we are interested in antiferromagnetic versions of hastatic order, where time-reversal symmetry is broken, but the ground state returns to itself under time-reversal followed by a lattice symmetry operation. 

\begin{figure}[h]\centering
	\includegraphics[width=\columnwidth]{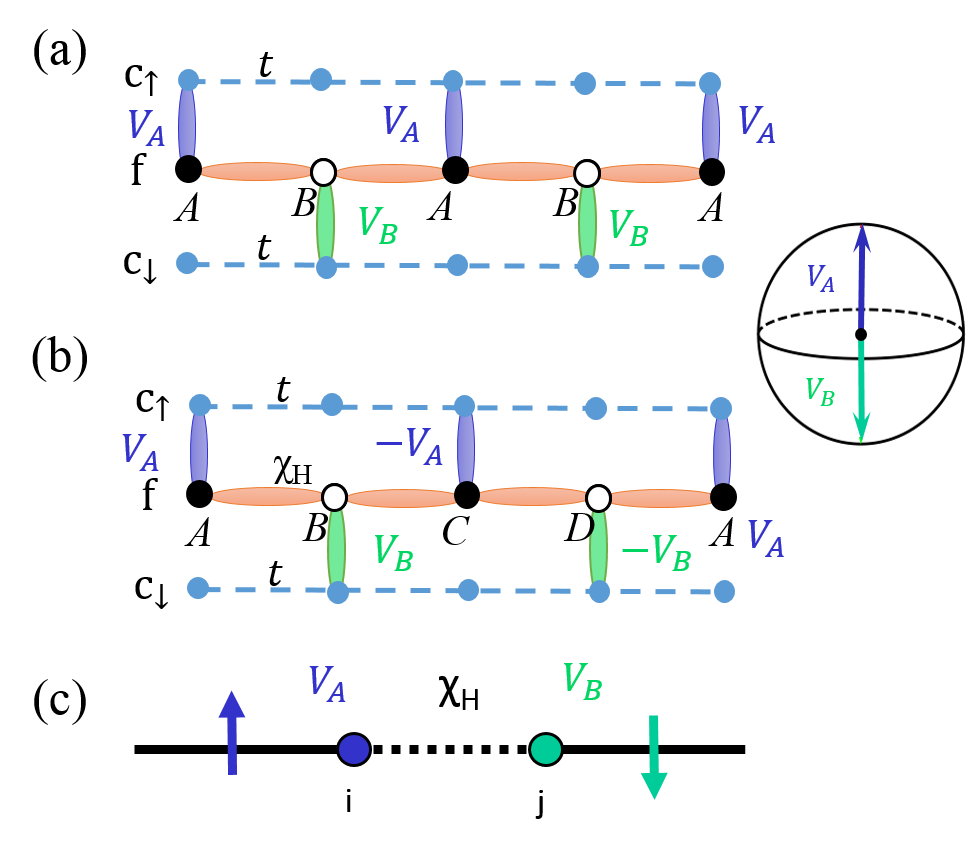}
\caption{A one dimensional cartoon of (a) two-sublattice (2SL) antiferrohastatic order, where the hybridization on sublattice $B$ is the time reverse of that on sublattice $A$; (b) four-sublattice(4SL) antiferrohastatic order where the hybridizations on the four sublattices are related by time reversal symmetry as $V_B = \theta V_A, V_C = \theta V_B$ and $V_D = \theta V_C$. (c) is a schematic illustration of the spin flip hopping of conduction electrons moving from a site in sublattice $A$ to a site in sublattice $B$. At $A$, a spin-up conduction electron hybridizes with the local $f$ moment. It then hops, as an $f$-electron to $B$, where it converts back to a \emph{spin-down} conduction electron.}
\label{fig:staggered_cartoon}
\end{figure}

One might naively expect that we can produce a N\'{e}el-like staggered hybridization by separating our lattice into two sublattices, defining the hybridization on sublattice A as $V_A$, and the hybridization on sublattice B as the time-reversed object, $V_B = \theta V_A$, as in Fig. \ref{fig:staggered_cartoon} (a). However, the spinor nature of the hastatic order parameter plays an essential role, and our intuition from vector antiferromagnets fails. A second time-reversal operation takes $\theta^2 V_A = -V_A$. Indeed, it is only after four time-reversal operations that we recover $\theta^4 V_A = V_A$. In order to write down an ansatz invariant under a combination of time-reversal ($\theta$) and a lattice symmetry ($S$), $P =S \theta$, we require a four-sublattice ansatz, as in Fig. \ref{fig:staggered_cartoon} (b),
\begin{align}
V_B \!=\! \theta V_A, \; V_C \!=\! \theta^2 V_A \!=\! -V_A,\; V_D \!=\! \theta^3 V_A = -V_B.
\end{align}
We can, of course, remove the extra sign in $V_C$ and $V_D$ by performing a gauge transformation on C and D sites. If there is no $f$-electron hopping between sublattices ($\chi_H = 0$), the mean-field Hamiltonian is invariant under this transformation, and we can consider a two-sublattice ansatz where time-reversal symmetry is represented by the usual time-reversal, $\theta$ followed by a staggered gauge transformation. The requirement to combine symmetry and gauge operations to reveal the true symmetry of the ground state is analogous to the use of projective symmetry groups in spin liquids \cite{Wen02}. However, $f$-electron hopping between sublattices ($\chi_H \neq 0$) causes the two-sublattice ansatz to truly break time-reversal symmetry, albeit subtly via the signs of the hybridization spinors. While no single-site observables break time-reversal symmetry, the bandstructure must do so through an emergent spin-flip hopping. If a conduction electron hybridizes at a site on sublattice A, hops as an $f$-electron to site B, and turns back into a conduction electron via hybridization at site B, it will flip its spin, see Fig. \ref{fig:staggered_cartoon}(c). As all of the four sublattice cases break additional lattice symmetries if $\chi_H = 0$, and the two sublattice case breaks time-reversal, when there is $f$-electron hopping, an extra symmetry beyond translation \emph{must} be broken. We consider both two (2SL) and four sublattice (4SL) ansatzes, and both generically are found in the phase diagrams.

\begin{figure}[h]\centering
	\includegraphics[width=\columnwidth]{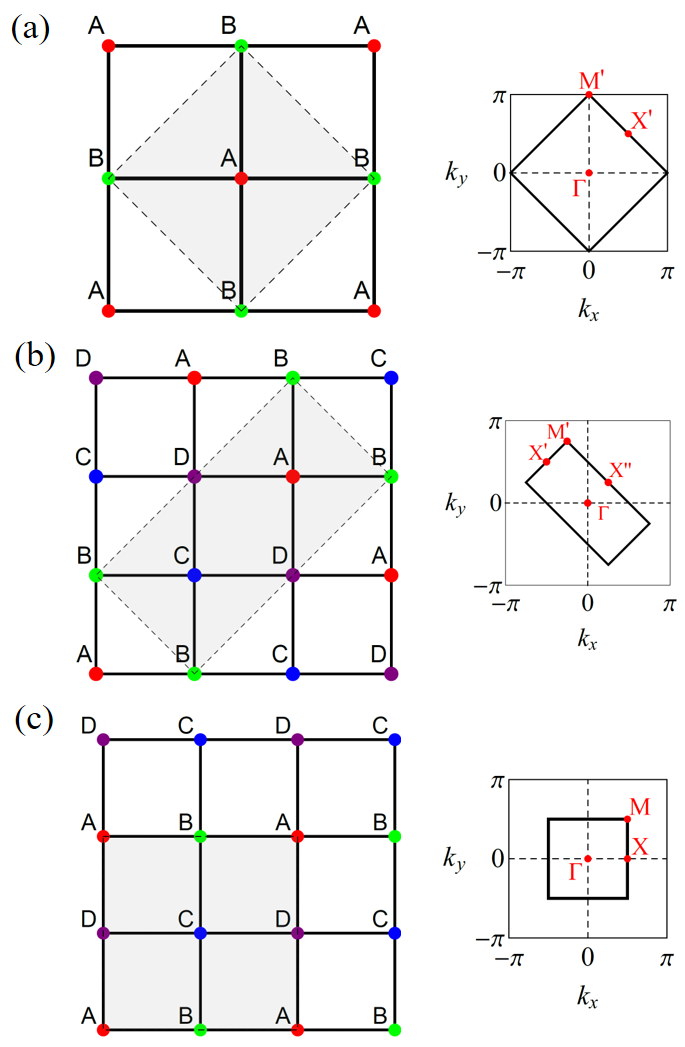}
\caption{The lattice structure (left) and Brillouin zone (right) for (a) two sublattice (2SL) staggered ansatz, which breaks time-reversal and lattice translation symmetry but preserves inversion and $C_4$ rotation symmetry; (b) four sublattice [4SL(1)] staggered ansatz, which breaks time-reversal, lattice translation and inversion symmetries; (c) four sublattice [4SL(2)] staggered ansatz, which breaks time-reversal, lattice translation symmetry and rotation symmetry, but preserves inversion symmetry. }
\label{fig:staggered}
\end{figure}

In 2D, there are two ways of arranging the four sublattices (ABCD) such that the hybridization moments, $V\dg\vec{\sigma}V$ form the same N\'{e}el order, but the signs either alternate or form uniform stripes along the $\hat x$ direction. We discuss the 3D cases in section \ref{sec:dimensionality}. The first ansatz, which we call 4SL(1) is shown in Fig. \ref{fig:staggered}(b), with a unit cell that is quadrupled along the $\hat x$ direction. This ansatz breaks time-reversal and lattice translation symmetry, but is invariant under time-reversal followed by translation by one site along $\hat x$. The Bravais lattice is rectangular, with a rotated and compressed Brillouin zone, as shown in Fig. \ref{fig:staggered}(b). The ansatz breaks inversion symmetry subtly due to the relative signs of the hybridization spinors. The second ansatz [4SL(2)] places ABCD around a single plaquette, as shown in Fig. \ref{fig:staggered}(c). The unit-cell is doubled along both $\hat x$ and $\hat y$, and the Brillouin zone remains square, as shown in Fig. \ref{fig:staggered} (c). Here, the ansatz is invariant under time-reversal followed by a four-fold rotation, but breaks translation and rotation symmetries, while preserving inversion.

\subsubsection{Kramers degeneracy}

Before hybridization, there are two Kramers degenerate conduction electron bands ($\sigma = \uparrow,\downarrow$, $\alpha = \pm$), and two non-Kramers ``singlet'' f-bands ($\alpha = \pm$). Hybridization mixes these Kramers and non-Kramers bands; however, if time-reversal is preserved in some fashion, the total number of Kramers degenerate bands must be preserved. The 2SL ansatz really does break time-reversal, and thus the Kramers degeneracy of the bands is lost, even at the $\Gamma$ point. The 4SL ansatzes preserve the Kramers degeneracy, however the Kramers pairs are not co-located in momentum space. While the 4SL ansatzes break time reversal symmetry locally, they preserve an anti-unitary time-reversal-like symmetry, $P = S\theta$, with $S$ being a lattice transformation. By way of analogy, in a simple square N\'{e}el antiferromagnet, $S$ is a translation by one site along $x$. The presence of corresponding $P$ symmetries for 4SL(1) and 4SL(2) imply Kramers degenerate eigenstates at time-reversal invariant momenta like the $\Gamma$ point. Away from these special points, the Kramers pair of a state at $\bk$ lies at $S\bk$,
and so for generic momenta the degeneracy at fixed $\bk$ is lifted. A simple antiferromagnet has doubly degenerate bands throughout the Brillouin zone as $P\bk = -\bk$, which is then mapped back to $\bk$ by inversion symmetry. For 4SL(1), the $S$ operation is again translation by one site along $x$, and so $P\bk = -\bk$; as 4SL(1) lacks inversion symmetry, there is no way to map this state back to $\bk$, and so the bands are not doubly degenerate at generic momenta. For 4SL(2), $S$ is a $C_4$ rotation about the middle of a plaquette, which means $P\bk =-R_{\pi/2} \bk$, where $R_{\pi/2}$ is a $C_4$ rotation matrix. Therefore, while the 4SL(2) ansatz has inversion symmetry, it still does not have a distinct unitary operation that can map $-R_{\pi/2} \bk$ back to $\bk$, and thus does not have Kramers degenerate bands. Note that the above discussion holds for generic $\chi_H \neq 0$, but for $\chi_H=0$, both 4SL ansatzes are equivalent to the two sublattice one via a gauge transformation. This version has inversion, and the $S$ operation is the same as in a simple antiferromagnet, so the conduction-electron-like bands are Kramers degenerate throughout the Brillouin zone.

\subsubsection{Two sublattice hastatic order (2SL)}

The 2SL ansatz may be represented in real space, as discussed above, or in momentum space, where the hybridization mixes bands with $\bk$ and $\bk +\bQ$, where $\bQ = (\pi, \pi)$. If the hybridization at site $A$ is $V_{A\sigma}=\left( V_{\uparrow}, V_{\downarrow} \right)^T$, the real space hybridization is,
\begin{align}
& V_{j\sigma}= V^{(1)}_\sigma+ V^{(2)}_\sigma \mathrm{e}^{i \bQ\cdot\bR_j} \cr
& V^{(1)}_\sigma = \frac{1}{2} (V_\sigma -\tilde{\sigma} V^\ast_{-\sigma}),\quad
V^{(2)}_\sigma = \frac{1}{2} (V_\sigma +\tilde{\sigma} V^\ast_{-\sigma}),
\end{align}
where $\tilde\sigma=\mathrm{sgn}(\sigma)$. The momentum space Hamiltonian is
\begin{multline}
H= \frac{1}{\mathcal{N}_s}\sum_{\bk}\Big[\epsilon_{\bk \alpha} c\dg_{\bk \alpha \sigma} c_{\bk \alpha \sigma}+\epsilon_{\bk+\bQ \alpha} c\dg_{\bk+\bQ \alpha \sigma} c_{\bk+\bQ \alpha \sigma}\\+ \left( V^{(1)}_\sigma c\dg_{\bk\alpha\sigma} f_{\bk \alpha}+ V^{(2)}_\sigma c\dg_{\bk+\bQ, \alpha\sigma} f_{\bk\alpha} +H.c. \right) \\
+ \epsilon_{f\bk} f\dg_{\bk \alpha}f_{\bk\alpha} + \epsilon_{f\bk+\bQ} f\dg_{\bk+\bQ \alpha}f_{\bk+\bQ\alpha} \Big] \\
+ \frac{N }{J_K} \sum_\sigma|V_{\sigma}|^2+  \frac{zN}{2J_H} |\chi_{H}|^2 -N\lambda +\frac{N}{2}\mu n_c,
\end{multline}
where the momentum sum is over the original Brillouin zone.
The calculation of the energy eigenvalues for the antiferrohastatic ansatzes proceeds by representing the corresponding Hamiltonians in matrix form, with $\bk$ ranging over the appropriate reduced Brillouin zones. Since the ferrohastatic ansatz contains six bands (four conduction, two f), the 2SL ansatz has twelve bands. In general, unless $\chi_H = \lambda = \mu = 0$, the antiferrohastatic Hamiltonians cannot be diagonalized analytically, and we rely on numerical results. In general, we solve the mean-field equations,
\begin{equation}
\frac{\partial F}{\partial \lambda} = 0,\; \frac{\partial F}{\partial V} = 0,\; \frac{\partial F}{\partial \chi_H} = 0,\; \mathrm{and}\; \frac{\partial F}{\partial \mu} = 0,
\end{equation}
to find the mean-field parameters, $\lambda$, $V$, $\chi_H$ and $\mu$ for a particular ansatz, where $V$ is the overall magnitude of the hybridization spinor; without loss of generality, we assume $V_A = (V,0)$, as we have $SU(2)$ spin (channel) symmetry. Note that if the $f$-electron hopping is zero, both of the four-sublattice ansatzes reduce to this two-sublattice Hamiltonian. Also note that since all the bands hybridize, there is a full hybridization gap, and we find hastatic Kondo insulators when $n_c = 1,3$ and the Fermi energy sits in the hybridization gap. As the $f$-electron bands are doubled, these Kondo insulators will always be trivial rather than topological insulators, as the parity of doubled bands cannot change \cite{dzero10}.

The band structure is invariant under $SU(2)$ spin-rotation and gauge transformations of $V_\sigma \rarrow V_\sigma \mathrm{e}^{i\phi}$. The eigenvectors, however, are not invariant, which leads to the magnetic moments discussed in section \ref{sec:multipolarmoments}.

\begin{figure}[h]\centering
	\includegraphics[width=\columnwidth]{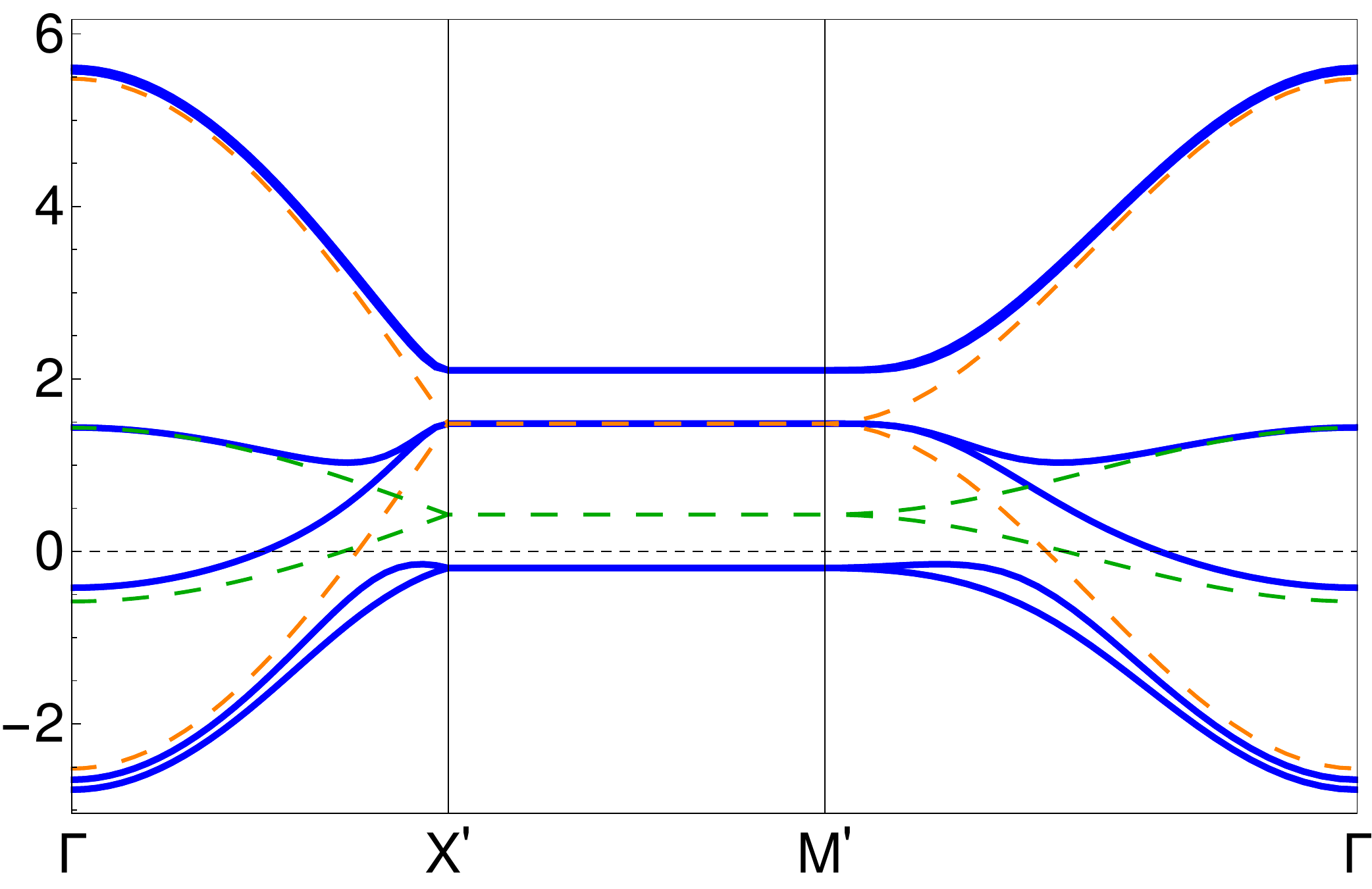}
\caption{Band structure along high symmetry lines in the first Brillouin zone for the 2D 2SL ansatz. Before hybridization, we have a single four-fold degenerate conduction electron band (orange dashed lines show bare conduction electron bands at $\bk$ and $\bk+\bQ$), and two doubly degenerate $f$-electron bands (green dotted lines show the unhybridized $f$-bands at $\bk$ and $\bk+\bQ$). After hybridization, all bands (blue) are hybridized and now doubly-degenerate due to the $\alpha$ pseudospin degeneracy; Kramers degeneracy is completely lost, even at the $\Gamma$ point. The parameters used were found self-consistently for $n_c = 1.2$, $J_K = 3t$, $J_H/J_K = 0.4$, $\eta = 1$.}
\label{fig:BandStructure_2SL}
\end{figure}

The bandstructure for the 2SL ansatz with nonzero $\chi_H$ is shown in Fig. \ref{fig:BandStructure_2SL}, where the parameters are found self-consistently for $n_c = 1.2$ and $J_H/J_K = 0.4$, which is in a region of the phase diagram where the 2SL ansatz has the lowest energy. The key signature of time-reversal symmetry breaking in 2SL order is that \emph{all} of the bands at the $\Gamma$ point are channel singlets. As we have two-fold pseudospin ($\alpha$) degeneracy, each band is only two-fold degenerate. The splitting can be clearly seen in the lowest conduction band; the highest conduction band is also split, but as it is far from the Fermi surface, the splitting is too small to resolve in the figure.

\subsubsection{Type 1 four sublattice hastatic order [4SL(1)]}

The 4SL(1) staggered ansatz can be written in momentum space as a hybridization between both states at $\bk$ and at $\bk \pm \bQ$, with $\bQ=(\pi/2,\pi/2)$. The hybridization at site $j$ is then,
\begin{align}
V_{j\sigma}=V^{(1)}_\sigma \mathrm{e}^{-i\bQ\cdot \bR_j}+ V^{(2)}_\sigma \mathrm{e}^{i \bQ\cdot\bR_j}
\end{align}
where we define,
\begin{align}
V^{(1)}_\sigma = \frac{1}{2} (V_\sigma -i\tilde{\sigma} V^\ast_{-\sigma}),\quad
V^{(2)}_\sigma = \frac{1}{2} (V_\sigma +i\tilde{\sigma} V^\ast_{-\sigma}).
\end{align}
The Hamiltonian in momentum space becomes,
\begin{multline}
H= \frac{1}{\mathcal{N}_s}\sum_{\bk}\Big[\epsilon_{\bk \alpha} c\dg_{\bk \alpha \sigma} c_{\bk \alpha \sigma}+V^{(1)}_\sigma c\dg_{\bk+\bQ,\alpha\sigma} f_{\bk \alpha} \\
 + V^{(2)}_\sigma c\dg_{\bk-\bQ, \alpha\sigma} f_{\bk\alpha}
 + \epsilon_{f\bk} f\dg_{\bk \alpha}f_{\bk\alpha}+H.c\Big] \\
 +  \frac{N }{J_K}\sum_\sigma|V_{\sigma}|^2+  \frac{ zN　}{2J_H} |\chi_{H}|^2 -N\lambda + N/2\mu n_c
\label{4SL(1)keqn}
\end{multline}
where $\bk$ ranges over the original unhybridized Brillouin zone. This 4SL ansatz has 24 bands.

\begin{figure}[h]\centering
	\includegraphics[width=\columnwidth]{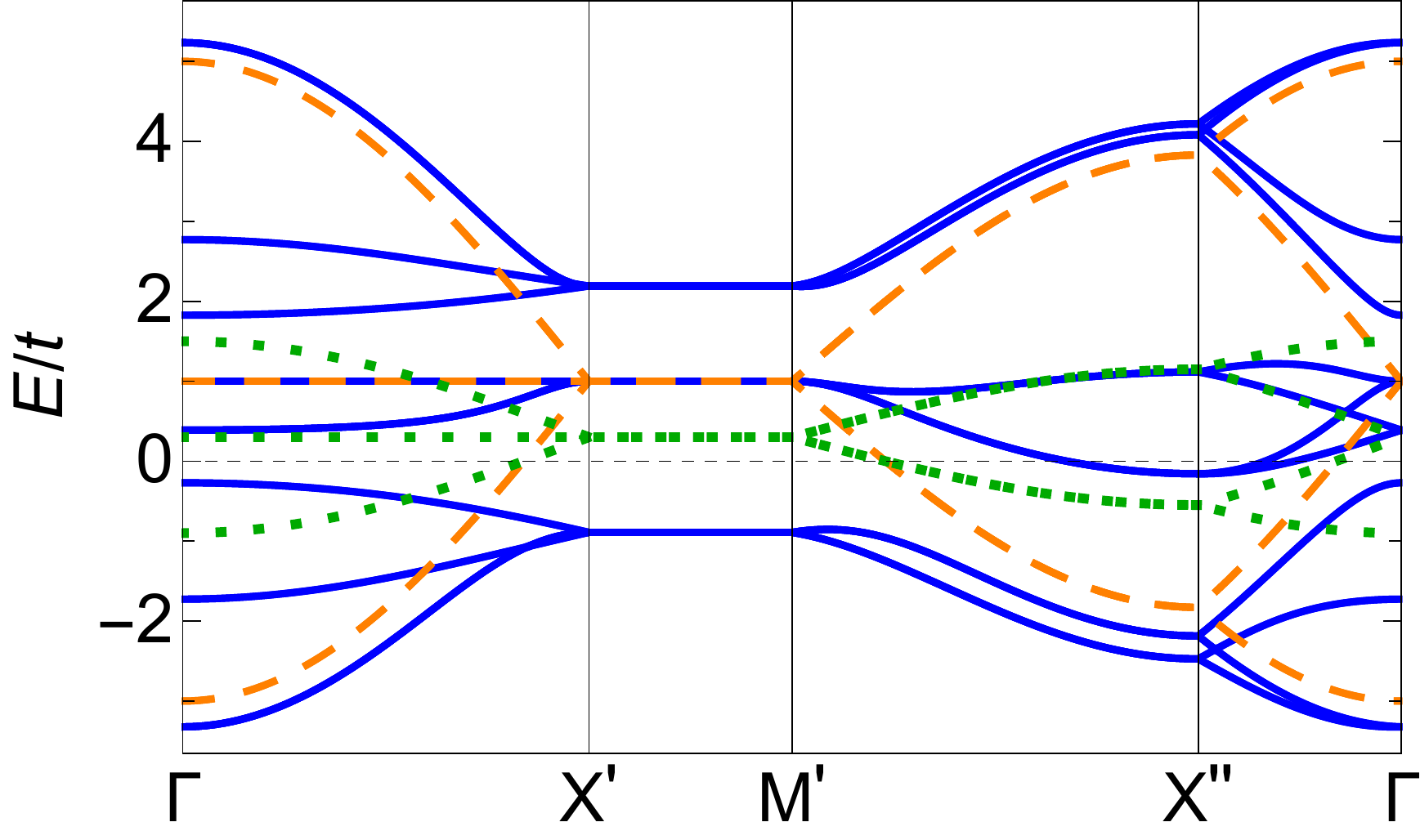}
\caption{Band structure along high symmetry lines in the first Brillouin zone for the 2D 4SL(1) ansatz. Before hybridization, we have four four-fold degenerate conduction electron bands (orange dashed lines), and four doubly degenerate $f$-electron bands (green dotted lines). After hybridization, all bands (blue lines) are hybridized, and the Kramers degeneracy at the $\Gamma$ point is preserved. Plotted for $V_\uparrow=1.5, V_\downarrow=0, \lambda=0.3,\chi=-0.3,\mu=-1$, $\eta =1$. }
\label{fig:BandStructure_4SL(1)}
\end{figure}

An example band structure for the 4SL(1) ansatz is shown in Fig. \ref{fig:BandStructure_4SL(1)}. For simplicity, we use the $\eta = 1$ structure which is always doubly degenerate in $\alpha$; the 4SL(1) ansatz does not appear in the mean-field phase diagram for $\eta = 1$, although it does for other values of $\eta$. We note a few important features. Unlike the ferrohastatic case, all the conduction electron bands hybridize at generic $\bk$ points. Unlike the 2SL case, the Kramers degeneracy is preserved at the $\Gamma$ point, leaving four four-fold degenerate bands and four two-fold degenerate bands. Away from the $\Gamma$ point, the spin-degeneracy is fully broken, and there are 12 doubly-degenerate bands, although the splitting is difficult to resolve in the figure. Note that the broken inversion symmetry is not immediately apparent in the band structure, which is invariant under $\bk \rarrow -\bk$ due to the time-reversal symmetry. The lack of inversion symmetry is responsible for the loss of spin-degenerate bands, as discussed above. Furthermore, the band structure is invariant under $SU(2)$ spin rotations, although the eigenvectors do reflect the broken symmetry, ultimately leading to $SU(2)$ symmetry-breaking staggered moments.

\subsubsection{Type 2 four sublattice hastatic order [4SL(2)]}

The 4SL(2) ansatz can be written in momentum space using hybridization between states at $\bk$ and at $\bk + \bQ_{1,2}$, where $\bQ_1 = (\pi,0)$ and $\bQ_2 = (0,\pi)$. The hybridization on site $j$ is,
\begin{align}
V_{j\sigma}=V^{(1)}_\sigma \mathrm{e}^{-i\bQ_1\cdot \bR_j} + V^{(2)}_\sigma \mathrm{e}^{-i \bQ_2\cdot\bR_j}
\end{align}
where we define
\begin{align}
V^{(1)}_\sigma \equiv \frac{1}{2} (V_\sigma+\tilde\sigma V^\ast_{-\sigma}),\quad V^{(2)}_\sigma \equiv \frac{1}{2} (V_\sigma-\tilde\sigma V^\ast_{-\sigma}).
\end{align}
The Hamiltonian becomes,
\begin{multline}\label{4SL(2)keqn}
H= \frac{1}{\mathcal{N}_s}\sum_{\bk}\Big[\epsilon_{\bk \alpha} c\dg_{\bk \alpha \sigma} c_{\bk \alpha \sigma}+ V^{(1)}_\sigma  c\dg_{\bk+\bQ_1, \alpha\sigma} f_{\bk\alpha} \\
+V^{(2)}_\sigma  c\dg_{\bk+\bQ_2,\alpha\sigma} f_{\bk\alpha}+ \epsilon_{f\bk} f\dg_{\bk \alpha}f_{\bk\alpha}+H.c\Big] \\
+ \frac{N}{J_K} \sum_\sigma|V_{\sigma}|^2+  \frac{zN　}{2J_H} |\chi_{H}|^2 -N\lambda+\frac{N}{2}\mu n_c.
\end{multline}
where $\bk$ ranges over the original unhybridized Brillouin zone. This 4SL ansatz also has 24 bands.

\begin{figure}[h]\centering
	\includegraphics[width=\columnwidth]{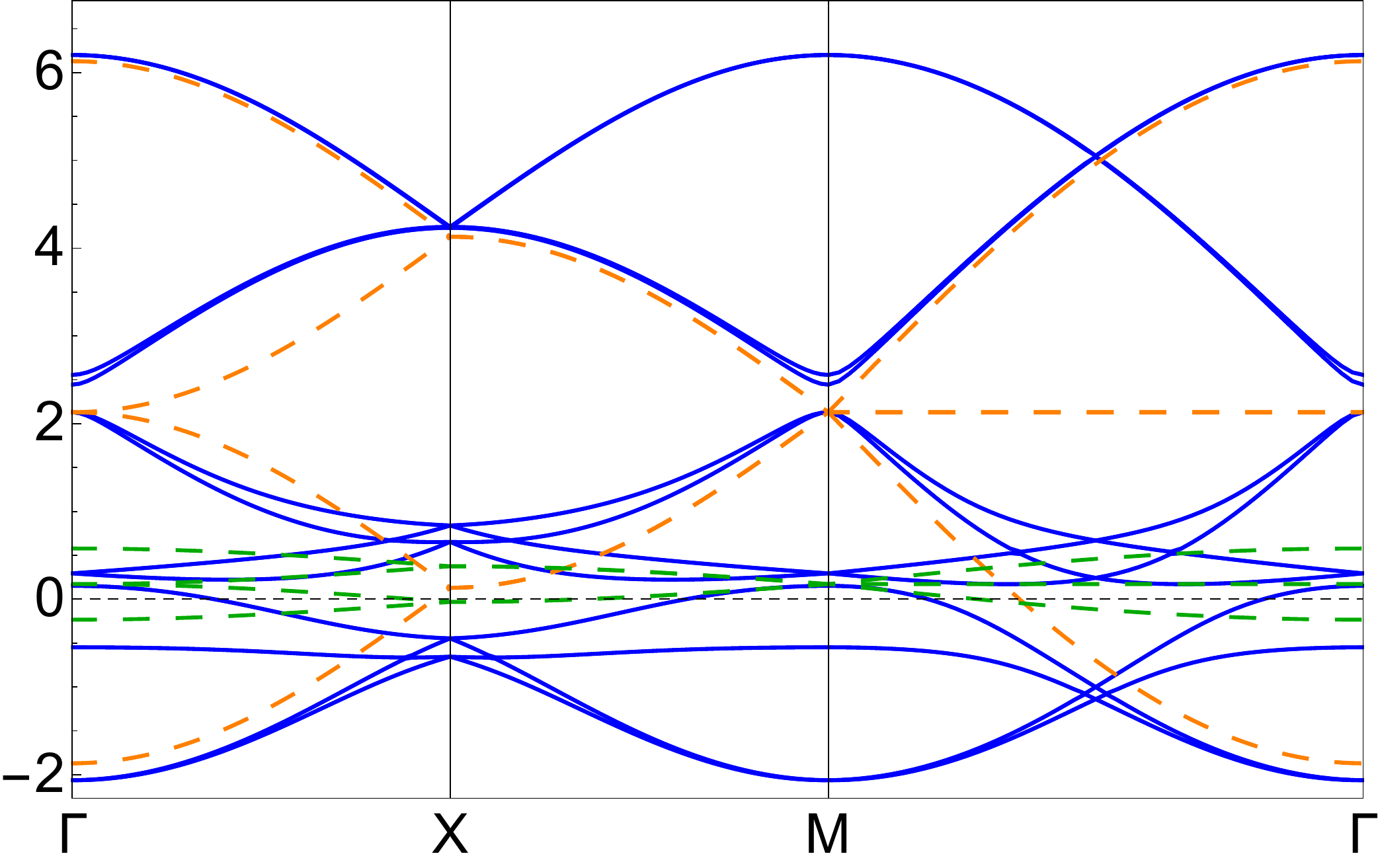}
\caption{Band structure along high symmetry lines in the first Brillouin zone for the 2D 4SL(2) ansatz. Before hybridization, we have four- and eight-fold degenerate conduction electron bands(orange dashed lines), and two- and four- degenerate $f$-electron bands(green dotted lines). After hybridization, all the bands (blue lines) are hybridized. At $\Gamma$, the highest and lowest bands that originate from $\bk$ and $\bk+\bQ_1+\bQ_2$ are four-fold degenerate, while all other bands are pseudospin doublets and spin singlets. Plotted for the self-consistent solution with $n_c = 0.8$, $J_K = 3t$, $J_H/J_K = 0.4$, $\eta = 1$. }
\label{fig:BandStructure_4SL(2)}
\end{figure}

An example band structure for the 4SL(2) ansatz is shown in Fig. \ref{fig:BandStructure_4SL(2)}, where the parameters are found self-consistently for $n_c = 0.8$ and $J_H/J_K = 0.4$, which is in a region of the phase diagram where the 4SL(2) ansatz has the lowest energy. Again, all conduction bands hybridize. Before hybridization, at the $\Gamma$ point there are two four-fold and one eight-fold degenerate conduction bands from $\bk$, $\bk+\bQ_1$, $\bk +\bQ_2$, and $\bk +\bQ_1 +\bQ_2$, as well as two doubly-degenerate and one fold-fold degenerate f-bands. After hybridization, the bands originating from $\bk$ and $\bk+\bQ_1+\bQ_2$ remain four-fold degenerate, while the other two groups split into doublets, as $\bQ_1$ and $\bQ_2$ are not invariant under the time-reversal-like symmetry, $P =  R_{\pi/2}\theta$. As before, the band structure is unchanged by $SU(2)$ spin rotations, with the eigenvectors reflecting the broken symmetry and leading to $SU(2)$ symmetry-breaking staggered moments.

\subsection{Canted hastatic ansatz}\label{sec:canted}

In addition to the ferro- and antiferrohastatic phases, we also consider a canted phase that combines features of both. The hastatic spinor behaves like a tiny magnetic moment in many ways, and so we expect it to cant in applied magnetic field. As such, we consider a hastatic spinor with both uniform and staggered components that are perpendicular to one another. This state both mimics a canted antiferromagnet and preserves the translation symmetry for the total hybridization on each site, $|V_j| = |V|$. We take the uniform component to be parallel to the external field, taken along $\hat z$, and the staggered component along $\hat x$. When the antiferrohastatic phases are placed in magnetic field, the canted phase develops, although it is not present in zero field. We therefore begin with any 4SL staggered phase and introduce a uniform component $\delta V$ as,
\begin{align}
& V_{A\sigma}= \left(\begin{array}{c}
V+\delta V \\
V
\end{array}\right),
 \quad
V_{B\sigma}=\left(\begin{array}{c}
-V-\delta V \\
V
\end{array}\right), \cr
& V_{C\sigma}=- V_{A\sigma},
 \quad
V_{D\sigma}= -V_{B\sigma}.
\end{align}
Here, $V$ and $\delta V$ are the staggered and uniform components, respectively. When $V \rarrow 0$, the uniform ansatz will have a staggered sign that may be removed by a gauge transformation even in the presence of $f$-hopping. If we redefine $V_\uparrow \equiv V+\delta V, V_\downarrow \equiv V$, we can continue to use the 4SL Hamiltonians, (\ref{4SL(1)keqn}) or (\ref{4SL(2)keqn}).

\begin{figure}[h]\centering
    \includegraphics[width=\columnwidth]{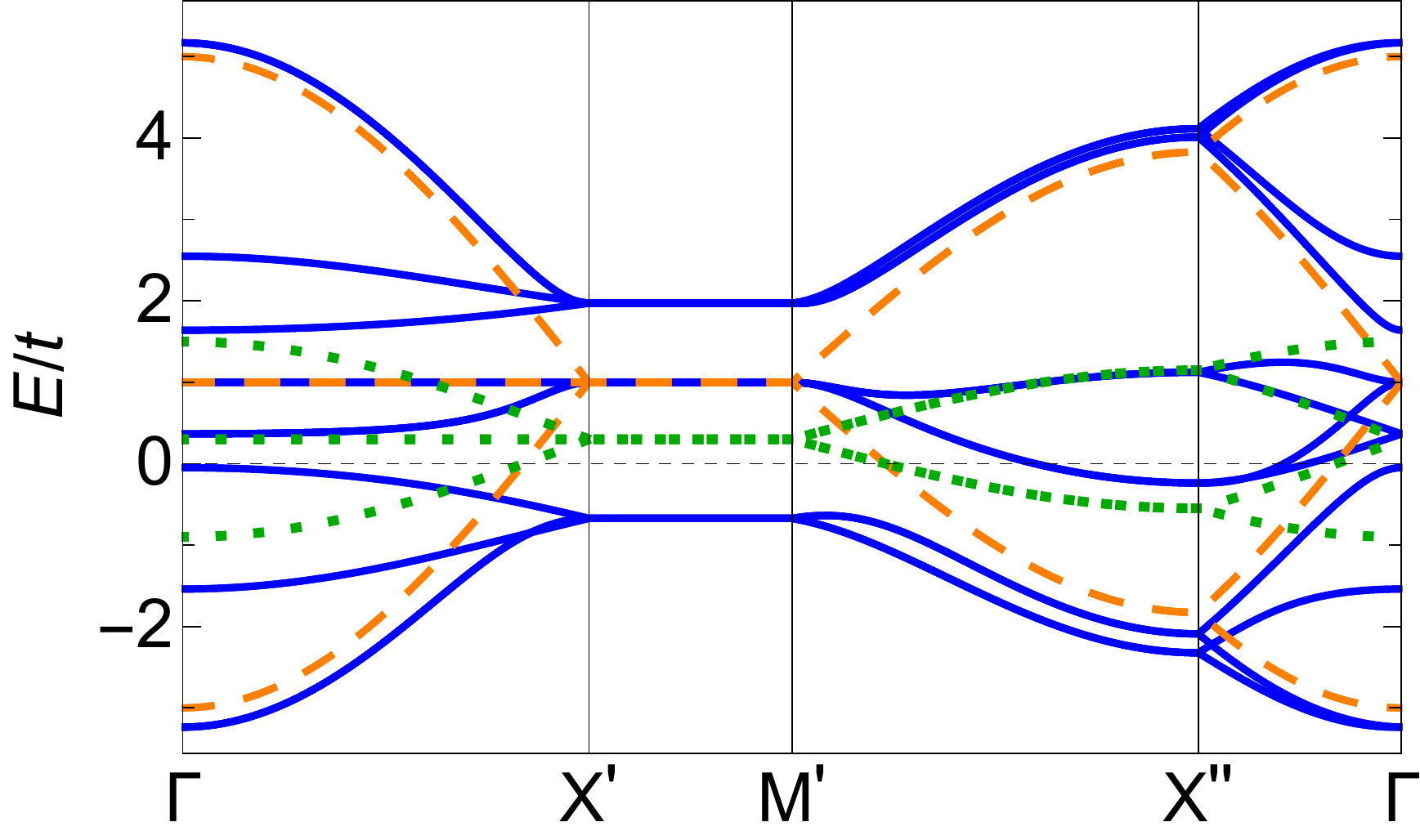}
\caption{An example bandstructure for canted 4SL(1) hastatic order with $V_\uparrow=1.5, V_\downarrow=1, \lambda=0.3, \chi=-0.3, \mu=-1$. }
\label{fig:BandStructure_canted}
\end{figure}

As the canted phase includes both uniform and staggered hybridization, all conduction bands hybridize, albeit unequally between the spin components, and the band structure qualitatively resembles the 4SL phase; an example canted 4SL(1) band structure is shown in Fig. \ref{fig:BandStructure_canted}.

\subsection{Non-hastatic phases: paraquadrupolar and quadrupolar liquid}\label{sec:non-hastatic}

In addition to various hastatic ansatzes, we also consider two different unhybridized states: the disordered high temperature ``paraquadrupolar'' state, and the quadrupolar liquid phase favored by large $J_H$. The paraquadrupolar state has $V = \lambda = \chi_H = 0$, and describes the Curie gas phase of the quadrupoles. It cannot be the ground state in the absence of field or strain due to its $R \ln 2$ entropy per site. 
In field and strain, the $\Gamma_3$ doublet splits, and the paraquadrupolar phase becomes partially or fully polarized, and can be the ground state.

The quadrupolar liquid is a spin liquid phase of the local moments ($V = \lambda =0, \chi_H \neq 0$), totally decoupled from the conduction electrons; as these are quadrupolar moments, we call it a quadrupolar liquid. Our mean-field ansatz limits us to neutral spinons hopping on the square lattice to form a spinon Fermi surface. Of course, beyond the mean-field limit, the quadrupole moments are much more likely to order at low temperatures than to form a spin liquid state. Our quadrupolar liquid phase captures the short-range quadrupolar order at high temperatures, and acts as a proxy to allow us to treat both $f$-electron hopping arising from beyond mean-field effects and the competition between hastatic and quadrupolar order. The quadrupolar liquid develops out of the paraquadrupolar phase via a second order phase transition at $T_{\mathrm{QL}}=\frac{J_H}{4}$.

\subsection{The Kondo temperature}

\begin{figure}[h]\centering
    \includegraphics[width=.65\columnwidth]{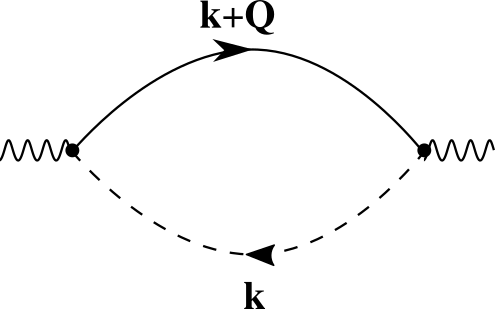}
\caption{Tree-level Feynman diagram for calculating the hastatic Kondo temperature for wave-vector $\bQ$; this diagram is the tree-level hybridization self-energy. Solid (dashed) lines indicate the bare c- and $f$-electron propagators, respectively. As the $f$-electron propagator is $\bk$ independent, the $\bQ$ dependence of this diagram can be removed.}
\label{fig:feynman}
\end{figure}

Hastatic order develops out of the paraquadrupolar state via a second order phase transition at $T_{K}$. This transition temperature is independent of the nature of the hastatic order, which can be seen straightforwardly by taking the action in terms of fermions, $c_{j\sigma \alpha}$ and $f_{j\alpha}$ and Hubbard-Stratonovich bosons, $V_{j\sigma}$, with the Hamiltonian given by equation (\ref{H_final}), and integrating out the fermions. Hastatic order develops when the dispersion for the bosons becomes negative at some $\bQ$ value and the bosons condense. As the free bosons above $T_K$ have no $\bQ$ dependence, this dispersion can be found by evaluating the boson self-energy, $\Sigma_{V\sigma}(i\nu_n,\bQ)$, where we are interested in ordering at high temperatures and so set $i \nu_n = 0$. As the vertex $V c\dg f$ is of order unity, this calculation is in principle extremely complicated. However, here we consider $\chi_H = 0$, such that the $f$-electrons have no $\bk$ dependence, $\mathcal{G}_{f0}^{-1}(i\omega_n,\bk) = i \omega_n$. As the bosons also have no $\bk$ dependence, the tree-level diagram shown in Fig. \ref{fig:feynman} can trivially have its $\bQ$-dependence removed by redefining $\bk$,
\begin{equation}
\Sigma_{V\sigma}(0,\bQ) = T \sum_{i\omega_n} \sum_{\bk} \mathcal{G}_{c0,\sigma}(i\omega_n,\bk+\bQ)\mathcal{G}_{f0}(i\omega_n).
\end{equation}
Any higher order corrections can similarly have their $\bQ$ dependence removed. Interactions between the bosons are required to differentiate the types of hastatic order.

\begin{figure}[h]\centering
     \includegraphics[width=0.9\columnwidth]{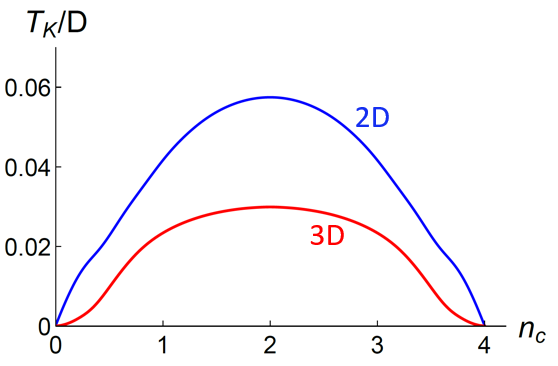}
\caption{Kondo temperature for hastatic order as a function of conduction electron filling in 2D(blue) and 3D(red). $T_K$ is the same for all hastatic orders. $T_K$ for 3D is renormalized by the bandwidth of bare conduction electron bands. Here, $t = 1$ and $J_K =3$. Note that $T_K \sim 0.05 D $, where $D = 8t, 12t$ is the bandwidth for the conduction electrons, which is significantly larger than in most rare-earth materials, but leads to better numerical convergence.}
\label{fig:TKU}
\end{figure}

As the Kondo temperature is independent of $\bQ$, we can explicitly calculate it from the ferrohastatic mean-field equations,
\begin{align}
\frac{1}{V}\frac{\partial F}{\partial V}\B|_{V,\lambda,\chi \rightarrow 0}= 0;\quad
\frac{\partial F}{\partial \mu}\B|_{V,\lambda,\chi \rightarrow 0}= 0,
\end{align}
where the second equation fixes the conduction electron filling. The free energy is
\begin{eqnarray}
F & = & - T \sum_{\eta \alpha} \int_\bk \ln (1+ \mathrm{e}^{-\beta \omega_{\bk \alpha \eta} } ) +  \frac{N }{J_K} (|V_{\uparrow}|^2+|V_{\downarrow}|^2) \cr
& & +  \frac{z N　}{2J_H} |\chi_{H}|^2 -\lambda N -\frac{N}{2}\mu n_c
\end{eqnarray}
where $\eta$ labels the three energy branches in eq. \eqref{uniform bands}.
Assuming the conduction electron filling is fixed, the Kondo temperature is thus determined by,
\begin{align}
\sum_{\bk\alpha} \frac{\tanh(\frac{\epsilon_{\bk\alpha}}{2T_{K}})}{\epsilon_{\bk \alpha}} =\frac{2N}{J_K}.
\end{align}
As can be seen in Fig. \ref{fig:TKU}, $T_{K}$ is particle-hole symmetric and vanishes smoothly for $n_c \rarrow 0, 4$, where there are no conduction electrons, with a maximum at half-filling. This scenario is quite different from the development of itinerant magnetism, where Fermi surface nesting enhances the ordering temperature at the ordering wave-vector. Here, all hastatic orders have the same transition temperature, and lower temperatures are required to select one particular order. For larger $J_H$, hastatic order can emerge out of the quadrupolar liquid, where the $f$-electron dispersion can lead to different $T_K(\bQ)$.

\section{Moments, Susceptibilities and $g$-factors}\label{sec:multipolarmoments}

As all hastatic orders break some symmetries, we expect nonzero moments and symmetry-breaking susceptibilities. While we can calculate these analytically for ferrohastatic order, we cannot generically do so for the antiferrohastatic cases. Therefore, we turn to numerical calculations. We can calculate arbitrary moments and susceptibilities numerically by introducing appropriate conjugate fields that couple only to the moments of interest, and taking numerical derivatives of the free energy. For instance, we calculate the staggered conduction electron moment along $\hat z$ with,
\begin{align}
H \; & \rarrow H -  B^z_{cs} \; \frac{1}{\mathcal{N}_s}\sum_{\bk \alpha} \tilde{\sigma} c\dg_{\bk \alpha \sigma} c_{\bk + \bQ \alpha \sigma}\cr
m_{cs}^z & = -\left.\frac{\partial F}{\partial B_{cs}^z}\right\vert_{B_{cs}^z \rarrow 0}.
\end{align}
Such calculations were done for uniform and staggered fields coupling to the magnetic and quadrupolar moments of the $c$-electrons, and the quadrupolar moments of the $f$-electrons. Susceptibilities were calculated via second derivatives with respect to the conjugate fields. 

\subsection{Multipolar moments}

\begin{figure}[h]\centering
	\includegraphics[width=\columnwidth]{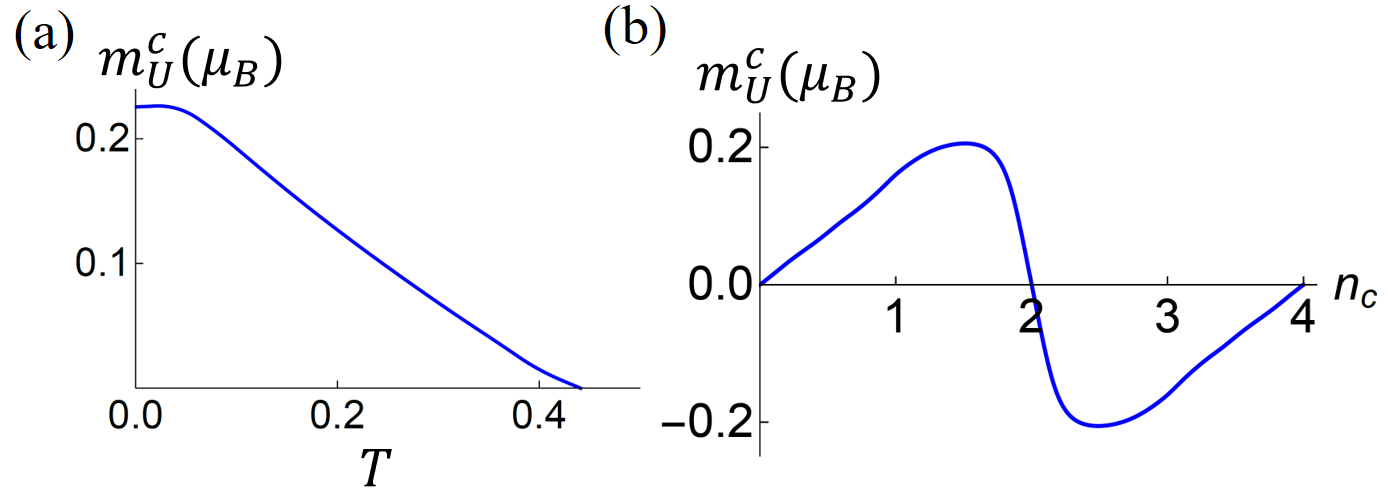}
\caption{Ferrohastatic order contains nonzero uniform conduction electron moments parallel to the hastatic spinor. Here, we show the moment (a) as a function of temperature for $n_c=1.5, J_H/J_K=1/30$; (b) as a function of conduction electron filling $n_c$ at low temperature for $J_H/J_K=1/30$. $m_U^c$ is linear in $T$ around $T_K$ and is particle-hole anti-symmetric. Both figures assume two degenerate conduction bands and are calculated self-consistently. Note that the magnitude is proportional to $T_K/D$, which we take to be quite large here, and realistic systems will have moments several orders of magnitude smaller. In our calculation, we fix $J_K=3t$.}
\label{fig:Mu}
\end{figure}

The ferrohastatic phase has a single nonzero moment: the conduction electron moment parallel to the direction of the hastatic spinor. This moment is plotted in Fig. \ref{fig:Mu} as a function of temperature $T$ and conduction electron filling $n_c$. As the order parameter is the hybridization spinor $V_\sigma$, the moment develops linearly in temperature. It is particle-hole antisymmetric and vanishes at half-filling, as found previously \cite{hoshino11}. The magnitude of these moments is proportional to $T_K/D$, where $D$ is the conduction electron bandwidth. This calculation was done self-consistently in the ferrohastatic phase, where $T_K/D \sim 0.05$ and the maximum moment is $\sim 0.2 \mu_B$. Realistic Pr-based systems typically have significantly smaller values of $T_K/D$, and will have similarly smaller hastatic moments.

\begin{figure}[h]\centering
    \includegraphics[width=\columnwidth]{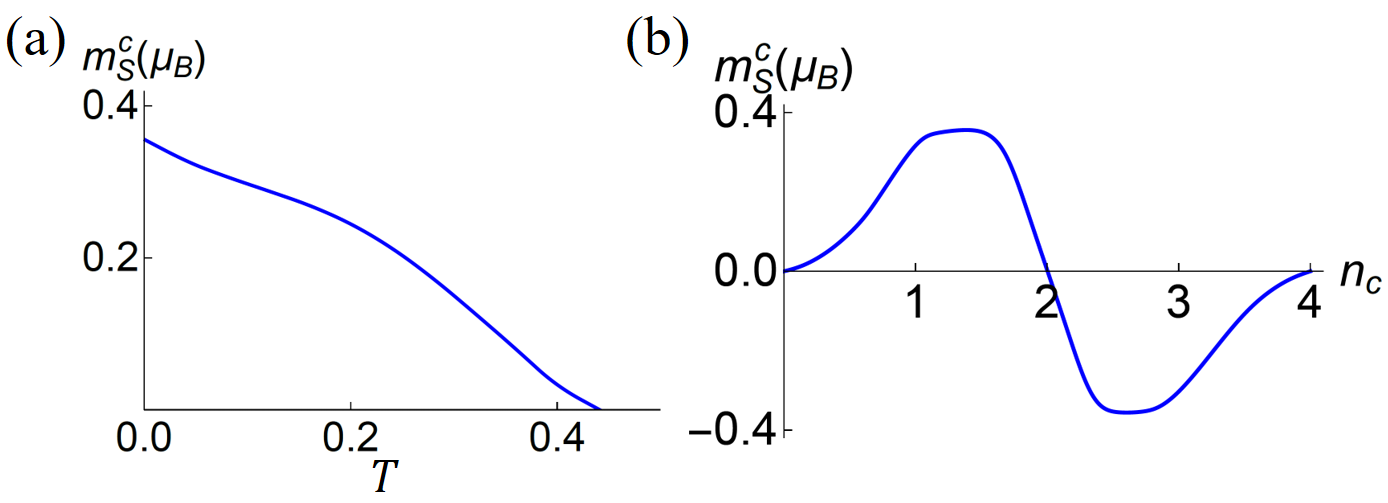}
\caption{Antiferrohastatic order has a nonzero staggered conduction electron moment, here shown (a) as a function of temperature for $n_c=1.5, J_H/J_K=1/30$; (b) as a function of conduction electron filling $n_c$ at low temperature for $J_H/J_K=1/30$. Both figures assume two degenerate conduction bands and are calculated self-consistently. Note that the magnitude is proportional to $T_K/D$, which we take to be quite large here, and realistic systems will have moments several orders of magnitude smaller. }
\label{fig:Ms_staggered}
\end{figure}

In the four sublattice antiferrohastatic phases, the only nonzero moments are staggered conduction electron dipole moments along the direction of the hastatic spinor, as expected. There are no nonzero quadrupolar moments of any kind. The staggered moment, like the ferrohastatic moment, develops linearly in temperature, and is particle-hole anti-symmetric as shown in Fig. \ref{fig:Ms_staggered}; again, the magnitude is proportional to $T_K/D$, with a maximum $\sim 0.4\mu_B$. None of the moments or susceptibilities reflect the additional broken symmetries of the four sublattice phases, and there is no qualitative distinction between the 4SL(1) and 4SL(2) moments.

The two-sublattice phase requires more careful treatment, as at first it appears to host both uniform and staggered moments. However, the uniform moments are gauge dependent, in that they depend on the overall phase of $V_\sigma$. All other quantities, including the staggered moments and the bandstructure are gauge independent. If $V_\sigma || \hat z$, with the complex phase $\phi$, the uniform moments will be in the basal plane, with $\phi$ dependence $m_\perp \propto (\cos \phi, \sin \phi)$. Any physical quantity must be gauge-independent, and indeed these moments vanish once we average over the possible gauge choices. The gauge invariant staggered moments are qualitatively similar to the 4SL(1) and 4SL(2) staggered moments, and there is no way to resolve between any of the antiferrohastatic phases based on moments alone.

\subsection{Susceptibility anisotropy}

We are primarily interested in symmetry-breaking susceptibilities that develop with the onset of hastatic order; these include magnetic, strain, and magnetostrictive susceptibilities, in principle. The susceptibilities are found by taking the second derivative of $F$ with respect to the appropriate combination of conjugate fields. The conduction electron magnetic susceptibilities have a $(g \mu_B)^2$ constant of proportionality, while the strain and magnetostrictive susceptibilities have materials dependent constants of proportionality. As we are interested in the symmetry breaking, rather than the absolute magnitudes, we set these constants of proportionality to one.

\begin{figure}[!htbp]\centering
  \includegraphics[width=0.7\columnwidth]{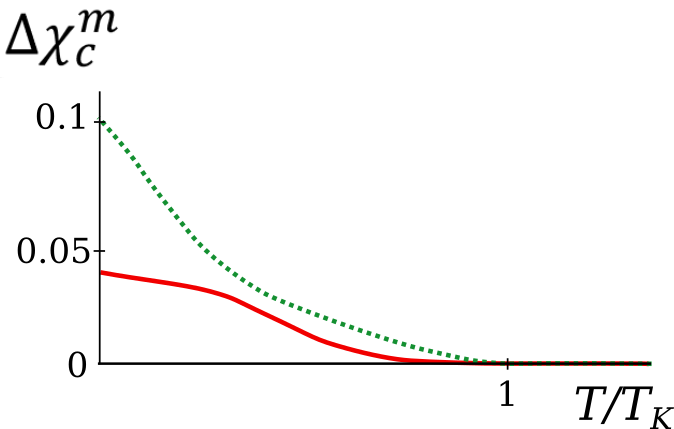}
\caption{Conduction electron magnetic susceptibility anisotropy as a function of temperature for the ferrohastatic (solid red line) and antiferrohastatic (dashed green line) orders. This figure is calculated for $n_c=1.5, J_H/J_K=1/30$ self-consistently, although the antiferrohastatic state is only metastable here. We define the dimensionless $\Delta\chi_c^m \equiv (\chi_c^{zz}-\chi_c^{xx})/(\chi_c^{xx}+\chi_c^{zz})$; both hastatic orders have enhanced susceptibility along the direction of the hastatic spinor. }
\label{fig:DeltaChi}
\end{figure}

The magnetic susceptibilities of ferrohastatic and antiferrohastatic phases behave similarly, with an enhancement of the susceptibility $\chi_c^{zz}$ along the direction of the hastatic spinor below $T_K$, developing as $(T-T_K)^2$, as shown in Fig. \ref{fig:DeltaChi}. Here, this symmetry breaking is simply a consequence of the magnetic moments, and also occurs in a normal magnet. The 2SL in-plane magnetic susceptibilities additionally have a gauge-dependent contribution due to the gauge dependent moments; this contribution again vanishes after gauge averaging.  

We similarly calculate the strain and magnetostrictive susceptibilities, but find that these do not break any symmetries
-- the strain susceptibility tensor has cubic symmetry, and the magnetostrictive susceptibilities vanish uniformly. As there is no spin-lattice coupling, this absence is not surprising, but might change in a spin-orbit coupled Anderson model treatment. Both $c$ and $f$ electron strain susceptibilities behave similarly.

\subsection{Coupling to magnetic field: $g$-factor}

As the conduction electrons hybridize with non-magnetic $f$-electrons, we might expect the $g$-factor of the resulting heavy electrons to be much reduced from the high temperature $g = 2$; this reduced $g$-factor would be a key indication that the conduction electrons were hybridizing with a non-magnetic doublet \cite{chandra13}. The $g$-factor in heavy fermion materials can be measured by looking at the Fermi surface magnetization via de Haas-van Alphen (dHvA), where this magnetization is a periodic function of the ratio of the Zeeman splitting and cyclotron frequencies \cite{Altarawneh12}. The Zeeman splitting, and thus the $g$-factor can be very sensitively measured by looking at the ``spin-zeros'' where this magnetization passes through zero. Measuring the $g$-factor this way requires doubly-degenerate bands everywhere in the Brillouin zone in order to define their splitting in magnetic field, and so we consider only antiferrohastatic order with $\chi_H = 0$, where all of our bands are doubly degenerate. 
For any antiferrohastatic order with nonzero $f$-electron hopping ($\chi_H \neq 0$), the bands are no longer doubly degenerate, and the $g$-factor is not well-defined. Instead of looking at the $g$-factor, we can look at how these bands move in magnetic field in order to see their magnetic content; the bands shift primarily as $B^2$, with a small linear in $B$ component proportional to $T_K/D$. However, small non-Zeeman splittings due to realistic $c$ and $f$ bandwidths are unlikely to seriously affect the measured spin-zeros.

\begin{figure}[h]\centering
	\includegraphics[width=\columnwidth]{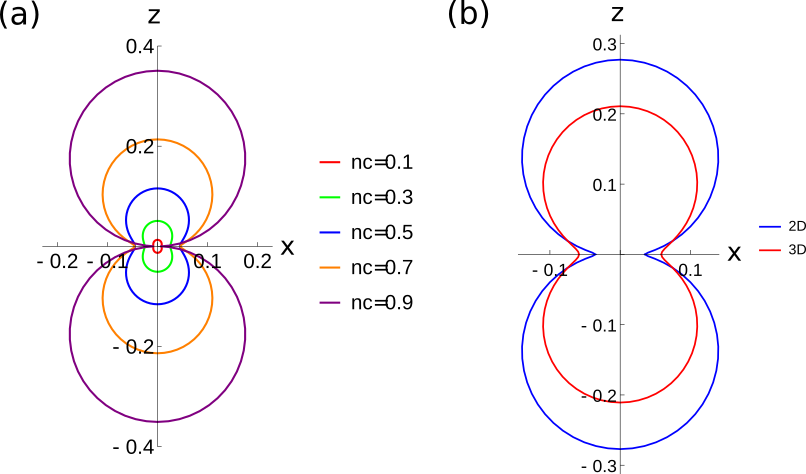}
\caption{Angle-dependent $g-$factor calculated for doubly degenerate bands in the antiferrohastatic ansatz ($J_H/J_K=0$, $\chi = 0$), with $V || \hat z$. (a) A polar plot of the $g$-factor for several $n_c$ within the staggered phase. The overall magnitude of $g(\theta)$ is proportional to $T_K(n_c)/D$, while the anisotropy increases with increasing $n_c$, up to $n_c = 1$. (b) A comparison of the $g$-factor in two and three dimensions, for $n_c = 0.8$. The 3D $g$-factor is similar, but slightly less anisotropic.}
\label{fig:g_factor}
\end{figure}

We calculate the $g$-factor by introducing the coupling $- g\mu_B \vec{B} \cdot \sum_{\bk \alpha \sigma} c\dg_{\bk \alpha \sigma} \vec{\sigma} c_{\bk \alpha \sigma}$ and examining how the Kramers-degenerate hybridized bands split in field. For simplicity, we take the hybridization $V_A$ to point along $\hat z$, and define the magnetic field direction in terms of the angle between the hybridization spinor and magnetic field, $\theta$, and the angle $\phi$ in the plane perpendicular to the hybridization spinor. The hybridized bands ($\omega_{\bk\eta\sigma}$) split linearly, as $\Delta E_{\bk\eta} = |\omega_{\bk\eta\uparrow}-\omega_{\bk\eta\downarrow}| = g_{\bk \eta}(\theta,\phi)B$, with
\begin{equation}
g_{\bk \eta}(\theta, \phi) = \left \vert \frac{d\Delta E_{\bk \eta }}{dB}\right\vert_{B\rightarrow 0}.
\end{equation}  
We are interested in the Fermi surface average,
\begin{equation}
g (\theta,\phi ) = \frac{
\sum_{\bk \eta } g_{\bk \eta } (\theta,\phi )\delta (\omega_{\bk\eta })
}{\sum_{\bk \eta } \delta (\omega_{\bk\eta })
}.
\end{equation}
$g(\theta,\phi)$ is independent of $\phi$, but has a $\theta$ dependence that is more pronounced for larger $n_c$. The maximum $g$-factor occurs for fields aligned with the hybridization spinor, as seen in Fig. \ref{fig:g_factor}. Note that we have fixed the hybridization spinor, while in reality it will be weakly pinned and may well rotate to follow the spinor, keeping this maximum value for all angles. The overall magnitude of the $g$-factor is suppressed from $g = 2$ by approximately $T_K/D$, although the details of the anisotropy depend on the conduction electron filling $n_c$. 

\section{Zero Temperature Phase diagram}\label{sec:PD}

To investigate the competition between ferro- and antiferrohastatic orders, and their competition or cooperation with quadrupolar order, we examine the zero temperature phase diagram for three different models: the two dimensional phase diagram, both for perfectly degenerate conduction bands and for non-degenerate, but symmetry related, bands, and the three dimensional phase diagram for degenerate bands. All three phase diagrams are qualitatively similar, with the main difference being the relative stabilities of the different antiferrohastatic phases. These phase diagrams were obtained by finding saddle point solutions for each ansatz, and taking that with the lowest energy. Again, note that we neglect more complicated hastatic orders, as well as superconductivity. The phase diagrams are found in the $(n_c, J_H/J_K)$ plane, where $n_c \in (0,4)$ is the conduction electron density. In each case, we fix $t = 1$, $J_K = 3t$ and vary $J_H$.

\begin{figure}[!htb]\centering
\includegraphics[width=0.95\linewidth]{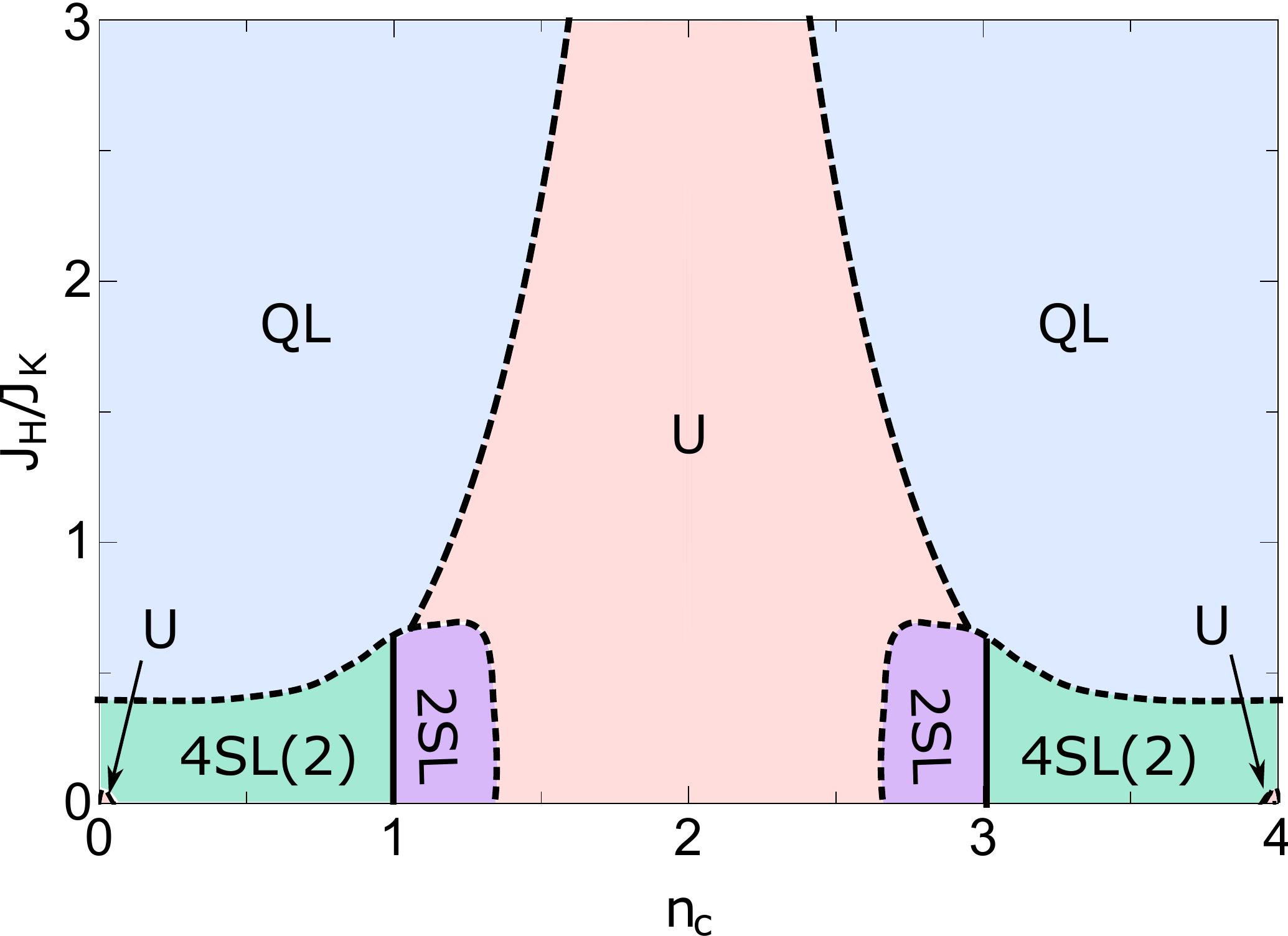}
\caption{Low temperature phase diagram for 2D in the $(n_c, J_H/J_K)$ plane, where the red region (U) indicates ferrohastatic order, purple indicates 2SL staggered order, teal indicates 4SL(2) staggered order, and blue represents the quadrupolar liquid (QL). The solid/dashed black lines indicate second/first order transitions between the phases.}
\label{fig:PD_2D}
\end{figure}

First, we discuss the 2D phase diagram for degenerate bands, as shown in Fig. \ref{fig:PD_2D}. As our conduction electron bands are particle-hole symmetric, so is our phase diagram. The ferrohastatic phase is favored near half-filling ($n_c = 2$), where it extends to $J_H/J_K \rarrow \infty$, and in very small pockets near $n_c = 0,4$. Generically, for finite $J_H$, the ferrohastatic phase also has $\chi_H \neq 0$. The infinite extent at half-filling is due to the perfect nesting of the conduction electron band structure. While the transition between ferrohastatic order and the quadrupolar liquid is always first order, we can see the expanded stability of the ferrohastatic phase by following the line where $d^2F/dV^2\vert_{V=0} = 0$. In the quadrupolar liquid, $V = 0, \lambda = 0, \chi_H = 2 J_H/\pi^2$, and so,
\begin{align}\label{dFdV2}
\frac{\partial F^2}{\partial V^2}\B|_{V\rightarrow 0} \!=\! \frac{2N}{J_K} -\sum_{\bk\alpha} \frac{\tanh(\frac{\epsilon_{f\bk}}{2T}) \!-\! \tanh(\frac{\epsilon_{\bk\alpha}}{2T}) }{\epsilon_{f\bk}-\epsilon_{\bk\alpha}}=0.
\end{align}
For $\mu = 0$, the above integral is proportional to $\int_\bk \frac{1}{|\cos(k_x)+\cos(k_y)|}$, which diverges logarithmically; therefore for sufficiently small $\mu$, $V = 0$ is not a stable minimum for any $J_H$. Note that the critical $(J_H/J_K)_c$ where $d^2F/dV^2\vert_{V=0}$ changes sign is not usually a second order transition in this case, as the quadrupolar liquid is already an excited metastable state relative to ferrohastatic order by this $(J_H/J_K)_c$.

Away from half-filling, the ferrohastatic region gives way to a dome of antiferrohastatic order peaked around quarter filling, again via a first order phase transition. The 2SL phase is stable for $n_c > 1$, while the 4SL(2) phase is stable for $n_c < 1$, with $\chi_H \neq 0$ for all finite $J_H$. Exactly at $n_c = 1$, $\chi_H$ vanishes smoothly and the two phases are equivalent. This line is a second order phase transition, and forms a Kondo insulator in which the bands regain the full four-fold degeneracy. Otherwise, these phases are generically metallic and lack Kramers degeneracy.

\subsection{Breaking conduction electron degeneracy}

\begin{figure}[!htb]\centering
  \includegraphics[width=\linewidth]{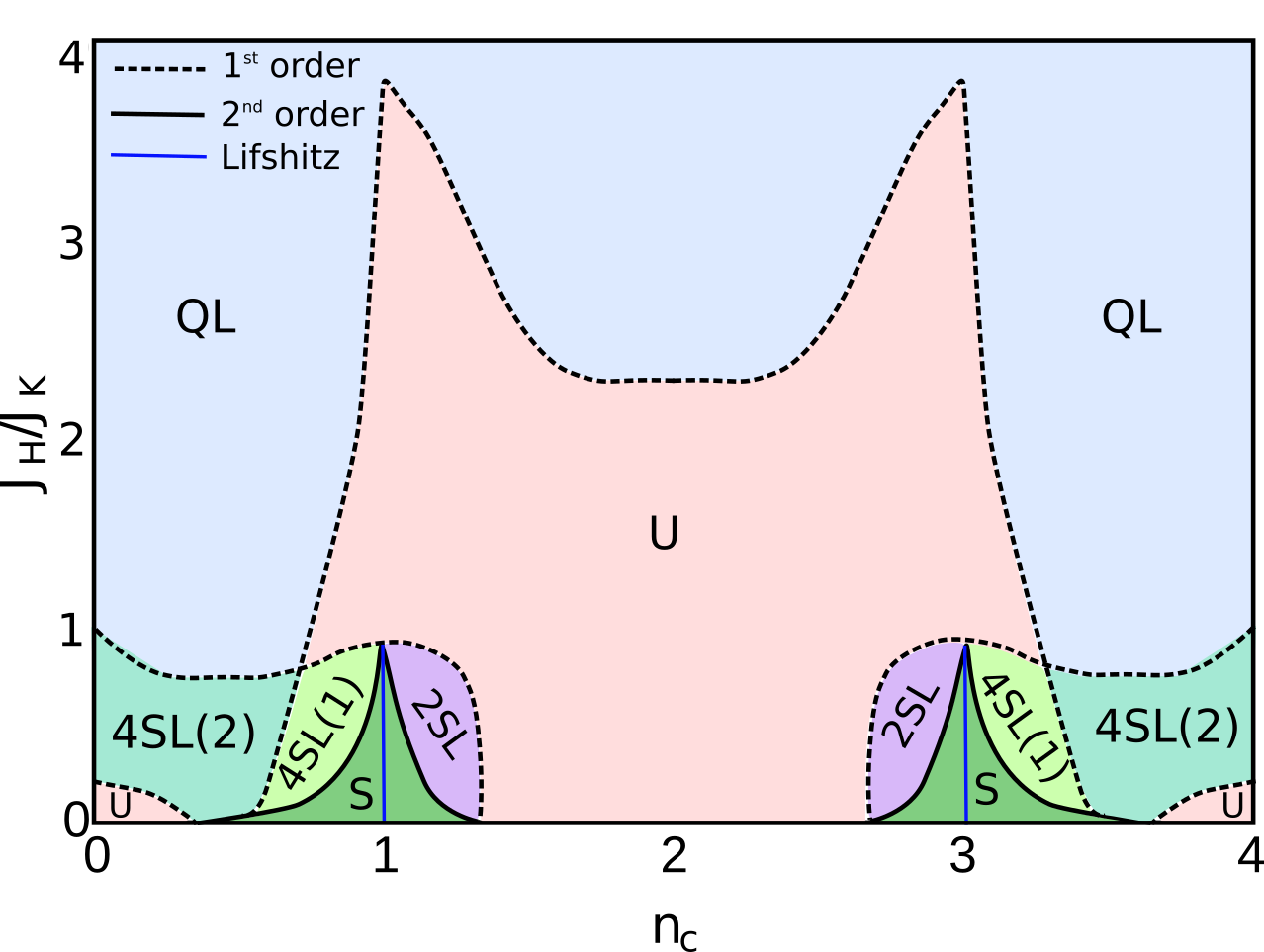}
\caption{Low temperature phase diagram for 2D in the $(n_c, J_H/J_K)$ plane, for $\eta = 1/3$. The red region (U) indicates ferrohastatic order, purple indicates 2SL staggered order, teal indicates 4SL(2) staggered order, light green 4SL(1), and dark green (S) the $\chi_H = 0$ antiferrohastatic order, while blue represents the quadrupolar liquid (QL). Dashed/solid black lines indicate first/second order transitions, while the blue solid lines indicate Lifshitz transitions within the antiferrohastatic order at the Kondo insulator lines for $n_c = 1, 3$.}
\label{fig:JH_nc_2D_dt}
\end{figure}

Next, we consider the effect of breaking the band degeneracy. Recall that the two conduction electron bands are still related by symmetry and are degenerate at the $\Gamma$ point. In Fig. \ref{fig:JH_nc_2D_dt}, we present an example phase diagram for $\eta = 1/3$. This phase diagram is qualitatively similar to the degenerate case: it is particle-hole symmetric, with the ferrohastatic phase favored at very low and half-filling, and antiferrohastatic order favored about the quarter-filling limit. Here, however, the band structure is no longer perfectly nested at half-filling, and so the ferrohastatic phase extends up only to a finite $(J_H/J_K)_c$, and now peaks at quarter-filling for both the ferro- and antiferrohastatic orders. The antiferrohastatic dome is more complex: again the 2SL phase is stable for $n_c > 1$, and the 4SL phases are stable for $n_c < 1$. However, about quarter-filling there is now a dome of $\chi_H = 0$ staggered phase where the three ansatzes are identical. Both 4SL(1) and 4SL(2) appear, with the pocket of 4SL(1) closer to quarter-filling. The ferrohastatic pockets at low filling are also substantially larger.

\begin{figure}[!htb]\centering
  \includegraphics[width=0.9\linewidth]{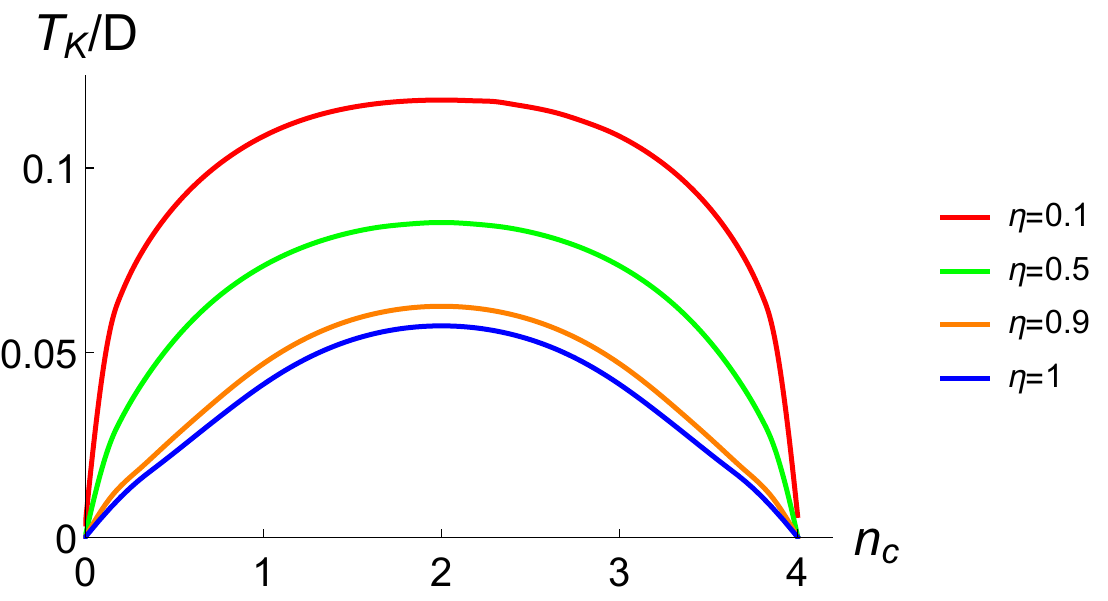}
\caption{Kondo temperature as a function of conduction filling for different $\eta$. $T_K$ is renormalized by the bandwidth of bare conduction electron bands. }
\label{fig:TKU_dt}
\end{figure}

In part, breaking the band degeneracy allows us to explore the effect of a different bandstructure; it clearly is not detrimental to hastatic order, nor does it seriously affect the competition between ferro- and antiferrohastatic order. As $\eta$ decreases from one, the bands become more one-dimensional, enhancing the density of states and hastatic order slightly, as shown in Fig. \ref{fig:TKU_dt}. Here, we plot the Kondo temperature as a function of $n_c$ for several anisotropies, showing that as the anisotropy increases, $T_K/D$ both increases in magnitude and flattens out more as $n_c$ approaches half-filling.

\begin{figure}[!htb]\centering
	\includegraphics[width=\columnwidth]{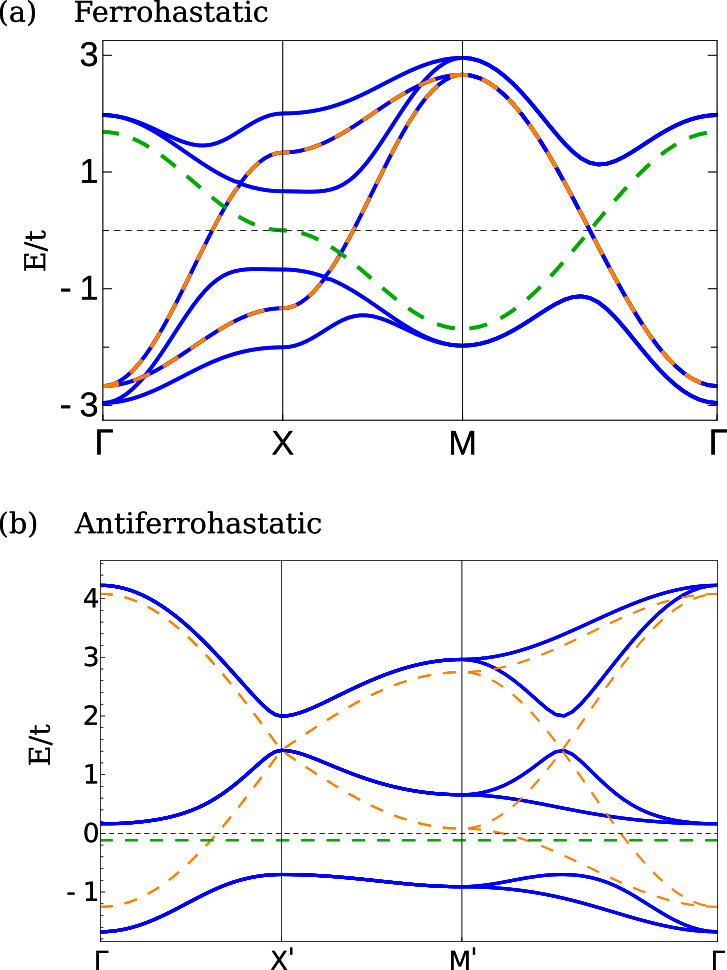}
\caption{Bandstructures of hastatic order for nondegenerate bands, with $\eta=1/3$. (a) bandstructure of ferrohastatic order(blue lines) for $n_c=2, J_H/J_K=1$. (b) bandstructure of antiferrohastatic order(blue lines) for $n_c=1, J_H/J_K=0.4$ where the system is a Kondo insulator and $\chi_H = 0$. Here the orange dashed lines represent free conduction electron bands and the green dotted lines are the free $f$-electron bands. Parameters were obtained self-consistently, with $J_K = 3t$.}
\label{fig:BandStructure_dt}
\end{figure}

Broken band degeneracy implies that we generically have two sets of doubly-degenerate bare conduction bands in addition to the doubly-degenerate bare f-bands. In ferrohastatic order, we now find two non-degenerate unhybridized bands and four non-degenerate hybridized bands, as shown in Fig. \ref{fig:BandStructure_dt} (a). Antiferrohastatic order shows few qualitative changes; see Fig. \ref{fig:BandStructure_dt} (b). The signatures of hastatic order all remain qualitatively the same, with only the relative stability of the hastatic and quadrupolar liquid phases being modified by the removal of the band-degeneracy, likely due to the enhanced density of states. For most of the paper, we consider the simpler, completely degenerate case, and mention only the key differences between the two cases.

\subsection{Effect of dimensionality}\label{sec:dimensionality}

\begin{figure}[!htbp]\centering
  \includegraphics[width=0.8\columnwidth]{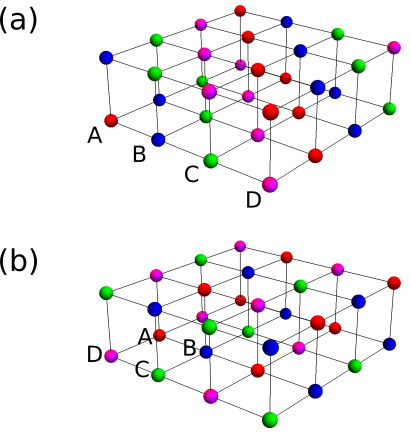}
\caption{Illustrations of the crystal structures for the 3D 4SL ansatzes: (a) rhombohedral 4SL(1), with ordering vector ($\pi/2$,$\pi/2$,$\pi/2$); (b) orthorhombic 4SL(2), with ordering vectors ($\pi$,0,0), (0,$\pi$,0), and (0,0,$\pi$).}
\label{fig:cartoon3D}
\end{figure}

We can also consider the effect of changing the dimensionality from two to three dimensions. As we are strictly in the mean-field limit, the difference here is not substantial, since the fluctuations that typically destroy long-range order in two-dimensions are absent in our calculations. The difference between 2D and 3D in our calculations is more a difference of the details; the van Hove singularity in the conduction electron density of states is removed, as it is in the non-degenerate band case, and the staggered unit cell becomes significantly more complicated, as we now have to consider the arrangements of ABCD in the $\hat z$-direction as well. In the following we consider 3D analogues of the 4SL(1) and 4SL(2) phases. The inversion symmetry-breaking 4SL(1) ansatz is naturally generalized to a rhombohedral structure in which planes of each sublattice are stacked along the [111] direction of the underlying cubic crystal, with the wavevector ($\pi/2$,$\pi/2$,$\pi/2$) [see Fig.\ref{fig:cartoon3D}(a)]. The 4SL(2) ansatz can be generalized in a number of a different ways. Here we have taken a 2D plane of ABCD sites in arranged clockwise in square plaquettes, and stacked it in the $z$ direction with a second layer having the plaquettes rotated by 90 degrees [see Fig.\ref{fig:cartoon3D}(b)]. These two types of layers are then repeated periodically along the $z$ direction; this pattern preserves inversion symmetry like the 2D 4SL(2) ansatz, but does require now eight sites per unit cell, and thus has 48 total bands.

\begin{figure}[!htb]\centering
\includegraphics[width=0.95\linewidth]{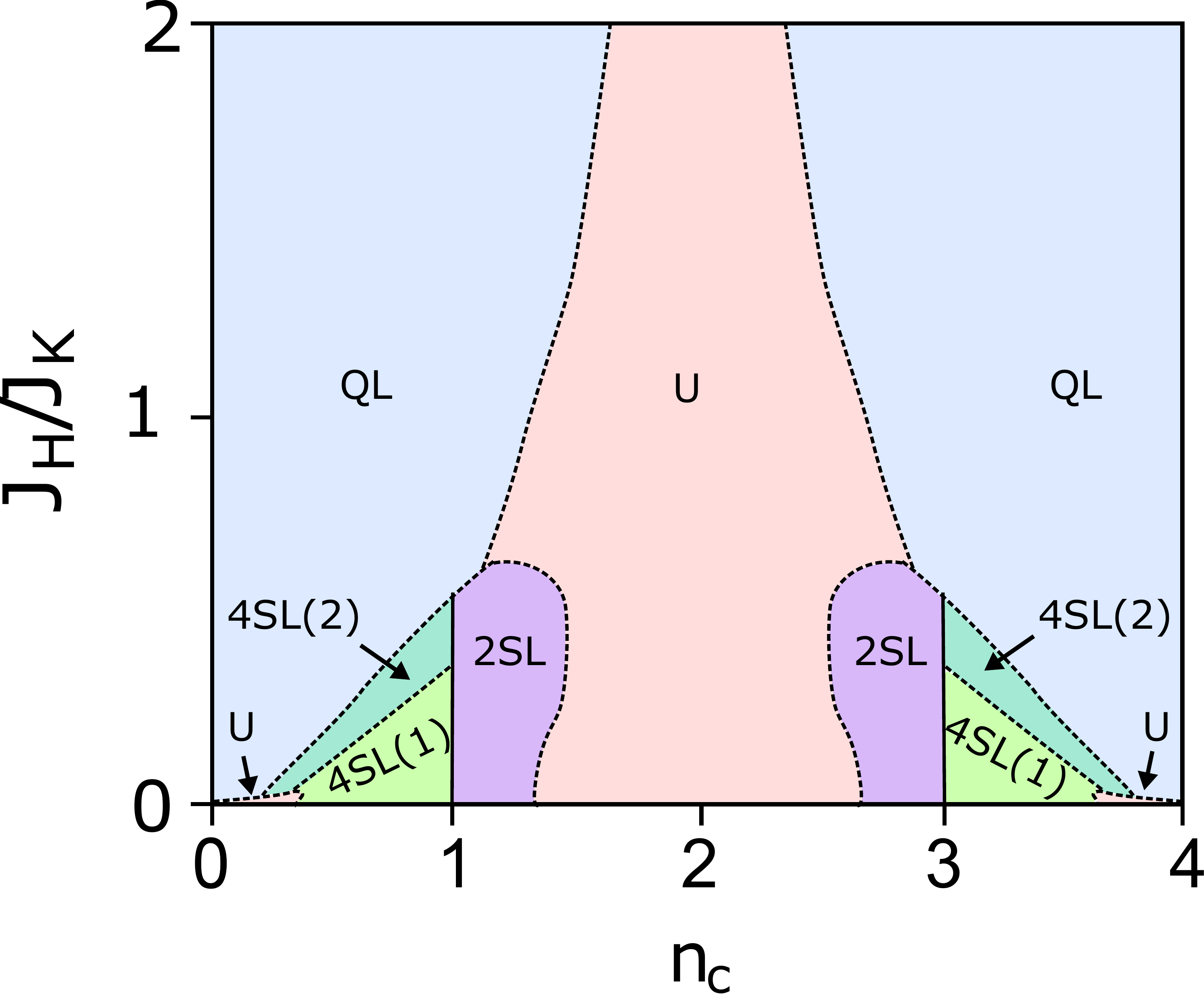}
\caption{Low temperature phase diagram for 3D in the $(n_c, J_H/J_K)$ plane, where the red region (U) indicates ferrohastatic order, teal represents 4SL(2) antiferrohastatic order, light green 4SL(1) antiferrohastatic order, and blue the quadrupolar liquid (QL). The dashed lines indicate first order transitions, while the solid line shows the second order transition between the staggered ansatzes for which $\chi_H=0$. }
\label{fig:PD_3D}
\end{figure}

The phase diagram in the $(n_c, J_H/J_K)$ plane is similar to the 2D cases, with ferrohastatic order around half-filling and antiferrohastatic order moving away from this limit, as shown in Fig. \ref{fig:PD_3D}. One also finds that both versions of the 4SL ansatz are realized here, as in the band non-degenerate 2D case. However, the region with $\chi_H = 0$ is confined to the $n_c=1$ line in the 3D phase diagram. From Fig. \ref{fig:TKU}, one can see that the Kondo temperature is suppressed for all conduction electron fillings compared with 2D. In magnetic field, the $g$-factor is still independent of azimuthal angle but has smaller anisotropy than in 2D (see Fig. \ref{fig:g_factor}).

\section{Finite Temperature}\label{sec:PD_T}

\begin{figure}[!htb]\centering
     \includegraphics[width=\linewidth]{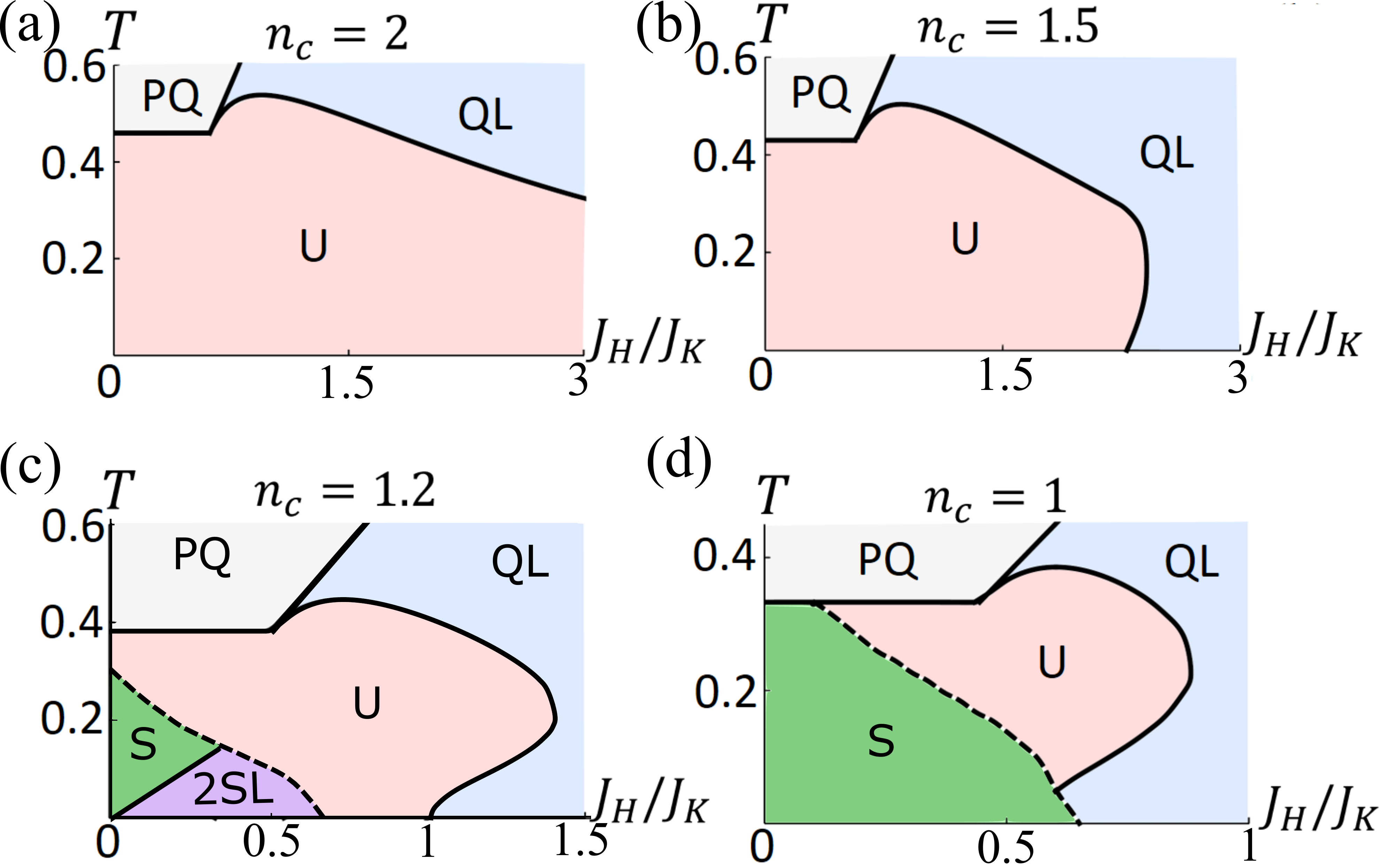}
\caption{Representative phase diagrams in the $(J_H/J_K, T)$ plane for (a) $n_c=2$; (b) $n_c=1.5$; (c) $n_c=1.2$; (d) $n_c=1$. Solid (dashed) lines represent second (first) order transitions. S indicates the antiferrohastatic phase with $\chi_H = 0$.}
\label{fig:PD_JHT}
\end{figure}

In this section, we present representative finite temperature phase diagrams for the 2D model. We have seen that the transition temperatures out of the paraquadrupolar state for all hastatic phases are identical, with clear first order transitions between them at zero temperature. Here, we find that the finite temperature phase diagrams can be similarly complex, with different types of hastatic order favored at different temperatures. We pick four representative conduction electron fillings: $n_c =1$, $n_c = 1.2$, $n_c = 1.5$ and $n_c = 2$, which span ground states from the Kondo insulator to 2SL antiferrohastatic to ferrohastatic order at small $J_H$, and plot the temperature-$J_H/J_K$ phase diagrams in Fig. \ref{fig:PD_JHT}.

First we discuss the effect of increasing $J_H/J_K$ on ferrohastatic order. For small $J_H/J_K$, the transition at $T_K$ into hastatic order is generically second order, and independent of $J_H/J_K$. The transition into the QL occurs at $T_Q = J_H/4$. After this line intersects $T_K$, hastatic order develops out of the quadrupolar liquid, typically still via a second order transition that is initially enhanced by $J_H$, but then suppressed. Generically, we obtain reentrant phase transitions between the ferrohastatic and quadrupole liquid phases, which we believe to be an artifact of the mean-field theory. While slave particle theories typically work quite well for capturing low temperature properties, they can fail at higher temperatures, particularly in capturing the nature of phase transitions \cite{desilva}. In addition, we neglect superconductivity in this paper, but it is well known that the single-channel Kondo-Heisenberg model gives rise to superconducting dome completely concealing the phase transition between heavy Fermi liquid and magnetic order \cite{coleman89, andrei89}. Here, the evolution of our ferrohastatic phase should be identical to the one-channel model, and so we expect a dome of quadrupolar resonating valence bond superconductivity to conceal these phase transitions. 

Increasing $J_H/J_K$ clearly favors ferrohastatic order over the antiferrohastatic orders. The first order antiferrohastatic transition temperature decreases monotonically, while the ferrohastatic temperature initially rises. This is true even when ferrohastatic order is never the ground state for a particular $n_c$, as for $n_c = 1$. Unsurprisingly, increasing $J_H$ increases the transition temperature at which $\chi_H$ turns on inside the antiferrohastatic phase, here the boundary between 2SL and the $\chi_H = 0$ antiferrohastatic phase for $n_c = 1.2$. The antiferrohastatic case is also likely unstable to superconductivity.

\section{Channel symmetry breaking: effect of magnetic field}\label{sec:magnetic}

Magnetic field is a channel symmetry breaking field that, in the isolated $\Gamma_3$ limit, couples only to the conduction electrons. In this artificial limit, $\vec{B}$ only favors ferrohastatic order, on account of its ferromagnetic moment, and disfavors antiferrohastatic order. In this section, we consider the more realistic case discussed in section \ref{section:mag_field}, where the $f$-electrons couple to $B^2$, and all types of hastatic order are suppressed for sufficiently large $\vec{B}$ due to the suppression of the Kondo effect. At intermediate fields, these two effects compete to give ferrohastatic order a slight dome in field. 

\begin{figure}[!htb]\centering
	\includegraphics[width=\columnwidth]{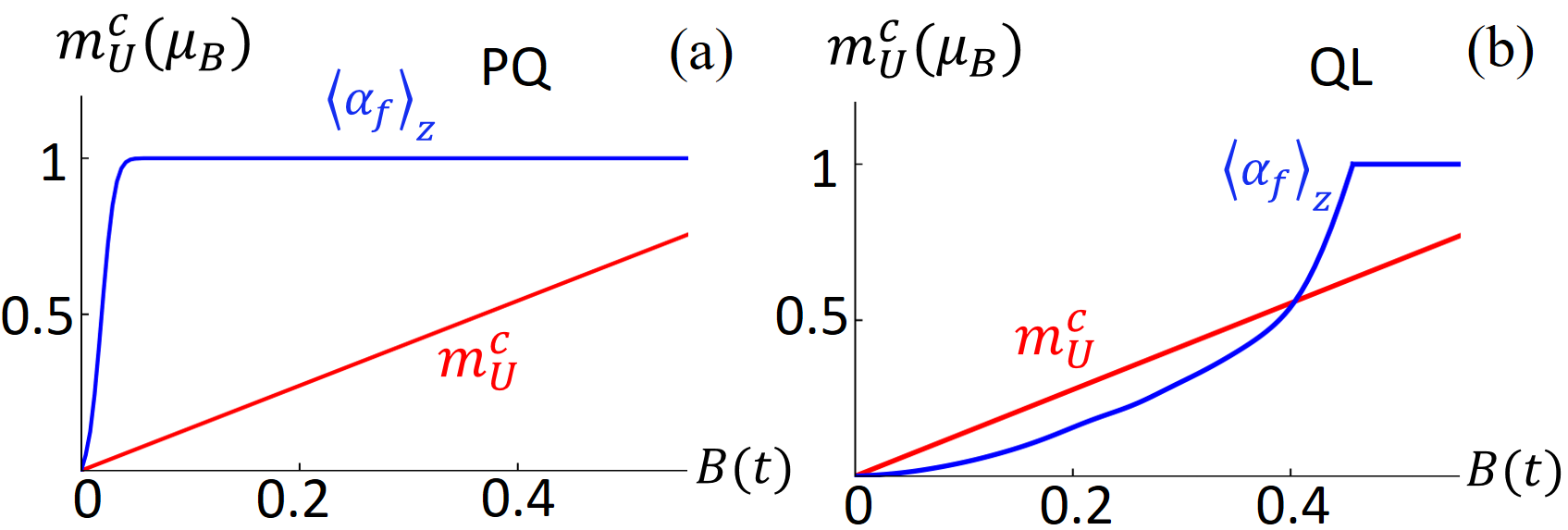}
\caption{Bare conduction electron moments (red) and $f$-electron polarization $\langle \alpha_z \rangle_f$ (blue) for $B||z$ in the (a) paraquadrupolar and (b) quadrupolar liquid phases for $n_c=1.2,J_H/J_K=2/3$ at low, but nonzero, temperature; $B$ is measured in units of $t$. In both phases, the conduction electron moment grows linearly in $B$; while $\langle \alpha_z \rangle_f$ increases as a modified Brillouin function in the paraquadrupolar phase, and grows quadratically in the quadrupolar liquid phase until the short range quadrupolar order is destroyed when the moments saturate. Note that we plot $\langle \alpha_z\rangle$, which is predominantly a quadrupolar moment for small fields, with a dipole moment growing linearly in $B$. }
\label{fig:B_T_N_moments}
\end{figure}

First, we discuss the effect of magnetic field on the competing non-hastatic phases. As there is no coupling of the moment direction to the lattice, we take $\vec{B} || \hat z$ without loss of generality.  The $f$-electron dipole moments do coupled more weakly to fields along $[110]$ or $[111]$, which will cause some quantitative, but no qualitative differences.
Magnetic field will generically induce both $c$- and $f$-electron dipole moments. In both the quadrupolar liquid and paraquadrupolar phases, the conduction electron moment simply grows linearly in $B$, as it would in a normal metal. The $f$-electron dipole moment convolves two effects: the induced dipole content of the $f$-electron doublet pseudospin, $\langle J_z\rangle$, and the polarization of the pseudospin, $\langle \vec{\alpha}\rangle$.
In Fig.~\ref{fig:B_T_N_moments}, we plot the conduction electron moment and pseudospin polarization $\langle \alpha_z \rangle$ versus $B$. The $f$-electron pseudospin moments are predominantly quadrupolar at low fields, with a dipolar contribution growing linearly in field, as shown in Fig. \ref{momentsinfield}; these moments, $\langle J_z \rangle$ will continue to evolve in field even after $\langle \alpha_z \rangle$ is fully polarized.

In the paraquadrupolar phase, $\langle \alpha_z \rangle$ follows a Brillouin function, adjusted for the $B^2$ nature of the coupling and saturating to unity. Once the $f$-electrons are polarized, this phase is equivalent to ferroquadrupolar order, and will be a polarized light Fermi liquid.  

The quadrupolar liquid phase is suppressed by magnetic field as the polarization of the local moments competes with antiferroquadrupolar correlations,
\begin{equation}
T_{QL} = \frac{J_H}{4}\mathrm{sech}^2\B(\frac{3B^2}{2T_{QL}\Delta}\B),\; B_{QL} = \frac{1}{3}\sqrt{\frac{2J_H \Delta}{3}},
\end{equation}
where $\Delta$ is the splitting to $\Gamma_4$, which we set equal to $t$ here. $T_{QL}$ is a second order phase transition derived from $d^2F/d\chi^2=0$ without any solution beyond some finite $B_{QL}$; $B_{QL}$ therefore indicates a first order phase transition. This suppression is also shown by the gradually increasing $\langle \alpha_z \rangle$, which grows much more slowly than in the paraquadrupolar phase. Again, the $f$-electron pseudospin has both uniform quadrupolar and dipolar components. Here, the staggered pseudospin moments remain uniformly zero, although in true antiferroquadrupolar order they would initially be large due to the quadrupolar order, and gain dipolar components in field.

In this model, the hastatic spinor is not pinned to the lattice at all, and so we assume that the ferrohastatic spinor immediately aligns with the external field, while the antiferrohastatic spinor aligns in the perpendicular plane, so that it may cant along the field direction. Even in more realistic Anderson models, the pinning of the hastatic order remains extremely weak.

\begin{figure}[!htb]\centering
	\includegraphics[width=\linewidth]{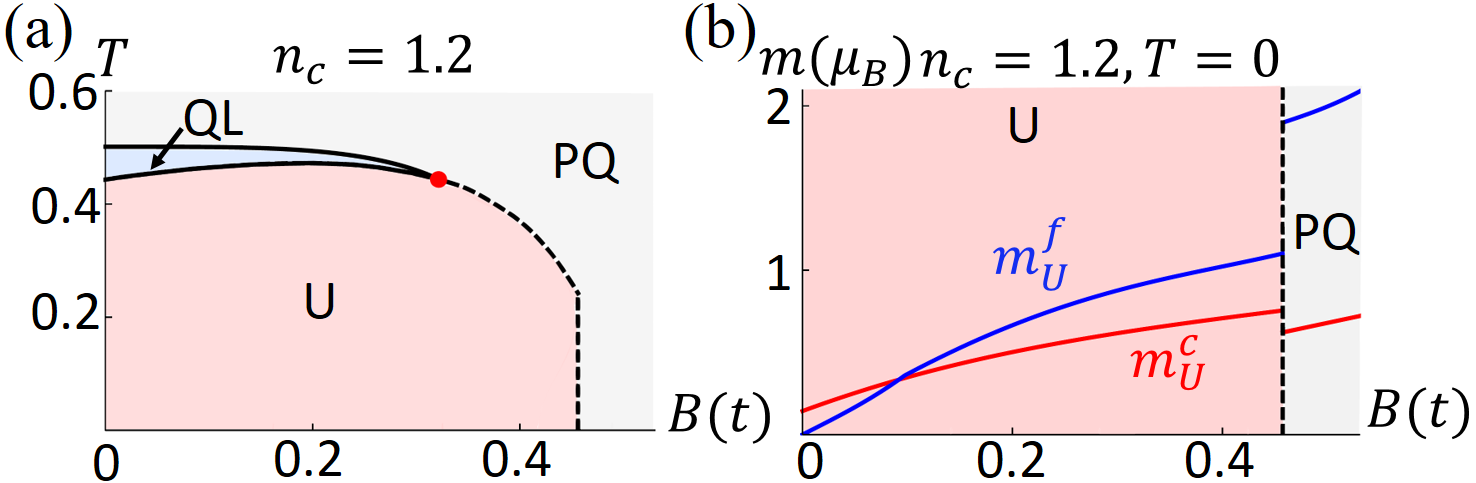}
\caption{(a) Phase diagram in magnetic field for $n_c=1.2,J_H/J_K=2/3$, with a wedge of quadrupolar liquid (QL) above ferrohastatic order (U) at small fields, and a first order transition to a fully polarized light Fermi liquid (PQ) at higher fields. (b) The dipole moments as a function of field at zero temperature, where $m_U^c$(red line) is the uniform conduction electron moment and $m_U^f$(blue line) is the uniform $f$-electron moment. Note that the kink around $B=0.1 t$ is due to a non-universal Lifshitz transition of a hybridized band. }
\label{fig:B_T_nc1dot2}
\end{figure}

To investigate how the hastatic phases respond to magnetic field, we examine two representative phase diagrams in field and temperature. For the first, we choose $n_c$ and $J_H/J_K$ such that in zero field, the quadrupolar liquid develops first, followed by a transition to ferrohastatic order at a lower temperature, as seen in Fig. \ref{fig:B_T_nc1dot2}(a). As magnetic field increases, the quadrupolar liquid is suppressed and ferrohastatic order is first enhanced and then suppressed, leading to a wedge of quadrupolar liquid above ferrohastatic order, the latter vanishing via a direct first-order transition to the polarized paraquadrupolar order at larger fields. The uniform $c$ and $f$-electron moments are shown in Fig. \ref{fig:B_T_nc1dot2}(b), where $m^c$ starts at a finite value and grows gradually, while the $f$-electron moment vanishes in zero field, but quickly surpasses $m^c$. Note that the polarization of the $f$-level also induces small quadrupolar moments.  



\begin{figure}[!htb]\centering
	\includegraphics[width=\linewidth]{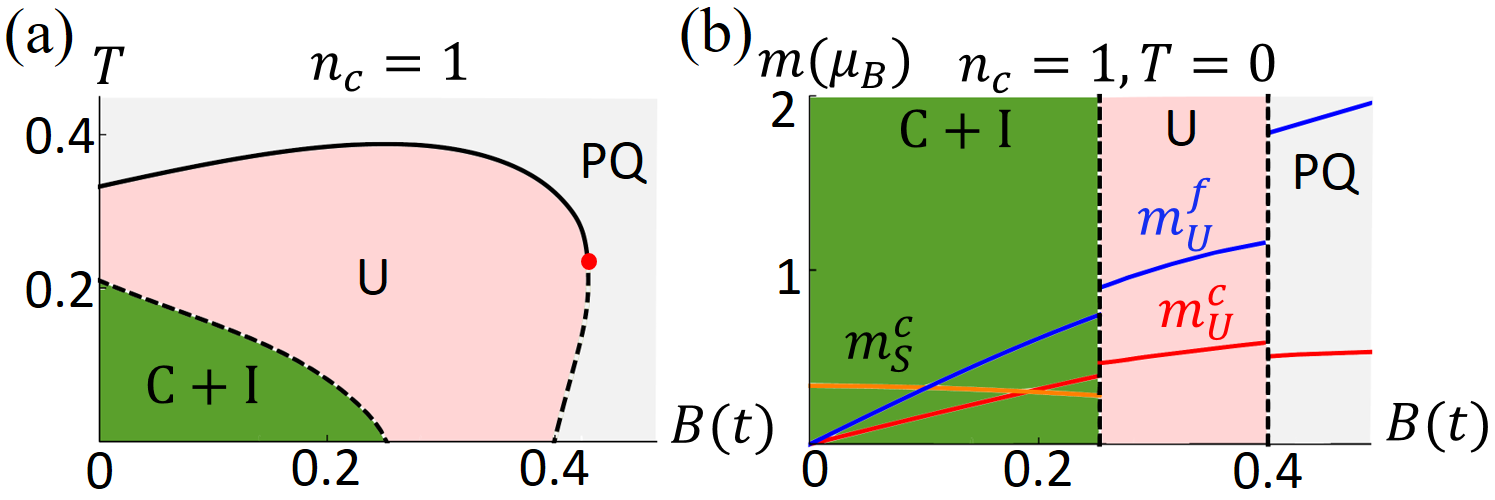}
\caption{(a) Phase diagram in magnetic field for $n_c=1,J_H/J_K=1/3$, which contains a low temperature canted Kondo insulator phase (C+I) uniformly suppressed in field, and a higher temperature region of ferrohastatic order (U) initially enhanced in field before a first order transition to a fully polarized light Fermi liquid (PQ). (b) The magnetic moments at zero temperature, where $m_S^c$ (orange line) is the staggered conduction electron moment, and $m_U^{(c/f)}$ (red/blue lines) are the uniform conduction/$f$-electron moments.}
\label{fig:B_T_nc1}
\end{figure} 

In the second representative phase diagram, show in Fig. \ref{fig:B_T_nc1}, we explored the effect on the competition between ferrohastatic and canted phases. We take quarter-filling ($n_c = 1$), where the hastatic ground state is always antiferrohastatic order with a full Kondo insulating gap and no $f$-electron correlations ($\chi_H = 0$), and then take sufficiently large $J_H/J_K$ such that the ferrohastatic phase, which can coexist with short range antiferroquadrupolar correlations, is favored at higher temperatures. Magnetic field will cause the antiferrohastatic moments to gradually cant; if initially $|V_\uparrow| = |V_\downarrow|$, then $V_\downarrow$ vanishes at the first order transition to ferrohastatic order at large magnetic field. In Fig. \ref{fig:B_T_nc1} (b), both the staggered conduction and the uniform conduction and $f$-electron moments are shown in the canted phase, with only the uniform moments appearing in the ferrohastatic phase, after the first order transition. Both the uniform $f$ and $c$ moments grow roughly linearly with field, while the staggered $c$ moment is slowly suppressed. There is never any staggered $f$-electron moment.

Finally, we note that magnetic field breaks all of the symmetries broken in ferrohastatic order, and so in field, the ferrohastatic spinor actually develops as a crossover. Effectively, the second order transition is broadened in field; however, as magnetic field couples to a tiny moment responsible for only $\sim |V|^2 R\log 2$ entropy, in contrast to the large entropy of condensation, $\sim R \log 2$, the broadening will be significantly less than for a purely ferromagnetic transition with the same entropy.

\section{Pseudospin symmetry breaking: the effect of strain}
\label{sec:strain field}

Strain is the primary pseudospin symmetry breaking field in the quadrupolar Kondo model, playing the role usually played by magnetic field in the single-channel Kondo model. Here we consider the strain components that couple linearly to the quadrupolar moments of the $\Gamma_3$ doublet: $\epsilon_{zz}$, which couples to $\alpha_z$ for both conduction and $f$-electrons, and $\epsilon_{xx-yy}$, which couples to $\alpha_x$. These are related by cubic symmetry, and so will behave identically. We neglect other strains, which require more complicated couplings. Both conduction and $f$-electron quadrupolar moments couple to strain with significantly different and materials dependent constants. Typically, the conduction electron strain for $d$-electrons is larger than that for $f$-electrons by one to two orders of magnitude\cite{Nakamura94, Hazama00}. In order to tease apart the two behaviors, we consider the coupling to conduction and $f$-electrons separately, setting the coupling constant $\kappa =1$ in each case, with the understanding that real materials will interpolate between the two.

\subsection{Coupling to conduction electrons}

\begin{figure}[!htb]\centering
 	\includegraphics[width=\linewidth]{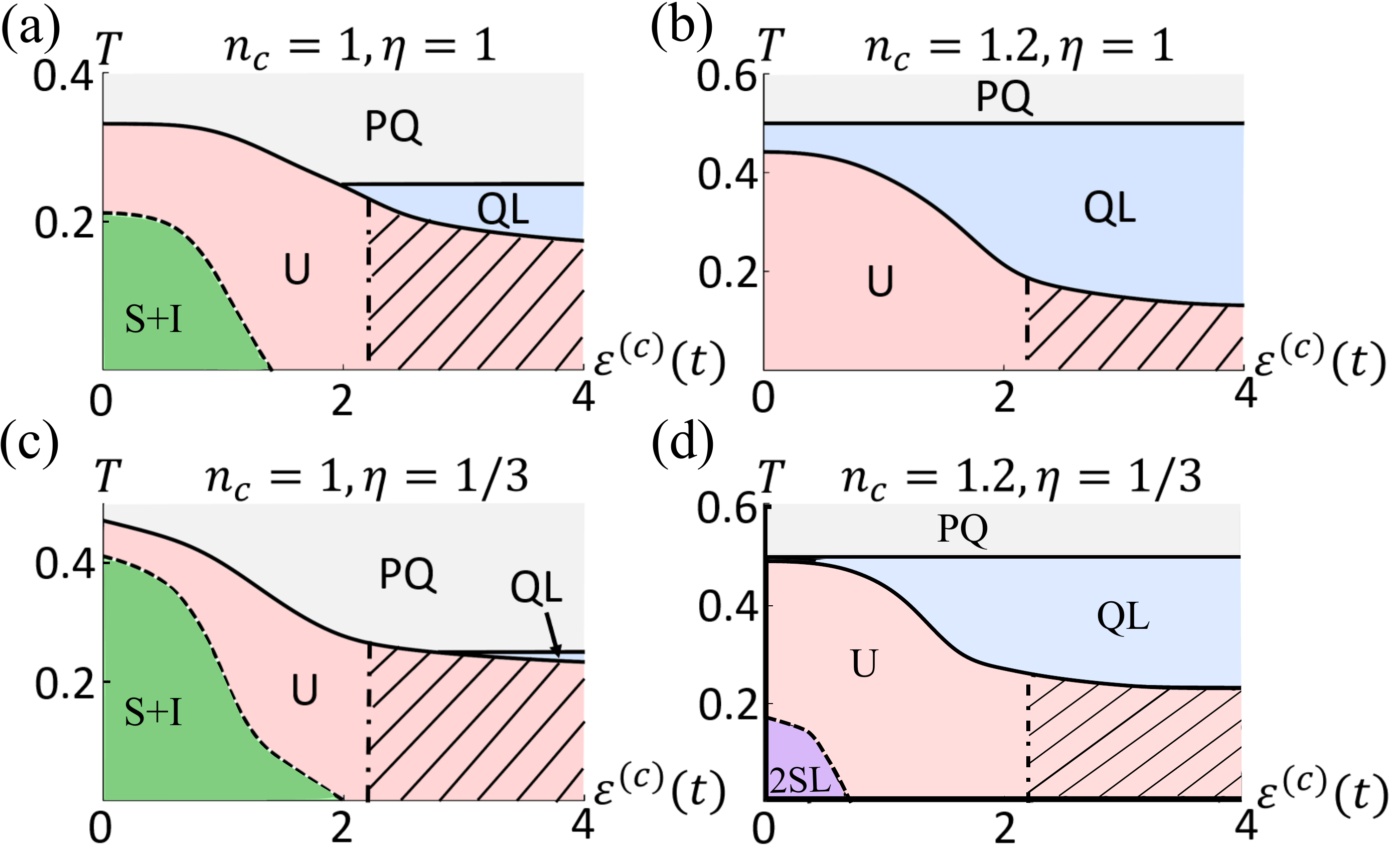}
\caption{Four example phase diagrams in strain, where strain couples only to the conduction electrons. (a) $n_c=1,J_H/J_K=1/3, \eta=1$; (b) $n_c=1.2,J_H/J_K=2/3, \eta=1$; (c) $n_c=1,J_H/J_K=1/3, \eta=1/3$; (d) $n_c=1.2,J_H/J_K=2/3, \eta=1/3$. We can see that both ferro- and antiferrohastatic order are suppressed, with ferrohastatic order suppressed more slowly. The dot-dashed line indicates the cross-over to the fully polarized conduction sea where the Kondo effect is absent. }
\label{fig:E_T_c}
\end{figure}

First, we consider the coupling to conduction electrons, where strain acts like a pseudo-magnetic field and splits the bands. For zero strain, the hastatic Kondo singlet is an equal superposition of $\langle c_{1\sigma}\dg f_1\rangle$ and $\langle c_{2\sigma}\dg f_2\rangle$, forming a quadrupolar particle-hole singlet that screens the $f$-electron moments. Strain breaks this pseudospin symmetry, $\langle c_{1\sigma}\dg f_1\rangle \neq \langle c_{2\sigma}\dg f_2\rangle$, and reduces the screening. There is a region of partial screening that persists until $\langle c_{2\sigma}\dg f_2\rangle = 0$, after which point the conduction electron sea is totally polarized, and the Kondo effect is no longer relevant; this region is indicated in Fig. \ref{fig:E_T_c} by hashmarks, with the transition indicating the development of the non-Kondo hybridization of $\langle c_{1\sigma}\dg f_1\rangle$.

All hastatic orders are suppressed by conduction electron strain, with ferrohastatic order suppressed more slowly. Example phase diagrams in temperature and strain ($\epsilon$) are shown in Fig. \ref{fig:E_T_c}, both with varying conduction electron density, $n_c$, and varying band anisotropy, $\eta$; the results are relatively parameter independent. In our mean-field calculation, the $f$-level is always pinned to the Fermi level, and so at least one ($\alpha$) conduction band will always overlap the $f$-level, even for large strain. As these two bands have the same symmetry, they can always hybridize, and so one of $\langle c_{\alpha\sigma}\dg f_\alpha\rangle$ with $\alpha = 1$ or $2$ will always be nonzero. This residual hybridization at large strain is an artifact of the mean-field theory, as in the absence of the Kondo effect, the $f$-level will not be pinned near the Fermi surface. Also note that, as we neglect the $f$-electron strain coupling here, the paraquadrupolar and quadrupolar liquid regions are unaffected. 

\subsection{Coupling to $f$-electrons}

\begin{figure}[!htb]\centering
	\includegraphics[width=\linewidth]{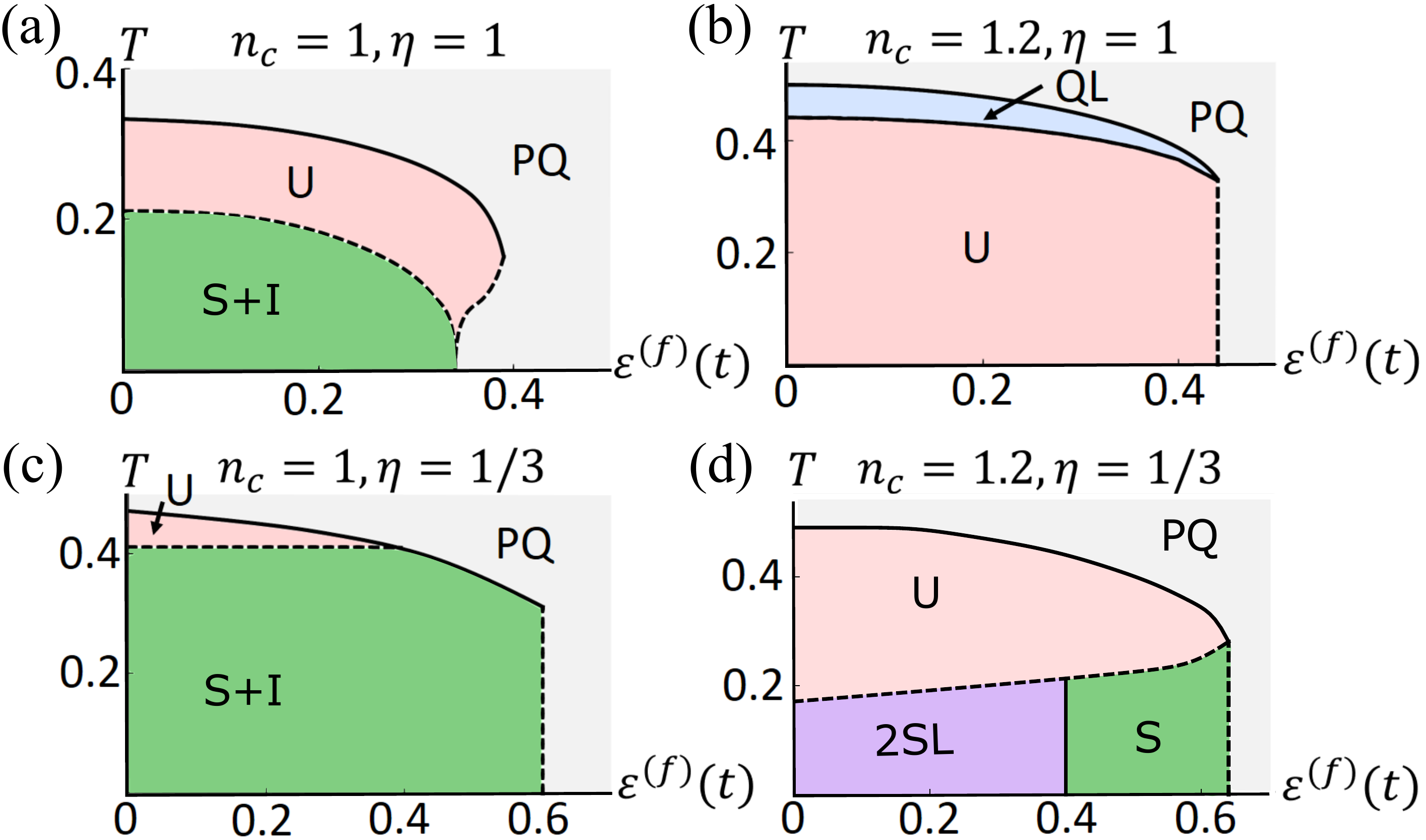}
\caption{Four example phase diagrams in strain, where strain couples only to the $f$-electrons: (a) $n_c=1,J_H/J_K=1/3, \eta=1$; (b) $n_c=1.2,J_H/J_K=2/3, \eta=1$; (c) $n_c=1,J_H/J_K=1/3, \eta=1/3$; (d) $n_c=1.2,J_H/J_K=2/3, \eta=1/3$. (a) and (b) show that for degenerate conduction bands, ferro- and antiferrohastatic orders are suppressed similarly; however, for non-degenerate conduction bands, antiferrohastatic order is favored over ferrohastatic order. }
\label{fig:StrainField_PD}
\end{figure}

Next we consider strain coupled only to the $f$-electrons, which acts like the magnetic field in the single-channel Kondo model and splits apart the $f$-level. This splitting suppresses hastatic order monotonically until the screening is totally lost. The transition is generically first order, as the paraquadrupolar phase is simultaneously enhanced by the $f$-electron quadrupolar moment polarization. The quadrupolar liquid is also suppressed, just as antiferroquadrupolar ordering would be suppressed, with $T_{QL} = J_H/4\cdot\mathrm{sech}^2(\varepsilon/(2T_{QL}))$. Example phase diagrams are shown in Fig. \ref{fig:StrainField_PD}. For perfectly degenerate conduction bands, both ferro- and antiferrohastatic orders are suppressed similarly, but for non-degenerate conduction bands, antiferrohastatic order is favored over ferrohastatic.  

\section{Experimental signatures of hastatic order}\label{sec:experiments}

Ultimately, hastatic order is a channel symmetry breaking heavy Fermi liquid, and as such its key signatures fall into two categories: heavy fermion formation, including heavy masses and hybridization gaps; and channel symmetry breaking, including symmetry-breaking moments. In terms of the hastatic spinor, we can associate these signatures with the development of a nonzero amplitude and a symmetry-breaking direction, respectively. In our mean-field treatment, both of these develop simultaneously at a phase transition, with the full $S(T_K) = R \ln 2$ entropy at the transition. Fluctuations will generically split these two features such that the non-symmetry breaking, heavy fermion characteristics develop at a higher crossover temperature, $T^*$, with the symmetry-breaking phase transition occurring at a lower temperature, $T_K$. This behavior can be understood by thinking of the hastatic spinor as a tiny magnetic moment in the excited Kramers doublet. As the temperature decreases below $T^*$, the ground state quadrupole moment and its associated $R \ln 2$ entropy is slowly quenched by Kondo screening. Simultaneously, the hastatic moment and its associated $n_V(T) R \ln 2$ entropy grow in amplitude, with $n_V(T) \propto |V|^2$. At $T_K$, this small hastatic moment orders. If these temperatures are well separated, $T_K$ may not have substantial signatures, due to the small associated entropy; if they are coincident, the signature at $T_K$ reflects the full entropy. Real systems will likely be somewhere in between. $T_K/T^*$ is likely be suppressed by lower dimensionality and frustration, as with any magnetic ordering; the Pr ions in the 1-2-20 materials sit on a diamond lattice, which is frustrated.

The main point is that, while heavy fermion features are a key signature of hastatic order, they may develop above the phase transition. Symmetry-breaking signatures must, by contrast, develop at the phase transition. For these, the key difficulty is distinguishing hastatic order from the competing quadrupolar order, especially as the associated zero-field magnetic moments of hastatic order are likely to be extremely small in praseodymium-based systems due to the small degrees of mixed valency; in-field measurements are then key to distinguish these orders. As magnetic field destabilizes antiferrohastatic order, such investigations will require relatively low fields.

\fg{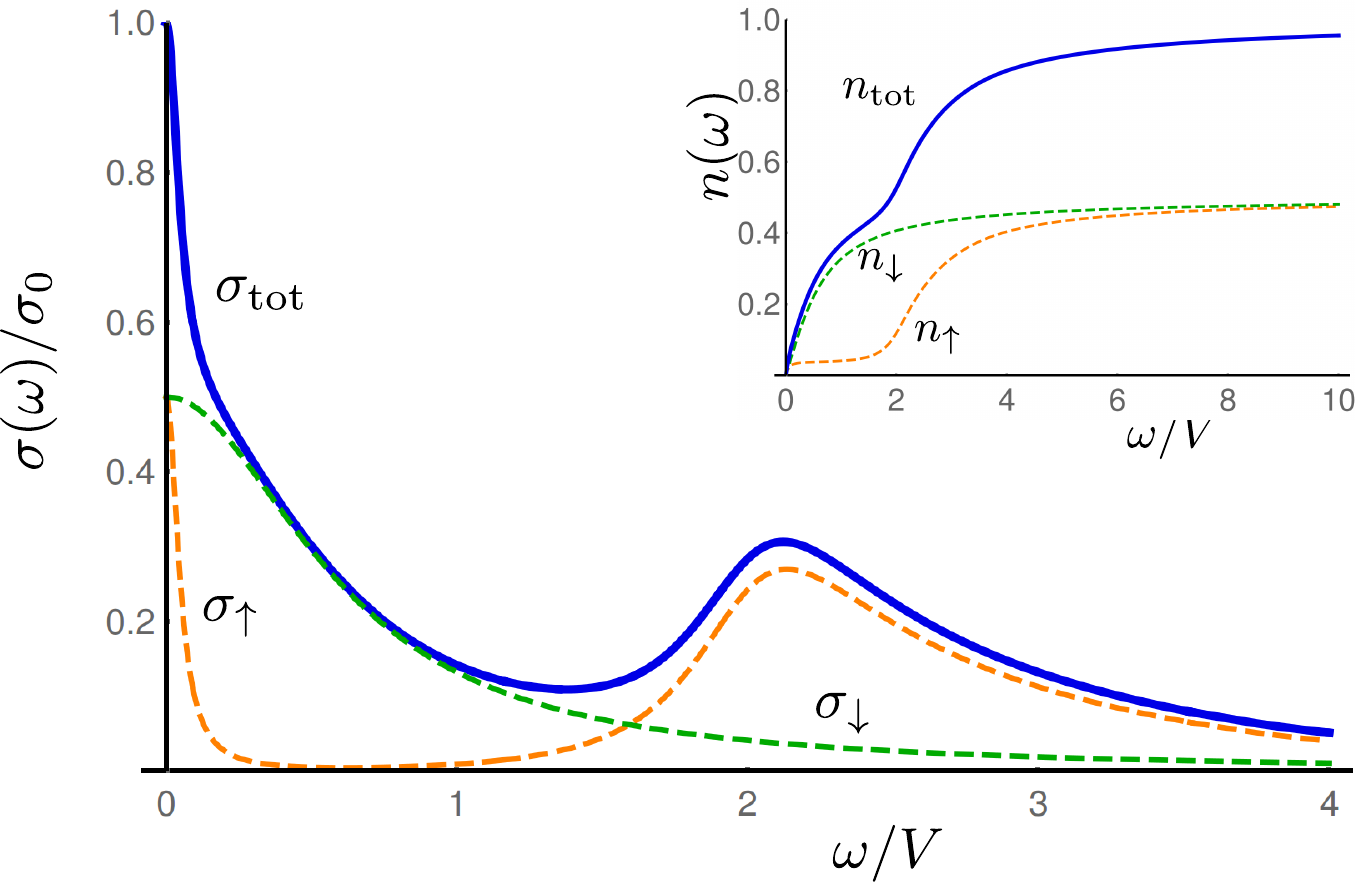}{optical}{Optical conductivity of ferrohastatic order. This figure is based on a simple Drude model for the unhybridized band, combined with the optical conductivity for a simple heavy fermion band\cite{Coleman_book}. (Inset): The integrated spectral weight, $n(\omega)$ shows a kink where $n(\omega) = n_{tot}/2$.}

\subsection{Ferrohastatic order}

Ferrohastatic order is characterized by spin polarized bands, where one band has significantly heavier masses than the other. These bands affect a number of experimental measurements.
\begin{itemize}
\item{{\bf Spin-polarized heavy effective masses:} Spin up and down bands will have significantly different effective masses, $m^*_\uparrow \gg m^*_\downarrow$. This splitting should be seen in quantum oscillations, although resolving the spin-polarization of the heavy/light bands would require a technique like spin-resolved angle-resolved photo-emission (ARPES) or spin-resolved scanning tunneling microscopy (STM).}
\item{{\bf Heavy Fermi liquid signatures:} The heavy bands strongly affect the thermodynamics properties, and ferrohastatic order should behave like a conventional heavy Fermi liquid, with enhanced Sommerfeld coefficient $C/T$, Pauli susceptibility, and $AT^2$ coefficient in the resistivity, among other signatures. There will be a jump in the Hall conductivity across $T_K$, as the Fermi surface volume jumps from $n_c$ to $n_c + 1$ \cite{Coleman01, Paschen04}, which could potentially be observed as $T_K$ is suppressed in field.}
\item{{\bf Half-hybridization gap:} Half of the relevant high temperature bands develop a hybridization gap, leading to additional structure in the optical conductivity. The optical conductivity is the sum of a simple Drude model for the unhybridized band $\sigma(\omega+i\delta) = \frac{ne^2}{m}\frac{1}{\Gamma-i \omega}$ with a typical heavy fermion optical conductivity for the hybridized band \cite{Coleman_book},
\begin{align}
&\sigma_{H}(\omega+i \delta)  = \frac{ne^2}{2m}\frac{1}{-i \omega+\Gamma}\Big[1+\cr
&\;\;\left.\frac{V^2}{(\omega+i \Gamma)(z_+-z_-)}\left(\ln \frac{z_+ +\frac{\omega}{2}}{z_+ -\frac{\omega}{2}} - \ln \frac{z_- +\frac{\omega}{2}}{z_- -\frac{\omega}{2}}\right)\right],\cr
& \;\mathrm{where}\; z_{\pm} = \lambda \pm \sqrt{\left(\frac{\omega}{2}\right)^2+V^2\frac{\omega}{\omega+i \Gamma}}.
\end{align}
We plot the real part, $\sigma_1(\omega)$, in Fig. \ref{optical}; there is a sum of both light and heavy Drude peaks, plus an interband transition from the heavy band. The integrated spectral weight, $n(\omega) = \frac{2 m}{e^2} \int_0^\omega d\omega' \sigma_1(\omega')$ has structure at $n_{tot}/2$.}
\item{{\bf Magnetic moments in field:} There are small conduction electron moments, $|m_c| \sim T_K/D$ in zero field; these are weakly pinned, and so domains will align quickly and grow in external field. $f$-electron dipole moments are generated in field, and can become substantial. Ferrohastatic moments should be contrasted with those present in ferroquadrupolar order, where $f$-electron dipole moments are also induced in field. Ferrohastatic moments are \emph{always} parallel to the field direction, while ferroquadrupolar moments induced in field will have perpendicular components for some field directions\cite{Shiina97,taniguchi16}. While $T_K$ is a phase transition in zero field, it will broaden out into a crossover in finite fields, as magnetic field breaks the symmetries of ferrohastatic order. However, as the field couples only to the small conduction electron moments, not the large hybridization, this broadening should be smaller than for a comparable ferromagnet.}
\end{itemize} 

A number of Pr compounds with $\Gamma_3$ doublets may develop ferrohastatic order. PrTi$_2$Al$_{20}$ develops $O_2^0$ ferroquadrupolar order at ambient pressure, detected by induced dipole moments perpendicular to an applied field $B || [111]$ \cite{matsubayashi12,taniguchi16}. However, under pressure, $T_Q$ is suppressed, and hidden beneath a superconducting dome \cite{matsubayashi12, Nakatsuji_talk}. This pressure phase diagram is reminiscent of the Doniach phase diagram for one-channel Kondo materials \cite{doniach77}, which leads us to expect that hastatic order will be revealed at still higher pressures, as long as the pressure is sufficiently hydrostatic to avoid splitting the $\Gamma_3$ doublets.

In addition, several compounds, including PrV$_2$Al$_{20}$, Pr(Ir,Rh)$_2$Zn$_{20}$ and PrPb$_3$, contain new heavy Fermi liquid regions in intermediate magnetic fields \cite{onimaru16b, yoshida17, sato10}. These regions have enhanced Sommerfeld coefficients, $C/T$, and resistivity $AT^2$ coefficients. This behavior is consistent with ferrohastatic order, as discussed above, and initially suggested in Ref. \onlinecite{onimaru16a}. Measurement of the in-field moments along different field directions via neutron diffraction or NMR in these intermediate phases could resolve whether these are hastatic order or new multipolar phases. Spectroscopic measurements, such as optical conductivity, ARPES, or STM, could provide further evidence for hastatic order through the detection of the half-hybridization gap and heavy quasiparticle bands. Hall conductivity measurements could detect changes in Fermi surface volume as the $f$-electrons hybridize with the conduction bands and participate in the Fermi sea. Similarly, quantum oscillations could see the spin splitting of effective masses, as has already been seen in dHvA on PrPb$_3$, whose high field phase shows a difference in effective mass for different spin bands\cite{endo02}.

\subsection{Antiferrohastatic order}

Antiferrohastatic order is a fully hybridized phase, with no net moments. In cubic systems, the conduction electrons hybridize with non-magnetic local moments, and so the hybridized bands lose much of their sensitivity to magnetic field. The key signatures are:

\begin{itemize}
\item{{\bf Full hybridization gap and heavy quasiparticles:} As all conduction electron bands hybridize, there will be a full hybridization gap that could be measured in optical conductivity, ARPES, or STM. Antiferrohastatic order should exhibit all the usual thermodynamic signatures of heavy Fermi liquids.}
\item{{\bf Band response to magnetic field:} The heavy bands are fairly insensitive to magnetic field, and will generically shift linearly in $B$ with a very small coefficient. If the $f$-electron hopping is negligible, and the bands thus Kramers degenerate, the $g$-factor will be suppressed by a factor of $T_K/D$. Unfortunately, as antiferrohastatic order is quickly destroyed in field, the quantum oscillations measurements that worked well for URu$_2$Si$_2$ \cite{Ohkuni99, Altarawneh12} will likely not work here; a measurement like electron spin resonance (ESR) might be more successful, although ESR can be difficult in heavy fermions \cite{Abrahams08}.} 
\item{{\bf Phase evolution in magnetic field:} Antiferrohastatic order is destroyed by moderate magnetic fields, even as it develops both uniform conduction and $f$-electron moments due to canting. These moments allow a clear distinction from antiferroquadrupolar order, which develops both uniform and staggered magnetic dipole moments in field. Antiferrohastatic order develops no staggered $f$-electron dipolar moments, but in the cubic case will generically cant to develop both uniform $f$ and $c$ dipole moments.}
\end{itemize}

Of the known Pr compounds with $\Gamma_3$ doublets, the zero field phases of PrTi$_2$Al$_{20}$, PrPb$_3$ and PrIr$_2$Zn$_{20}$ have been positively identified as quadrupolar order via neutron diffraction \cite{sato12,onimaru05,iwasa17} or NMR \cite{taniguchi16}. PrV$_2$Al$_{20}$ exhibits a double transition into two unknown phases, where no in-field moments have been reported \cite{tsujimoto14}. These have been proposed to be quadrupolar and octupolar orders \cite{freyer17}, but alternately either or both of the phases could be some form of hastatic order. Similarly, no moments have yet been reported for PrRh$_2$Zn$_{20}$, and it remains a potential candidate. Optical conductivity or tunneling measurements of the hybridization gap could positively identify antiferrohastatic order. Differentiating between different types of antiferrohastatic order, 2SL and 4SL, is likely only possible by examining the splitting of different bands at the $\Gamma$ point with a measurement like quasiparticle interference (QPI) in STM, although a more detailed microscopic theory may make the additional symmetry breaking manifest in other experimental quantities, like symmetry-breaking hybridization gaps.

\section{Connection to previous theoretical results}\label{sec:theory}

\subsection{Hastatic order in URu$_2$Si$_2$}

Hastatic order was initially introduced to explain the hidden order in the tetragonal compound URu$_2$Si$_2$ \cite{chandra13, chandra15}. In this section, we discuss the key differences between that model and our current treatment.

The major physical difference is between cubic and tetragonal non-Kramers doublets. The cubic ($\Gamma_3$) non-Kramers doublet is completely non-magnetic, while the tetragonal ($\Gamma_5$) one is an Ising doublet: magnetic along $\hat z$ and quadrupolar in the plane. Thus in cubic systems, there are two independent $SU(2)$ symmetries: the $\Gamma_3$ pseudospin, and the excited magnetic doublet. In tetragonal symmetry, the symmetries protecting the two doublets are not fully independent. The physical consequences of hastatic order in cubic and tetragonal symmetries are similar; the main distinction is that here our moments are all parallel to the hastatic spinor and the susceptibility only develops anisotropy along the hastatic spinor direction, while in URu$_2$Si$_2$, there were moments perpendicular to the hastatic spinor, and a symmetry breaking $\chi_{xy}$. These are a consequence of tetragonal symmetry, and the entanglement of the excited and ground state pseudospin symmetries.

There are several key theoretical differences. First, we treat a simplified Kondo-Heisenberg model that ignores spin orbit coupling and any momentum dependence of the hybridization, as well as disallowing non-integral valence. This simplification means that we miss some of the complicated bandstructure effects seen in the URu$_2$Si$_2$ model, like symmetry-breaking hybridization gaps. However, these features are restored in a more realistic Anderson model treatment \cite{vandyke18}. Second, we explicitly include two ($\alpha$) degenerate conduction bands, while the previous model used a single conduction band that hybridized in two distinct symmetries. Our Kondo model is nonsensical without two symmetry-related conduction bands, but an Anderson model allowing for non-local hybridization could explore the difference between symmetry-related and non-symmetry related bands.

Finally, hastatic order in URu$_2$Si$_2$ strictly considered antiferrohastatic order with the hastatic spinor restricted to the basal plane, and with $f$-electron hopping that in principle breaks time-reversal symmetry. However, time-reversal can be restored when the hastatic spinor is in the basal plane, via a gauge transformation that absorbs the spinor into the $f$-electron definition. In cubic symmetry, no such generic gauge transformation exists, and the cubic case is more similar to the tetragonal case with the hastatic spinor along the $\hat z$ direction.

\subsection{Other results on channel-symmetry breaking heavy Fermi liquids}

As our results hold strictly only in the large-$N$ limit, it is important to compare to other methods to see whether or not hastatic order is a competitive ground state of the two-channel $N=2$ Kondo lattice model away from strong coupling. Several studies, both in one and infinite dimensions do show channel symmetry breaking for some parts of phase space.

The one-dimensional two-channel Kondo lattice model was treated with density matrix renormalization group, with algebraic antiferrohastatic order found at quarter filling for sufficiently strong $J_K/t$ \cite{moreno01,schauerte05}. Hastatic order was not detected at other fillings, as $J_K/t$ was too weak, but further studies would be valuable.

The infinite dimensional two-channel Kondo lattice model was studied with dynamical mean-field theory (DMFT) early on by Cox, Jarrell and collaborators, where they found non-Fermi liquid behavior at high temperatures \cite{jarrell96,cox96}, and both odd-frequency superconducting and antiferromagnetic ground states at low temperatures \cite{jarrell97,anders99}; however, channel symmetry breaking was not examined in these early studies. Recently, Hoshino and collaborators have studied this problem extensively in the infinite-dimensional limit using continuous time quantum Monte Carlo (CTQMC) and DMFT to treat the hypercubic two-channel Kondo lattice \cite{hoshino11,hoshino142,kuramoto14,kusunose16}. They find odd-frequency composite pair superconductivity over much of the phase diagram, but also antiferromagnetism, and both uniform and staggered channel orders, which they term diagonal composite order\cite{hoshino11}. The uniform diagonal composite (ferrohastatic) order was found to be stable near half-filling and also characterized by a particle-hole antisymmetric conduction electron magnetic moment, however it was hidden by more stable antiferromagnetic order for the chosen parameters. A staggered diagonal composite (antiferrohastatic) order was found to be stable near quarter-filling \cite{hoshino12}. These infinite dimensional results are consistent with our two and three dimensional phase diagrams, and are promising for the relevance of these types of novel Kondo orders in more realistic models. 

\section{non-Kramers Doniach phase diagram}\label{sec:Doniach}

\begin{figure}[!htbp]\centering
  \includegraphics[width=0.99\columnwidth]{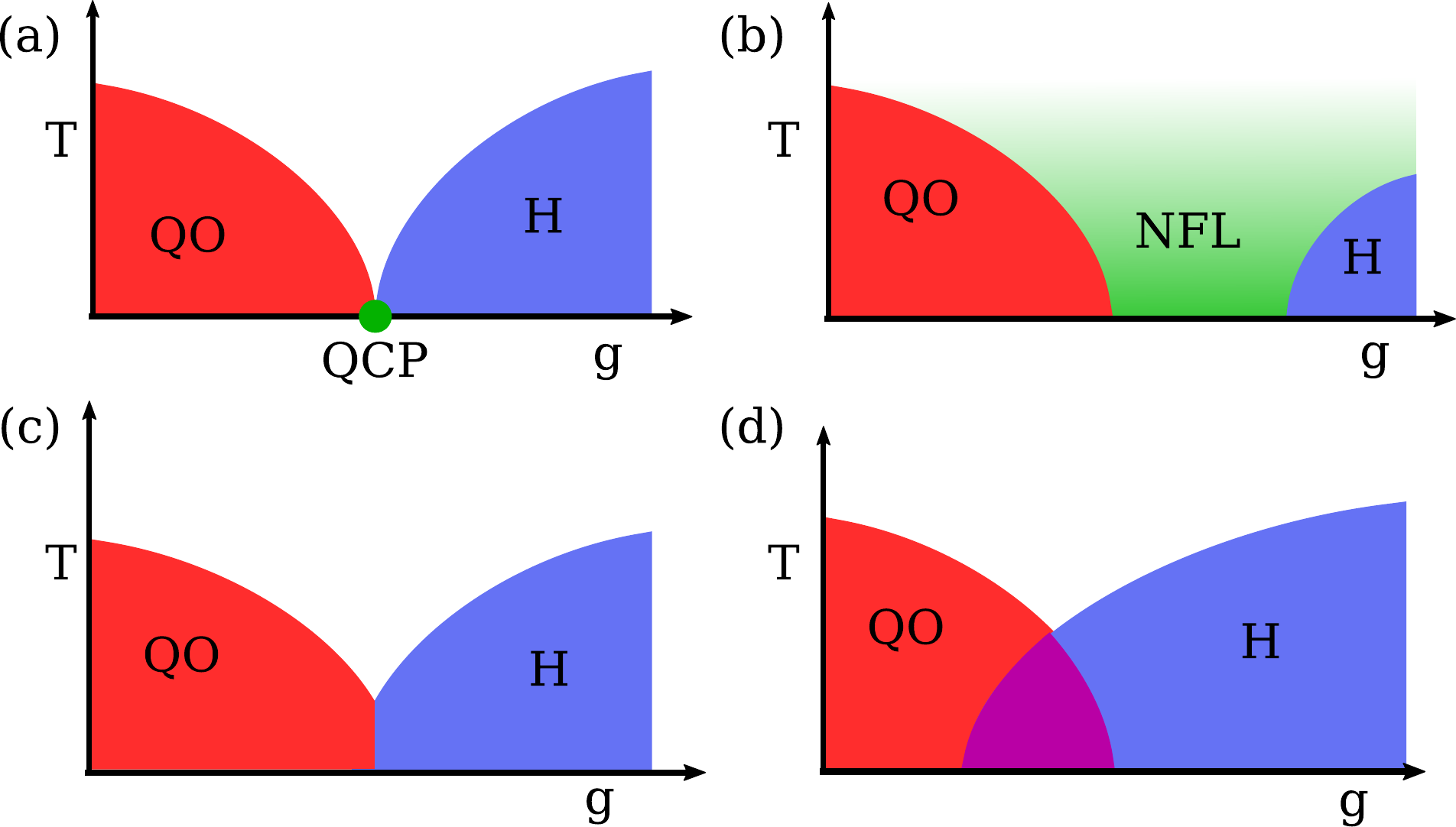}
\caption{Different possible manifestations of the non-Kramers Doniach phase diagram. Here, $g$ is a non-symmetry breaking parameter, like pressure, that suppresses quadrupolar order (QO), shown in red, and favors hastatic order (HO), shown in blue. Both of these are symmetry-breaking phases. There are four distinct possibilities for the transition between the two: (a) The two second order phase transitions can meet precisely at $T=0$, either requiring extreme fine-tuning or exhibiting some sort of deconfined criticality\cite{Senthil04science}. (b) The two end points can be separated in parameter space by a region of non-Fermi liquid behavior (shown in green). (c) The two phases meet at a first order phase transition, with no coexistence, or (d) the two phases can coexist, giving rise to ``small'' and ''large'' Fermi surface quadrupolar order\cite{si06,yamamoto07}.}
\label{fig:quad_crit}
\end{figure}

While our fermionic mean-field approach only partially captures the competition between hastatic and quadrupolar orders, we can speculate about the generic non-Kramers Doniach phase diagram. If quadrupolar order is the ground state at ambient conditions, we can suppress it via some non-symmetry-breaking parameter, $g$, like pressure. There are four distinct ways to destroy the quadrupolar order, shown in Fig. \ref{fig:quad_crit}. As hastatic order \emph{always} breaks symmetries, both quadrupolar and hastatic lines are phase transitions, and no non-symmetry breaking Fermi liquids are allowed at $T = 0$; in other words, the $R \ln 2$ entropy of the local moments must be quenched somehow, and this process \emph{must} break some symmetries. 

The conventional (single-channel) Doniach phase diagram contains two main scenarios for antiferromagnetic quantum criticality: conventional bosonic quantum criticality, where the magnetic order is suppressed independently of the Kondo physics\cite{hertz, millis, moriya}, and the Kondo breakdown scenario, where the N\'{e}el temperature and Kondo temperature go to zero at the same quantum critical point (QCP) \cite{si01, Coleman01}. For non-Kramers materials, the conventional bosonic criticality scenario is no longer allowed, while the Kondo breakdown scenario becomes similar to deconfined criticality\cite{Senthil04science}, with two second order phase transitions driven to zero at the same point [Fig. \ref{fig:quad_crit} (a)]. The hastatic and quadrupolar order quantum critical points can potentially be separated, but only by a non-Fermi liquid region, as in Fig. \ref{fig:quad_crit} (b). Alternately, the two transitions can overlap, either leading to a first order phase transition between the two [Fig. \ref{fig:quad_crit} (c)] or to coexistence [Fig. \ref{fig:quad_crit} (d)]. The phase diagram with coexistence contains two types of quadrupolar order: one without hybridization, and therefore containing a ``small'' Fermi surface; and one with hybridization, and thus a ``large'' Fermi surface -- this phase must break additional symmetries. In analogy with the small and large Fermi surface antiferromagnetic phases discussed for the single channel case \cite{si06, yamamoto07}, we can call these QO$_{L}$ and QO$_{S}$. This scenario provides an alternate explanation for the two phase transitions seen in PrV$_2$Al$_{20}$\cite{tsujimoto14}.

\section{Conclusions}\label{sec:conclusions}

In conclusion, we have used an $SU(N)$ mean-field treatment of the two-channel Kondo-Heisenberg model to explore the properties and stability of hastatic order in cubic systems. We studied both ferro- and antiferrohastatic orders, and showed that antiferrohastatic orders with $f$-electron hopping necessarily break additional symmetries. All hastatic phases have distinct signatures including hybridization gaps, heavy Fermi liquid behavior and tiny conduction electron magnetic moments; the band structure proves particularly useful in distinguishing different antiferrohastatic orders. 

We obtained the mean-field phase diagram in a few representative cases, and found all of the above phases to be stabilized in some region. We also explored the relative stability of these phases in temperature, and both channel and spin symmetry breaking fields. In particular, magnetic field favors ferrohastatic order, which might explain the intermediate field heavy Fermi liquid regions seen in PrV$_2$Al$_{20}$, Pr(Ir,Rh)$_2$Zn$_{20}$ and PrPb$_3$ \cite{vandyke18}.

As the model considered here is particularly simple, future work should incorporate the effect of strong spin-orbit coupling on the hybridization, as has been done for the ferrohastatic case in Ref. \onlinecite{vandyke18}, as well as more complicated hastatic spinor arrangements. We have additionally neglected any competition or cooperation with superconductivity, which is well known to be a competing ground state on the two-channel Kondo lattice \cite{jarrell97,flint08,hoshino142,hoshino143}.

\section{Acknowledgements}

We would like to thank Piers Coleman, Premala Chandra, and Peter Orth for helpful discussions. This work was supported by US Department of Energy Grant No. DE-SC0015891. R.F. also acknowledges the hospitality of the Aspen Center for Physics, supported by National Science Foundation Grant No. PHYS-1066293.

\bibliographystyle{apsrev4-1}
\bibliography{hastatic}

\begin{thebibliography}{103}%
\makeatletter
\providecommand \@ifxundefined [1]{%
 \@ifx{#1\undefined}
}%
\providecommand \@ifnum [1]{%
 \ifnum #1\expandafter \@firstoftwo
 \else \expandafter \@secondoftwo
 \fi
}%
\providecommand \@ifx [1]{%
 \ifx #1\expandafter \@firstoftwo
 \else \expandafter \@secondoftwo
 \fi
}%
\providecommand \natexlab [1]{#1}%
\providecommand \enquote  [1]{``#1''}%
\providecommand \bibnamefont  [1]{#1}%
\providecommand \bibfnamefont [1]{#1}%
\providecommand \citenamefont [1]{#1}%
\providecommand \href@noop [0]{\@secondoftwo}%
\providecommand \href [0]{\begingroup \@sanitize@url \@href}%
\providecommand \@href[1]{\@@startlink{#1}\@@href}%
\providecommand \@@href[1]{\endgroup#1\@@endlink}%
\providecommand \@sanitize@url [0]{\catcode `\\12\catcode `\$12\catcode
  `\&12\catcode `\#12\catcode `\^12\catcode `\_12\catcode `\%12\relax}%
\providecommand \@@startlink[1]{}%
\providecommand \@@endlink[0]{}%
\providecommand \url  [0]{\begingroup\@sanitize@url \@url }%
\providecommand \@url [1]{\endgroup\@href {#1}{\urlprefix }}%
\providecommand \urlprefix  [0]{URL }%
\providecommand \Eprint [0]{\href }%
\providecommand \doibase [0]{http://dx.doi.org/}%
\providecommand \selectlanguage [0]{\@gobble}%
\providecommand \bibinfo  [0]{\@secondoftwo}%
\providecommand \bibfield  [0]{\@secondoftwo}%
\providecommand \translation [1]{[#1]}%
\providecommand \BibitemOpen [0]{}%
\providecommand \bibitemStop [0]{}%
\providecommand \bibitemNoStop [0]{.\EOS\space}%
\providecommand \EOS [0]{\spacefactor3000\relax}%
\providecommand \BibitemShut  [1]{\csname bibitem#1\endcsname}%
\let\auto@bib@innerbib\@empty
\bibitem [{\citenamefont {Doniach}(1977)}]{doniach77}%
  \BibitemOpen
  \bibfield  {author} {\bibinfo {author} {\bibfnamefont {S.}~\bibnamefont
  {Doniach}},\ }\href@noop {} {\bibfield  {journal} {\bibinfo  {journal}
  {Physica B}\ }\textbf {\bibinfo {volume} {91}},\ \bibinfo {pages} {231}
  (\bibinfo {year} {1977})}\BibitemShut {NoStop}%
\bibitem [{\citenamefont {Coleman}\ and\ \citenamefont
  {Schofield}(2005)}]{coleman05}%
  \BibitemOpen
  \bibfield  {author} {\bibinfo {author} {\bibfnamefont {P.}~\bibnamefont
  {Coleman}}\ and\ \bibinfo {author} {\bibfnamefont {A.~J.}\ \bibnamefont
  {Schofield}},\ }\href {\doibase 10.1038/nature03279} {\bibfield  {journal}
  {\bibinfo  {journal} {Nature}\ }\textbf {\bibinfo {volume} {433}},\ \bibinfo
  {pages} {226} (\bibinfo {year} {2005})}\BibitemShut {NoStop}%
\bibitem [{\citenamefont {Gegenwart}\ \emph {et~al.}(2008)\citenamefont
  {Gegenwart}, \citenamefont {Si},\ and\ \citenamefont
  {Steglich}}]{gegenwart08}%
  \BibitemOpen
  \bibfield  {author} {\bibinfo {author} {\bibfnamefont {P.}~\bibnamefont
  {Gegenwart}}, \bibinfo {author} {\bibfnamefont {Q.}~\bibnamefont {Si}}, \
  and\ \bibinfo {author} {\bibfnamefont {F.}~\bibnamefont {Steglich}},\ }\href
  {\doibase 10.1038/nphys892} {\bibfield  {journal} {\bibinfo  {journal}
  {Nature Physics}\ }\textbf {\bibinfo {volume} {4}},\ \bibinfo {pages} {186}
  (\bibinfo {year} {2008})}\BibitemShut {NoStop}%
\bibitem [{\citenamefont {Stewart}(2017)}]{stewart17}%
  \BibitemOpen
  \bibfield  {author} {\bibinfo {author} {\bibfnamefont {G.~R.}\ \bibnamefont
  {Stewart}},\ }\href {\doibase 10.1080/00018732.2017.1331615} {\bibfield
  {journal} {\bibinfo  {journal} {Advances in Physics}\ }\textbf {\bibinfo
  {volume} {66}},\ \bibinfo {pages} {75} (\bibinfo {year} {2017})}\BibitemShut
  {NoStop}%
\bibitem [{\citenamefont {Dzero}\ \emph {et~al.}(2010)\citenamefont {Dzero},
  \citenamefont {Sun}, \citenamefont {Galitski},\ and\ \citenamefont
  {Coleman}}]{dzero10}%
  \BibitemOpen
  \bibfield  {author} {\bibinfo {author} {\bibfnamefont {M.}~\bibnamefont
  {Dzero}}, \bibinfo {author} {\bibfnamefont {K.}~\bibnamefont {Sun}}, \bibinfo
  {author} {\bibfnamefont {V.}~\bibnamefont {Galitski}}, \ and\ \bibinfo
  {author} {\bibfnamefont {P.}~\bibnamefont {Coleman}},\ }\href {\doibase
  10.1103/PhysRevLett.104.106408} {\bibfield  {journal} {\bibinfo  {journal}
  {Phys. Rev. Lett.}\ }\textbf {\bibinfo {volume} {104}},\ \bibinfo {pages}
  {106408} (\bibinfo {year} {2010})}\BibitemShut {NoStop}%
\bibitem [{\citenamefont {Kenzelmann}\ \emph {et~al.}(2008)\citenamefont
  {Kenzelmann}, \citenamefont {Str{\"a}ssle}, \citenamefont {Niedermayer},
  \citenamefont {Sigrist}, \citenamefont {Padmanabhan}, \citenamefont
  {Zolliker}, \citenamefont {Bianchi}, \citenamefont {Movshovich},
  \citenamefont {Bauer}, \citenamefont {Sarrao},\ and\ \citenamefont
  {Thompson}}]{qphase}%
  \BibitemOpen
  \bibfield  {author} {\bibinfo {author} {\bibfnamefont {M.}~\bibnamefont
  {Kenzelmann}}, \bibinfo {author} {\bibfnamefont {T.}~\bibnamefont
  {Str{\"a}ssle}}, \bibinfo {author} {\bibfnamefont {C.}~\bibnamefont
  {Niedermayer}}, \bibinfo {author} {\bibfnamefont {M.}~\bibnamefont
  {Sigrist}}, \bibinfo {author} {\bibfnamefont {B.}~\bibnamefont
  {Padmanabhan}}, \bibinfo {author} {\bibfnamefont {M.}~\bibnamefont
  {Zolliker}}, \bibinfo {author} {\bibfnamefont {A.~D.}\ \bibnamefont
  {Bianchi}}, \bibinfo {author} {\bibfnamefont {R.}~\bibnamefont {Movshovich}},
  \bibinfo {author} {\bibfnamefont {E.~D.}\ \bibnamefont {Bauer}}, \bibinfo
  {author} {\bibfnamefont {J.~L.}\ \bibnamefont {Sarrao}}, \ and\ \bibinfo
  {author} {\bibfnamefont {J.~D.}\ \bibnamefont {Thompson}},\ }\href {\doibase
  10.1126/science.1161818} {\bibfield  {journal} {\bibinfo  {journal}
  {Science}\ }\textbf {\bibinfo {volume} {321}},\ \bibinfo {pages} {1652}
  (\bibinfo {year} {2008})}\BibitemShut {NoStop}%
\bibitem [{\citenamefont {Nakatsuji}\ \emph {et~al.}(2006)\citenamefont
  {Nakatsuji}, \citenamefont {Machida}, \citenamefont {Maeno}, \citenamefont
  {Tayama}, \citenamefont {Sakakibara}, \citenamefont {Duijn}, \citenamefont
  {Balicas}, \citenamefont {Millican}, \citenamefont {Macaluso},\ and\
  \citenamefont {Chan}}]{pr2271}%
  \BibitemOpen
  \bibfield  {author} {\bibinfo {author} {\bibfnamefont {S.}~\bibnamefont
  {Nakatsuji}}, \bibinfo {author} {\bibfnamefont {Y.}~\bibnamefont {Machida}},
  \bibinfo {author} {\bibfnamefont {Y.}~\bibnamefont {Maeno}}, \bibinfo
  {author} {\bibfnamefont {T.}~\bibnamefont {Tayama}}, \bibinfo {author}
  {\bibfnamefont {T.}~\bibnamefont {Sakakibara}}, \bibinfo {author}
  {\bibfnamefont {J.~v.}\ \bibnamefont {Duijn}}, \bibinfo {author}
  {\bibfnamefont {L.}~\bibnamefont {Balicas}}, \bibinfo {author} {\bibfnamefont
  {J.~N.}\ \bibnamefont {Millican}}, \bibinfo {author} {\bibfnamefont {R.~T.}\
  \bibnamefont {Macaluso}}, \ and\ \bibinfo {author} {\bibfnamefont {J.~Y.}\
  \bibnamefont {Chan}},\ }\href {\doibase 10.1103/PhysRevLett.96.087204}
  {\bibfield  {journal} {\bibinfo  {journal} {Phys. Rev. Lett.}\ }\textbf
  {\bibinfo {volume} {96}},\ \bibinfo {pages} {087204} (\bibinfo {year}
  {2006})}\BibitemShut {NoStop}%
\bibitem [{\citenamefont {Machida}\ \emph {et~al.}(2009)\citenamefont
  {Machida}, \citenamefont {Nakatsuji}, \citenamefont {Onoda}, \citenamefont
  {Tayama},\ and\ \citenamefont {Sakakibara}}]{pr2272}%
  \BibitemOpen
  \bibfield  {author} {\bibinfo {author} {\bibfnamefont {Y.}~\bibnamefont
  {Machida}}, \bibinfo {author} {\bibfnamefont {S.}~\bibnamefont {Nakatsuji}},
  \bibinfo {author} {\bibfnamefont {S.}~\bibnamefont {Onoda}}, \bibinfo
  {author} {\bibfnamefont {T.}~\bibnamefont {Tayama}}, \ and\ \bibinfo {author}
  {\bibfnamefont {T.}~\bibnamefont {Sakakibara}},\ }\href
  {http://dx.doi.org/10.1038/nature08680} {\bibfield  {journal} {\bibinfo
  {journal} {Nature}\ }\textbf {\bibinfo {volume} {463}},\ \bibinfo {pages}
  {210 EP } (\bibinfo {year} {2009})}\BibitemShut {NoStop}%
\bibitem [{\citenamefont {Fritsch}\ \emph {et~al.}(2014)\citenamefont
  {Fritsch}, \citenamefont {Bagrets}, \citenamefont {Goll}, \citenamefont
  {Kittler}, \citenamefont {Wolf}, \citenamefont {Grube}, \citenamefont
  {Huang},\ and\ \citenamefont {L\"ohneysen}}]{cepdal}%
  \BibitemOpen
  \bibfield  {author} {\bibinfo {author} {\bibfnamefont {V.}~\bibnamefont
  {Fritsch}}, \bibinfo {author} {\bibfnamefont {N.}~\bibnamefont {Bagrets}},
  \bibinfo {author} {\bibfnamefont {G.}~\bibnamefont {Goll}}, \bibinfo {author}
  {\bibfnamefont {W.}~\bibnamefont {Kittler}}, \bibinfo {author} {\bibfnamefont
  {M.~J.}\ \bibnamefont {Wolf}}, \bibinfo {author} {\bibfnamefont
  {K.}~\bibnamefont {Grube}}, \bibinfo {author} {\bibfnamefont {C.-L.}\
  \bibnamefont {Huang}}, \ and\ \bibinfo {author} {\bibfnamefont {H.~v.}\
  \bibnamefont {L\"ohneysen}},\ }\href {\doibase 10.1103/PhysRevB.89.054416}
  {\bibfield  {journal} {\bibinfo  {journal} {Phys. Rev. B}\ }\textbf {\bibinfo
  {volume} {89}},\ \bibinfo {pages} {054416} (\bibinfo {year}
  {2014})}\BibitemShut {NoStop}%
\bibitem [{\citenamefont {Cox}\ and\ \citenamefont
  {Zawadowski}(1998)}]{zawadowski}%
  \BibitemOpen
  \bibfield  {author} {\bibinfo {author} {\bibfnamefont {D.~L.}\ \bibnamefont
  {Cox}}\ and\ \bibinfo {author} {\bibfnamefont {A.}~\bibnamefont
  {Zawadowski}},\ }\href {\doibase 10.1080/000187398243500} {\bibfield
  {journal} {\bibinfo  {journal} {Advances in Physics}\ }\textbf {\bibinfo
  {volume} {47}},\ \bibinfo {pages} {599} (\bibinfo {year} {1998})}\BibitemShut
  {NoStop}%
\bibitem [{\citenamefont {{Nozi\`eres, Ph.}}\ and\ \citenamefont {{Blandin,
  A.}}(1980)}]{blandin}%
  \BibitemOpen
  \bibfield  {author} {\bibinfo {author} {\bibnamefont {{Nozi\`eres, Ph.}}}\
  and\ \bibinfo {author} {\bibnamefont {{Blandin, A.}}},\ }\href {\doibase
  10.1051/jphys:01980004103019300} {\bibfield  {journal} {\bibinfo  {journal}
  {J. Phys. France}\ }\textbf {\bibinfo {volume} {41}},\ \bibinfo {pages} {193}
  (\bibinfo {year} {1980})}\BibitemShut {NoStop}%
\bibitem [{\citenamefont {Cox}(1987)}]{cox87}%
  \BibitemOpen
  \bibfield  {author} {\bibinfo {author} {\bibfnamefont {D.~L.}\ \bibnamefont
  {Cox}},\ }\href {\doibase 10.1103/PhysRevLett.59.1240} {\bibfield  {journal}
  {\bibinfo  {journal} {Physical Review Letters}\ }\textbf {\bibinfo {volume}
  {59}},\ \bibinfo {pages} {1240} (\bibinfo {year} {1987})}\BibitemShut
  {NoStop}%
\bibitem [{\citenamefont {Cox}(1988)}]{cox88}%
  \BibitemOpen
  \bibfield  {author} {\bibinfo {author} {\bibfnamefont {D.}~\bibnamefont
  {Cox}},\ }\href {\doibase https://doi.org/10.1016/0921-4534(88)90437-6}
  {\bibfield  {journal} {\bibinfo  {journal} {Physica C: Superconductivity}\
  }\textbf {\bibinfo {volume} {153-155}},\ \bibinfo {pages} {1642 } (\bibinfo
  {year} {1988})},\ \bibinfo {note} {proceedings of the International
  Conference on High Temperature Superconductors and Materials and Mechanisms
  of Superconductivity Part II}\BibitemShut {NoStop}%
\bibitem [{\citenamefont {Cox}\ and\ \citenamefont {Jarrell}(1996)}]{cox96}%
  \BibitemOpen
  \bibfield  {author} {\bibinfo {author} {\bibfnamefont {D.~L.}\ \bibnamefont
  {Cox}}\ and\ \bibinfo {author} {\bibfnamefont {M.}~\bibnamefont {Jarrell}},\
  }\href {http://stacks.iop.org/0953-8984/8/i=48/a=012} {\bibfield  {journal}
  {\bibinfo  {journal} {Journal of Physics: Condensed Matter}\ }\textbf
  {\bibinfo {volume} {8}},\ \bibinfo {pages} {9825} (\bibinfo {year}
  {1996})}\BibitemShut {NoStop}%
\bibitem [{\citenamefont {Jarrell}\ \emph {et~al.}(1996)\citenamefont
  {Jarrell}, \citenamefont {Pang}, \citenamefont {Cox},\ and\ \citenamefont
  {Luk}}]{jarrell96}%
  \BibitemOpen
  \bibfield  {author} {\bibinfo {author} {\bibfnamefont {M.}~\bibnamefont
  {Jarrell}}, \bibinfo {author} {\bibfnamefont {H.}~\bibnamefont {Pang}},
  \bibinfo {author} {\bibfnamefont {D.~L.}\ \bibnamefont {Cox}}, \ and\
  \bibinfo {author} {\bibfnamefont {K.~H.}\ \bibnamefont {Luk}},\ }\href
  {\doibase 10.1103/PhysRevLett.77.1612} {\bibfield  {journal} {\bibinfo
  {journal} {Phys. Rev. Lett.}\ }\textbf {\bibinfo {volume} {77}},\ \bibinfo
  {pages} {1612} (\bibinfo {year} {1996})}\BibitemShut {NoStop}%
\bibitem [{\citenamefont {Jarrell}\ \emph {et~al.}(1997)\citenamefont
  {Jarrell}, \citenamefont {Pang},\ and\ \citenamefont {Cox}}]{jarrell97}%
  \BibitemOpen
  \bibfield  {author} {\bibinfo {author} {\bibfnamefont {M.}~\bibnamefont
  {Jarrell}}, \bibinfo {author} {\bibfnamefont {H.}~\bibnamefont {Pang}}, \
  and\ \bibinfo {author} {\bibfnamefont {D.~L.}\ \bibnamefont {Cox}},\ }\href
  {\doibase 10.1103/PhysRevLett.78.1996} {\bibfield  {journal} {\bibinfo
  {journal} {Phys. Rev. Lett.}\ }\textbf {\bibinfo {volume} {78}},\ \bibinfo
  {pages} {1996} (\bibinfo {year} {1997})}\BibitemShut {NoStop}%
\bibitem [{\citenamefont {Ott}\ \emph {et~al.}(1983)\citenamefont {Ott},
  \citenamefont {Rudigier}, \citenamefont {Fisk},\ and\ \citenamefont
  {Smith}}]{ott83}%
  \BibitemOpen
  \bibfield  {author} {\bibinfo {author} {\bibfnamefont {H.~R.}\ \bibnamefont
  {Ott}}, \bibinfo {author} {\bibfnamefont {H.}~\bibnamefont {Rudigier}},
  \bibinfo {author} {\bibfnamefont {Z.}~\bibnamefont {Fisk}}, \ and\ \bibinfo
  {author} {\bibfnamefont {J.~L.}\ \bibnamefont {Smith}},\ }\href {\doibase
  10.1103/PhysRevLett.50.1595} {\bibfield  {journal} {\bibinfo  {journal}
  {Phys. Rev. Lett.}\ }\textbf {\bibinfo {volume} {50}},\ \bibinfo {pages}
  {1595} (\bibinfo {year} {1983})}\BibitemShut {NoStop}%
\bibitem [{\citenamefont {Sakai}\ and\ \citenamefont
  {Nakatsuji}(2011)}]{sakai11}%
  \BibitemOpen
  \bibfield  {author} {\bibinfo {author} {\bibfnamefont {A.}~\bibnamefont
  {Sakai}}\ and\ \bibinfo {author} {\bibfnamefont {S.}~\bibnamefont
  {Nakatsuji}},\ }\href {\doibase 10.1143/JPSJ.80.063701} {\bibfield  {journal}
  {\bibinfo  {journal} {Journal of the Physical Society of Japan}\ }\textbf
  {\bibinfo {volume} {80}},\ \bibinfo {pages} {063701} (\bibinfo {year}
  {2011})}\BibitemShut {NoStop}%
\bibitem [{\citenamefont {Tokunaga}\ \emph {et~al.}(2013)\citenamefont
  {Tokunaga}, \citenamefont {Sakai}, \citenamefont {Kambe}, \citenamefont
  {Sakai}, \citenamefont {Nakatsuji},\ and\ \citenamefont
  {Harima}}]{tokunaga13}%
  \BibitemOpen
  \bibfield  {author} {\bibinfo {author} {\bibfnamefont {Y.}~\bibnamefont
  {Tokunaga}}, \bibinfo {author} {\bibfnamefont {H.}~\bibnamefont {Sakai}},
  \bibinfo {author} {\bibfnamefont {S.}~\bibnamefont {Kambe}}, \bibinfo
  {author} {\bibfnamefont {A.}~\bibnamefont {Sakai}}, \bibinfo {author}
  {\bibfnamefont {S.}~\bibnamefont {Nakatsuji}}, \ and\ \bibinfo {author}
  {\bibfnamefont {H.}~\bibnamefont {Harima}},\ }\href {\doibase
  10.1103/PhysRevB.88.085124} {\bibfield  {journal} {\bibinfo  {journal}
  {Physical Review B}\ }\textbf {\bibinfo {volume} {88}},\ \bibinfo {pages}
  {085124} (\bibinfo {year} {2013})}\BibitemShut {NoStop}%
\bibitem [{\citenamefont {Onimaru}\ \emph {et~al.}(2010)\citenamefont
  {Onimaru}, \citenamefont {T.~Matsumoto}, \citenamefont {F.~Inoue},
  \citenamefont {Umeo}, \citenamefont {Saiga}, \citenamefont {Matsushita},
  \citenamefont {Tamura}, \citenamefont {Nishimoto}, \citenamefont {Ishii},
  \citenamefont {Suzuki},\ and\ \citenamefont {Takabatake}}]{onimaru10}%
  \BibitemOpen
  \bibfield  {author} {\bibinfo {author} {\bibfnamefont {T.}~\bibnamefont
  {Onimaru}}, \bibinfo {author} {\bibfnamefont {K.}~\bibnamefont
  {T.~Matsumoto}}, \bibinfo {author} {\bibfnamefont {Y.}~\bibnamefont
  {F.~Inoue}}, \bibinfo {author} {\bibfnamefont {K.}~\bibnamefont {Umeo}},
  \bibinfo {author} {\bibfnamefont {Y.}~\bibnamefont {Saiga}}, \bibinfo
  {author} {\bibfnamefont {Y.}~\bibnamefont {Matsushita}}, \bibinfo {author}
  {\bibfnamefont {R.}~\bibnamefont {Tamura}}, \bibinfo {author} {\bibfnamefont
  {K.}~\bibnamefont {Nishimoto}}, \bibinfo {author} {\bibfnamefont
  {I.}~\bibnamefont {Ishii}}, \bibinfo {author} {\bibfnamefont
  {T.}~\bibnamefont {Suzuki}}, \ and\ \bibinfo {author} {\bibfnamefont
  {T.}~\bibnamefont {Takabatake}},\ }\href {\doibase 10.1143/JPSJ.79.033704}
  {\bibfield  {journal} {\bibinfo  {journal} {Journal of the Physical Society
  of Japan}\ }\textbf {\bibinfo {volume} {79}},\ \bibinfo {pages} {033704}
  (\bibinfo {year} {2010})}\BibitemShut {NoStop}%
\bibitem [{\citenamefont {Onimaru}\ \emph {et~al.}(2011)\citenamefont
  {Onimaru}, \citenamefont {Matsumoto}, \citenamefont {Inoue}, \citenamefont
  {Umeo}, \citenamefont {Sakakibara}, \citenamefont {Karaki}, \citenamefont
  {Kubota},\ and\ \citenamefont {Takabatake}}]{onimaru11}%
  \BibitemOpen
  \bibfield  {author} {\bibinfo {author} {\bibfnamefont {T.}~\bibnamefont
  {Onimaru}}, \bibinfo {author} {\bibfnamefont {K.~T.}\ \bibnamefont
  {Matsumoto}}, \bibinfo {author} {\bibfnamefont {Y.~F.}\ \bibnamefont
  {Inoue}}, \bibinfo {author} {\bibfnamefont {K.}~\bibnamefont {Umeo}},
  \bibinfo {author} {\bibfnamefont {T.}~\bibnamefont {Sakakibara}}, \bibinfo
  {author} {\bibfnamefont {Y.}~\bibnamefont {Karaki}}, \bibinfo {author}
  {\bibfnamefont {M.}~\bibnamefont {Kubota}}, \ and\ \bibinfo {author}
  {\bibfnamefont {T.}~\bibnamefont {Takabatake}},\ }\href {\doibase
  10.1103/PhysRevLett.106.177001} {\bibfield  {journal} {\bibinfo  {journal}
  {Phys. Rev. Lett.}\ }\textbf {\bibinfo {volume} {106}},\ \bibinfo {pages}
  {177001} (\bibinfo {year} {2011})}\BibitemShut {NoStop}%
\bibitem [{\citenamefont {Sato}\ \emph {et~al.}(2012)\citenamefont {Sato},
  \citenamefont {Ibuka}, \citenamefont {Nambu}, \citenamefont {Yamazaki},
  \citenamefont {Hong}, \citenamefont {Sakai},\ and\ \citenamefont
  {Nakatsuji}}]{sato12}%
  \BibitemOpen
  \bibfield  {author} {\bibinfo {author} {\bibfnamefont {T.~J.}\ \bibnamefont
  {Sato}}, \bibinfo {author} {\bibfnamefont {S.}~\bibnamefont {Ibuka}},
  \bibinfo {author} {\bibfnamefont {Y.}~\bibnamefont {Nambu}}, \bibinfo
  {author} {\bibfnamefont {T.}~\bibnamefont {Yamazaki}}, \bibinfo {author}
  {\bibfnamefont {T.}~\bibnamefont {Hong}}, \bibinfo {author} {\bibfnamefont
  {A.}~\bibnamefont {Sakai}}, \ and\ \bibinfo {author} {\bibfnamefont
  {S.}~\bibnamefont {Nakatsuji}},\ }\href {\doibase 10.1103/PhysRevB.86.184419}
  {\bibfield  {journal} {\bibinfo  {journal} {Phys. Rev. B}\ }\textbf {\bibinfo
  {volume} {86}},\ \bibinfo {pages} {184419} (\bibinfo {year}
  {2012})}\BibitemShut {NoStop}%
\bibitem [{\citenamefont {Onimaru}\ \emph {et~al.}(2012)\citenamefont
  {Onimaru}, \citenamefont {Nagasawa}, \citenamefont {Matsumoto}, \citenamefont
  {Wakiya}, \citenamefont {Umeo}, \citenamefont {Kittaka}, \citenamefont
  {Sakakibara}, \citenamefont {Matsushita},\ and\ \citenamefont
  {Takabatake}}]{onimaru12}%
  \BibitemOpen
  \bibfield  {author} {\bibinfo {author} {\bibfnamefont {T.}~\bibnamefont
  {Onimaru}}, \bibinfo {author} {\bibfnamefont {N.}~\bibnamefont {Nagasawa}},
  \bibinfo {author} {\bibfnamefont {K.~T.}\ \bibnamefont {Matsumoto}}, \bibinfo
  {author} {\bibfnamefont {K.}~\bibnamefont {Wakiya}}, \bibinfo {author}
  {\bibfnamefont {K.}~\bibnamefont {Umeo}}, \bibinfo {author} {\bibfnamefont
  {S.}~\bibnamefont {Kittaka}}, \bibinfo {author} {\bibfnamefont
  {T.}~\bibnamefont {Sakakibara}}, \bibinfo {author} {\bibfnamefont
  {Y.}~\bibnamefont {Matsushita}}, \ and\ \bibinfo {author} {\bibfnamefont
  {T.}~\bibnamefont {Takabatake}},\ }\href {\doibase
  10.1103/PhysRevB.86.184426} {\bibfield  {journal} {\bibinfo  {journal} {Phys.
  Rev. B}\ }\textbf {\bibinfo {volume} {86}},\ \bibinfo {pages} {184426}
  (\bibinfo {year} {2012})}\BibitemShut {NoStop}%
\bibitem [{\citenamefont {Nagasawa}\ \emph {et~al.}(2012)\citenamefont
  {Nagasawa}, \citenamefont {Onimaru}, \citenamefont {Matsumoto}, \citenamefont
  {Umeo},\ and\ \citenamefont {Takabatake}}]{nagasawa12}%
  \BibitemOpen
  \bibfield  {author} {\bibinfo {author} {\bibfnamefont {N.}~\bibnamefont
  {Nagasawa}}, \bibinfo {author} {\bibfnamefont {T.}~\bibnamefont {Onimaru}},
  \bibinfo {author} {\bibfnamefont {K.~T.}\ \bibnamefont {Matsumoto}}, \bibinfo
  {author} {\bibfnamefont {K.}~\bibnamefont {Umeo}}, \ and\ \bibinfo {author}
  {\bibfnamefont {T.}~\bibnamefont {Takabatake}},\ }\href
  {http://stacks.iop.org/1742-6596/391/i=1/a=012051} {\bibfield  {journal}
  {\bibinfo  {journal} {Journal of Physics: Conference Series}\ }\textbf
  {\bibinfo {volume} {391}},\ \bibinfo {pages} {012051} (\bibinfo {year}
  {2012})}\BibitemShut {NoStop}%
\bibitem [{\citenamefont {Iwasa}\ \emph {et~al.}(2013)\citenamefont {Iwasa},
  \citenamefont {Kobayashi}, \citenamefont {Onimaru}, \citenamefont
  {T.~Matsumoto}, \citenamefont {Nagasawa}, \citenamefont {Takabatake},
  \citenamefont {Ohira-Kawamura}, \citenamefont {Kikuchi}, \citenamefont
  {Inamura},\ and\ \citenamefont {Nakajima}}]{iwasa13}%
  \BibitemOpen
  \bibfield  {author} {\bibinfo {author} {\bibfnamefont {K.}~\bibnamefont
  {Iwasa}}, \bibinfo {author} {\bibfnamefont {H.}~\bibnamefont {Kobayashi}},
  \bibinfo {author} {\bibfnamefont {T.}~\bibnamefont {Onimaru}}, \bibinfo
  {author} {\bibfnamefont {K.}~\bibnamefont {T.~Matsumoto}}, \bibinfo {author}
  {\bibfnamefont {N.}~\bibnamefont {Nagasawa}}, \bibinfo {author}
  {\bibfnamefont {T.}~\bibnamefont {Takabatake}}, \bibinfo {author}
  {\bibfnamefont {S.}~\bibnamefont {Ohira-Kawamura}}, \bibinfo {author}
  {\bibfnamefont {T.}~\bibnamefont {Kikuchi}}, \bibinfo {author} {\bibfnamefont
  {Y.}~\bibnamefont {Inamura}}, \ and\ \bibinfo {author} {\bibfnamefont
  {K.}~\bibnamefont {Nakajima}},\ }\href {\doibase 10.7566/JPSJ.82.043707}
  {\bibfield  {journal} {\bibinfo  {journal} {Journal of the Physical Society
  of Japan}\ }\textbf {\bibinfo {volume} {82}},\ \bibinfo {pages} {043707}
  (\bibinfo {year} {2013})}\BibitemShut {NoStop}%
\bibitem [{\citenamefont {Tsujimoto}\ \emph {et~al.}(2014)\citenamefont
  {Tsujimoto}, \citenamefont {Matsumoto}, \citenamefont {Tomita}, \citenamefont
  {Sakai},\ and\ \citenamefont {Nakatsuji}}]{tsujimoto14}%
  \BibitemOpen
  \bibfield  {author} {\bibinfo {author} {\bibfnamefont {M.}~\bibnamefont
  {Tsujimoto}}, \bibinfo {author} {\bibfnamefont {Y.}~\bibnamefont
  {Matsumoto}}, \bibinfo {author} {\bibfnamefont {T.}~\bibnamefont {Tomita}},
  \bibinfo {author} {\bibfnamefont {A.}~\bibnamefont {Sakai}}, \ and\ \bibinfo
  {author} {\bibfnamefont {S.}~\bibnamefont {Nakatsuji}},\ }\href {\doibase
  10.1103/PhysRevLett.113.267001} {\bibfield  {journal} {\bibinfo  {journal}
  {Physical Review Letters}\ }\textbf {\bibinfo {volume} {113}},\ \bibinfo
  {pages} {267001} (\bibinfo {year} {2014})}\BibitemShut {NoStop}%
\bibitem [{\citenamefont {Chandra}\ \emph {et~al.}(2013)\citenamefont
  {Chandra}, \citenamefont {Coleman},\ and\ \citenamefont {Flint}}]{chandra13}%
  \BibitemOpen
  \bibfield  {author} {\bibinfo {author} {\bibfnamefont {P.}~\bibnamefont
  {Chandra}}, \bibinfo {author} {\bibfnamefont {P.}~\bibnamefont {Coleman}}, \
  and\ \bibinfo {author} {\bibfnamefont {R.}~\bibnamefont {Flint}},\ }\href
  {http://dx.doi.org/10.1038/nature11820} {\bibfield  {journal} {\bibinfo
  {journal} {Nature}\ }\textbf {\bibinfo {volume} {493}},\ \bibinfo {pages}
  {621 EP } (\bibinfo {year} {2013})},\ \bibinfo {note} {article}\BibitemShut
  {NoStop}%
\bibitem [{\citenamefont {Andrei}\ and\ \citenamefont
  {Destri}(1984)}]{andrei84}%
  \BibitemOpen
  \bibfield  {author} {\bibinfo {author} {\bibfnamefont {N.}~\bibnamefont
  {Andrei}}\ and\ \bibinfo {author} {\bibfnamefont {C.}~\bibnamefont
  {Destri}},\ }\href {\doibase 10.1103/PhysRevLett.52.364} {\bibfield
  {journal} {\bibinfo  {journal} {Phys. Rev. Lett.}\ }\textbf {\bibinfo
  {volume} {52}},\ \bibinfo {pages} {364} (\bibinfo {year} {1984})}\BibitemShut
  {NoStop}%
\bibitem [{\citenamefont {Emery}\ and\ \citenamefont
  {Kivelson}(1992)}]{emery92}%
  \BibitemOpen
  \bibfield  {author} {\bibinfo {author} {\bibfnamefont {V.~J.}\ \bibnamefont
  {Emery}}\ and\ \bibinfo {author} {\bibfnamefont {S.}~\bibnamefont
  {Kivelson}},\ }\href {\doibase 10.1103/PhysRevB.46.10812} {\bibfield
  {journal} {\bibinfo  {journal} {Phys. Rev. B}\ }\textbf {\bibinfo {volume}
  {46}},\ \bibinfo {pages} {10812} (\bibinfo {year} {1992})}\BibitemShut
  {NoStop}%
\bibitem [{\citenamefont {Chandra}\ \emph {et~al.}(2015)\citenamefont
  {Chandra}, \citenamefont {Coleman},\ and\ \citenamefont {Flint}}]{chandra15}%
  \BibitemOpen
  \bibfield  {author} {\bibinfo {author} {\bibfnamefont {P.}~\bibnamefont
  {Chandra}}, \bibinfo {author} {\bibfnamefont {P.}~\bibnamefont {Coleman}}, \
  and\ \bibinfo {author} {\bibfnamefont {R.}~\bibnamefont {Flint}},\ }\href
  {\doibase 10.1103/PhysRevB.91.205103} {\bibfield  {journal} {\bibinfo
  {journal} {Phys. Rev. B}\ }\textbf {\bibinfo {volume} {91}},\ \bibinfo
  {pages} {205103} (\bibinfo {year} {2015})}\BibitemShut {NoStop}%
\bibitem [{\citenamefont {Hoshino}\ \emph {et~al.}(2011)\citenamefont
  {Hoshino}, \citenamefont {Otsuki},\ and\ \citenamefont
  {Kuramoto}}]{hoshino11}%
  \BibitemOpen
  \bibfield  {author} {\bibinfo {author} {\bibfnamefont {S.}~\bibnamefont
  {Hoshino}}, \bibinfo {author} {\bibfnamefont {J.}~\bibnamefont {Otsuki}}, \
  and\ \bibinfo {author} {\bibfnamefont {Y.}~\bibnamefont {Kuramoto}},\ }\href
  {\doibase 10.1103/PhysRevLett.107.247202} {\bibfield  {journal} {\bibinfo
  {journal} {Phys. Rev. Lett.}\ }\textbf {\bibinfo {volume} {107}},\ \bibinfo
  {pages} {247202} (\bibinfo {year} {2011})}\BibitemShut {NoStop}%
\bibitem [{\citenamefont {Sakurai}(1994)}]{sakurai}%
  \BibitemOpen
  \bibfield  {author} {\bibinfo {author} {\bibfnamefont {J.~J.}\ \bibnamefont
  {Sakurai}},\ }\enquote {\bibinfo {title} {Modern quantum mechanics},}\ \
  (\bibinfo  {publisher} {Addison-Wesley Publishing Company},\ \bibinfo {year}
  {1994})\ p.\ \bibinfo {pages} {277},\ \bibinfo {edition} {revised}\
  ed.\BibitemShut {Stop}%
\bibitem [{\citenamefont {Lea}\ \emph {et~al.}(1962)\citenamefont {Lea},
  \citenamefont {Leask},\ and\ \citenamefont {Wolf}}]{lealeaskwolf}%
  \BibitemOpen
  \bibfield  {author} {\bibinfo {author} {\bibfnamefont {K.}~\bibnamefont
  {Lea}}, \bibinfo {author} {\bibfnamefont {M.}~\bibnamefont {Leask}}, \ and\
  \bibinfo {author} {\bibfnamefont {W.}~\bibnamefont {Wolf}},\ }\href {\doibase
  https://doi.org/10.1016/0022-3697(62)90192-0} {\bibfield  {journal} {\bibinfo
   {journal} {Journal of Physics and Chemistry of Solids}\ }\textbf {\bibinfo
  {volume} {23}},\ \bibinfo {pages} {1381 } (\bibinfo {year}
  {1962})}\BibitemShut {NoStop}%
\bibitem [{\citenamefont {Freyer}\ \emph {et~al.}(2018)\citenamefont {Freyer},
  \citenamefont {Attig}, \citenamefont {Lee}, \citenamefont {Paramekanti},
  \citenamefont {Trebst},\ and\ \citenamefont {Kim}}]{freyer17}%
  \BibitemOpen
  \bibfield  {author} {\bibinfo {author} {\bibfnamefont {F.}~\bibnamefont
  {Freyer}}, \bibinfo {author} {\bibfnamefont {J.}~\bibnamefont {Attig}},
  \bibinfo {author} {\bibfnamefont {S.}~\bibnamefont {Lee}}, \bibinfo {author}
  {\bibfnamefont {A.}~\bibnamefont {Paramekanti}}, \bibinfo {author}
  {\bibfnamefont {S.}~\bibnamefont {Trebst}}, \ and\ \bibinfo {author}
  {\bibfnamefont {Y.~B.}\ \bibnamefont {Kim}},\ }\href {\doibase
  10.1103/PhysRevB.97.115111} {\bibfield  {journal} {\bibinfo  {journal} {Phys.
  Rev. B}\ }\textbf {\bibinfo {volume} {97}},\ \bibinfo {pages} {115111}
  (\bibinfo {year} {2018})}\BibitemShut {NoStop}%
\bibitem [{\citenamefont {Matsunami}\ \emph {et~al.}(2011)\citenamefont
  {Matsunami}, \citenamefont {Taguchi}, \citenamefont {Chainani}, \citenamefont
  {Eguchi}, \citenamefont {Oura}, \citenamefont {Sakai}, \citenamefont
  {Nakatsuji},\ and\ \citenamefont {Shin}}]{matsunami11}%
  \BibitemOpen
  \bibfield  {author} {\bibinfo {author} {\bibfnamefont {M.}~\bibnamefont
  {Matsunami}}, \bibinfo {author} {\bibfnamefont {M.}~\bibnamefont {Taguchi}},
  \bibinfo {author} {\bibfnamefont {A.}~\bibnamefont {Chainani}}, \bibinfo
  {author} {\bibfnamefont {R.}~\bibnamefont {Eguchi}}, \bibinfo {author}
  {\bibfnamefont {M.}~\bibnamefont {Oura}}, \bibinfo {author} {\bibfnamefont
  {A.}~\bibnamefont {Sakai}}, \bibinfo {author} {\bibfnamefont
  {S.}~\bibnamefont {Nakatsuji}}, \ and\ \bibinfo {author} {\bibfnamefont
  {S.}~\bibnamefont {Shin}},\ }\href {\doibase 10.1103/PhysRevB.84.193101}
  {\bibfield  {journal} {\bibinfo  {journal} {Phys. Rev. B}\ }\textbf {\bibinfo
  {volume} {84}},\ \bibinfo {pages} {193101} (\bibinfo {year}
  {2011})}\BibitemShut {NoStop}%
\bibitem [{\citenamefont {Cox}(1993)}]{cox93}%
  \BibitemOpen
  \bibfield  {author} {\bibinfo {author} {\bibfnamefont {D.}~\bibnamefont
  {Cox}},\ }\href {\doibase https://doi.org/10.1016/0921-4526(93)90563-L}
  {\bibfield  {journal} {\bibinfo  {journal} {Physica B: Condensed Matter}\
  }\textbf {\bibinfo {volume} {186-188}},\ \bibinfo {pages} {312 } (\bibinfo
  {year} {1993})}\BibitemShut {NoStop}%
\bibitem [{\citenamefont {Alexandrov}\ \emph {et~al.}(2013)\citenamefont
  {Alexandrov}, \citenamefont {Dzero},\ and\ \citenamefont
  {Coleman}}]{alexandrov13}%
  \BibitemOpen
  \bibfield  {author} {\bibinfo {author} {\bibfnamefont {V.}~\bibnamefont
  {Alexandrov}}, \bibinfo {author} {\bibfnamefont {M.}~\bibnamefont {Dzero}}, \
  and\ \bibinfo {author} {\bibfnamefont {P.}~\bibnamefont {Coleman}},\ }\href
  {\doibase 10.1103/PhysRevLett.111.226403} {\bibfield  {journal} {\bibinfo
  {journal} {Phys. Rev. Lett.}\ }\textbf {\bibinfo {volume} {111}},\ \bibinfo
  {pages} {226403} (\bibinfo {year} {2013})}\BibitemShut {NoStop}%
\bibitem [{\citenamefont {Baruselli}\ and\ \citenamefont
  {Vojta}(2014)}]{Vojta2014}%
  \BibitemOpen
  \bibfield  {author} {\bibinfo {author} {\bibfnamefont {P.~P.}\ \bibnamefont
  {Baruselli}}\ and\ \bibinfo {author} {\bibfnamefont {M.}~\bibnamefont
  {Vojta}},\ }\href {\doibase 10.1103/PhysRevB.90.201106} {\bibfield  {journal}
  {\bibinfo  {journal} {Phys. Rev. B}\ }\textbf {\bibinfo {volume} {90}},\
  \bibinfo {pages} {201106} (\bibinfo {year} {2014})}\BibitemShut {NoStop}%
\bibitem [{\citenamefont {Pang}\ and\ \citenamefont {Cox}(1991)}]{pang91}%
  \BibitemOpen
  \bibfield  {author} {\bibinfo {author} {\bibfnamefont {H.~B.}\ \bibnamefont
  {Pang}}\ and\ \bibinfo {author} {\bibfnamefont {D.~L.}\ \bibnamefont {Cox}},\
  }\href {\doibase 10.1103/PhysRevB.44.9454} {\bibfield  {journal} {\bibinfo
  {journal} {Phys. Rev. B}\ }\textbf {\bibinfo {volume} {44}},\ \bibinfo
  {pages} {9454} (\bibinfo {year} {1991})}\BibitemShut {NoStop}%
\bibitem [{\citenamefont {Shimura}\ \emph {et~al.}(2015)\citenamefont
  {Shimura}, \citenamefont {Tsujimoto}, \citenamefont {Zeng}, \citenamefont
  {Balicas}, \citenamefont {Sakai},\ and\ \citenamefont
  {Nakatsuji}}]{shimura15}%
  \BibitemOpen
  \bibfield  {author} {\bibinfo {author} {\bibfnamefont {Y.}~\bibnamefont
  {Shimura}}, \bibinfo {author} {\bibfnamefont {M.}~\bibnamefont {Tsujimoto}},
  \bibinfo {author} {\bibfnamefont {B.}~\bibnamefont {Zeng}}, \bibinfo {author}
  {\bibfnamefont {L.}~\bibnamefont {Balicas}}, \bibinfo {author} {\bibfnamefont
  {A.}~\bibnamefont {Sakai}}, \ and\ \bibinfo {author} {\bibfnamefont
  {S.}~\bibnamefont {Nakatsuji}},\ }\href {\doibase 10.1103/PhysRevB.91.241102}
  {\bibfield  {journal} {\bibinfo  {journal} {Phys. Rev. B}\ }\textbf {\bibinfo
  {volume} {91}},\ \bibinfo {pages} {241102} (\bibinfo {year}
  {2015})}\BibitemShut {NoStop}%
\bibitem [{\citenamefont {Ito}\ \emph {et~al.}(2011)\citenamefont {Ito},
  \citenamefont {Higemoto}, \citenamefont {Ninomiya}, \citenamefont {Luetkens},
  \citenamefont {Baines}, \citenamefont {Sakai},\ and\ \citenamefont
  {Nakatsuji}}]{ito11}%
  \BibitemOpen
  \bibfield  {author} {\bibinfo {author} {\bibfnamefont {T.~U.}\ \bibnamefont
  {Ito}}, \bibinfo {author} {\bibfnamefont {W.}~\bibnamefont {Higemoto}},
  \bibinfo {author} {\bibfnamefont {K.}~\bibnamefont {Ninomiya}}, \bibinfo
  {author} {\bibfnamefont {H.}~\bibnamefont {Luetkens}}, \bibinfo {author}
  {\bibfnamefont {C.}~\bibnamefont {Baines}}, \bibinfo {author} {\bibfnamefont
  {A.}~\bibnamefont {Sakai}}, \ and\ \bibinfo {author} {\bibfnamefont
  {S.}~\bibnamefont {Nakatsuji}},\ }\href {\doibase 10.1143/JPSJ.80.113703}
  {\bibfield  {journal} {\bibinfo  {journal} {Journal of the Physical Society
  of Japan}\ }\textbf {\bibinfo {volume} {80}},\ \bibinfo {pages} {113703}
  (\bibinfo {year} {2011})}\BibitemShut {NoStop}%
\bibitem [{\citenamefont {Taniguchi}\ \emph {et~al.}(2016)\citenamefont
  {Taniguchi}, \citenamefont {Yoshida}, \citenamefont {Takeda}, \citenamefont
  {Takigawa}, \citenamefont {Tsujimoto}, \citenamefont {Sakai}, \citenamefont
  {Matsumoto},\ and\ \citenamefont {Nakatsuji}}]{taniguchi16}%
  \BibitemOpen
  \bibfield  {author} {\bibinfo {author} {\bibfnamefont {T.}~\bibnamefont
  {Taniguchi}}, \bibinfo {author} {\bibfnamefont {M.}~\bibnamefont {Yoshida}},
  \bibinfo {author} {\bibfnamefont {H.}~\bibnamefont {Takeda}}, \bibinfo
  {author} {\bibfnamefont {M.}~\bibnamefont {Takigawa}}, \bibinfo {author}
  {\bibfnamefont {M.}~\bibnamefont {Tsujimoto}}, \bibinfo {author}
  {\bibfnamefont {A.}~\bibnamefont {Sakai}}, \bibinfo {author} {\bibfnamefont
  {Y.}~\bibnamefont {Matsumoto}}, \ and\ \bibinfo {author} {\bibfnamefont
  {S.}~\bibnamefont {Nakatsuji}},\ }\href {\doibase 10.7566/JPSJ.85.113703}
  {\bibfield  {journal} {\bibinfo  {journal} {Journal of the Physical Society
  of Japan}\ }\textbf {\bibinfo {volume} {85}},\ \bibinfo {pages} {113703}
  (\bibinfo {year} {2016})}\BibitemShut {NoStop}%
\bibitem [{\citenamefont {Iwasa}\ \emph {et~al.}(2017)\citenamefont {Iwasa},
  \citenamefont {Matsumoto}, \citenamefont {Onimaru}, \citenamefont
  {Takabatake}, \citenamefont {Mignot},\ and\ \citenamefont
  {Gukasov}}]{iwasa17}%
  \BibitemOpen
  \bibfield  {author} {\bibinfo {author} {\bibfnamefont {K.}~\bibnamefont
  {Iwasa}}, \bibinfo {author} {\bibfnamefont {K.~T.}\ \bibnamefont
  {Matsumoto}}, \bibinfo {author} {\bibfnamefont {T.}~\bibnamefont {Onimaru}},
  \bibinfo {author} {\bibfnamefont {T.}~\bibnamefont {Takabatake}}, \bibinfo
  {author} {\bibfnamefont {J.-M.}\ \bibnamefont {Mignot}}, \ and\ \bibinfo
  {author} {\bibfnamefont {A.}~\bibnamefont {Gukasov}},\ }\href {\doibase
  10.1103/PhysRevB.95.155106} {\bibfield  {journal} {\bibinfo  {journal} {Phys.
  Rev. B}\ }\textbf {\bibinfo {volume} {95}},\ \bibinfo {pages} {155106}
  (\bibinfo {year} {2017})}\BibitemShut {NoStop}%
\bibitem [{\citenamefont {Higashinaka}\ \emph {et~al.}(2011)\citenamefont
  {Higashinaka}, \citenamefont {Nakama}, \citenamefont {Ando}, \citenamefont
  {Watanabe}, \citenamefont {Aoki},\ and\ \citenamefont
  {Sato}}]{higashinaka11}%
  \BibitemOpen
  \bibfield  {author} {\bibinfo {author} {\bibfnamefont {R.}~\bibnamefont
  {Higashinaka}}, \bibinfo {author} {\bibfnamefont {A.}~\bibnamefont {Nakama}},
  \bibinfo {author} {\bibfnamefont {M.}~\bibnamefont {Ando}}, \bibinfo {author}
  {\bibfnamefont {M.}~\bibnamefont {Watanabe}}, \bibinfo {author}
  {\bibfnamefont {Y.}~\bibnamefont {Aoki}}, \ and\ \bibinfo {author}
  {\bibfnamefont {H.}~\bibnamefont {Sato}},\ }\href {\doibase
  10.1143/JPSJS.80SA.SA048} {\bibfield  {journal} {\bibinfo  {journal} {Journal
  of the Physical Society of Japan}\ }\textbf {\bibinfo {volume} {80}},\
  \bibinfo {pages} {SA048} (\bibinfo {year} {2011})}\BibitemShut {NoStop}%
\bibitem [{\citenamefont {Kubo}\ \emph {et~al.}(2015)\citenamefont {Kubo},
  \citenamefont {Kotegawa}, \citenamefont {Tou}, \citenamefont {Higashinaka},
  \citenamefont {Nakama}, \citenamefont {Aoki},\ and\ \citenamefont
  {Sato}}]{kubo15}%
  \BibitemOpen
  \bibfield  {author} {\bibinfo {author} {\bibfnamefont {T.}~\bibnamefont
  {Kubo}}, \bibinfo {author} {\bibfnamefont {H.}~\bibnamefont {Kotegawa}},
  \bibinfo {author} {\bibfnamefont {H.}~\bibnamefont {Tou}}, \bibinfo {author}
  {\bibfnamefont {R.}~\bibnamefont {Higashinaka}}, \bibinfo {author}
  {\bibfnamefont {A.}~\bibnamefont {Nakama}}, \bibinfo {author} {\bibfnamefont
  {Y.}~\bibnamefont {Aoki}}, \ and\ \bibinfo {author} {\bibfnamefont
  {H.}~\bibnamefont {Sato}},\ }\href {\doibase 10.7566/JPSJ.84.074701}
  {\bibfield  {journal} {\bibinfo  {journal} {Journal of the Physical Society
  of Japan}\ }\textbf {\bibinfo {volume} {84}},\ \bibinfo {pages} {074701}
  (\bibinfo {year} {2015})}\BibitemShut {NoStop}%
\bibitem [{\citenamefont {Matsubayashi}\ \emph {et~al.}(2012)\citenamefont
  {Matsubayashi}, \citenamefont {Tanaka}, \citenamefont {Sakai}, \citenamefont
  {Nakatsuji}, \citenamefont {Kubo},\ and\ \citenamefont
  {Uwatoko}}]{matsubayashi12}%
  \BibitemOpen
  \bibfield  {author} {\bibinfo {author} {\bibfnamefont {K.}~\bibnamefont
  {Matsubayashi}}, \bibinfo {author} {\bibfnamefont {T.}~\bibnamefont
  {Tanaka}}, \bibinfo {author} {\bibfnamefont {A.}~\bibnamefont {Sakai}},
  \bibinfo {author} {\bibfnamefont {S.}~\bibnamefont {Nakatsuji}}, \bibinfo
  {author} {\bibfnamefont {Y.}~\bibnamefont {Kubo}}, \ and\ \bibinfo {author}
  {\bibfnamefont {Y.}~\bibnamefont {Uwatoko}},\ }\href {\doibase
  10.1103/PhysRevLett.109.187004} {\bibfield  {journal} {\bibinfo  {journal}
  {Physical Review Letters}\ }\textbf {\bibinfo {volume} {109}},\ \bibinfo
  {pages} {187004} (\bibinfo {year} {2012})}\BibitemShut {NoStop}%
\bibitem [{\citenamefont {Onimaru}\ \emph {et~al.}(2016)\citenamefont
  {Onimaru}, \citenamefont {Izawa}, \citenamefont {Matsumoto}, \citenamefont
  {Yoshida}, \citenamefont {Machida}, \citenamefont {Ikeura}, \citenamefont
  {Wakiya}, \citenamefont {Umeo}, \citenamefont {Kittaka}, \citenamefont
  {Araki}, \citenamefont {Sakakibara},\ and\ \citenamefont
  {Takabatake}}]{onimaru16b}%
  \BibitemOpen
  \bibfield  {author} {\bibinfo {author} {\bibfnamefont {T.}~\bibnamefont
  {Onimaru}}, \bibinfo {author} {\bibfnamefont {K.}~\bibnamefont {Izawa}},
  \bibinfo {author} {\bibfnamefont {K.~T.}\ \bibnamefont {Matsumoto}}, \bibinfo
  {author} {\bibfnamefont {T.}~\bibnamefont {Yoshida}}, \bibinfo {author}
  {\bibfnamefont {Y.}~\bibnamefont {Machida}}, \bibinfo {author} {\bibfnamefont
  {T.}~\bibnamefont {Ikeura}}, \bibinfo {author} {\bibfnamefont
  {K.}~\bibnamefont {Wakiya}}, \bibinfo {author} {\bibfnamefont
  {K.}~\bibnamefont {Umeo}}, \bibinfo {author} {\bibfnamefont {S.}~\bibnamefont
  {Kittaka}}, \bibinfo {author} {\bibfnamefont {K.}~\bibnamefont {Araki}},
  \bibinfo {author} {\bibfnamefont {T.}~\bibnamefont {Sakakibara}}, \ and\
  \bibinfo {author} {\bibfnamefont {T.}~\bibnamefont {Takabatake}},\ }\href
  {\doibase 10.1103/PhysRevB.94.075134} {\bibfield  {journal} {\bibinfo
  {journal} {Phys. Rev. B}\ }\textbf {\bibinfo {volume} {94}},\ \bibinfo
  {pages} {075134} (\bibinfo {year} {2016})}\BibitemShut {NoStop}%
\bibitem [{\citenamefont {Yoshida}\ \emph {et~al.}(2017)\citenamefont
  {Yoshida}, \citenamefont {Machida}, \citenamefont {Izawa}, \citenamefont
  {Shimada}, \citenamefont {Nagasawa}, \citenamefont {Onimaru}, \citenamefont
  {Takabatake}, \citenamefont {Gourgout}, \citenamefont {Pourret},
  \citenamefont {Knebel},\ and\ \citenamefont {Brison}}]{yoshida17}%
  \BibitemOpen
  \bibfield  {author} {\bibinfo {author} {\bibfnamefont {T.}~\bibnamefont
  {Yoshida}}, \bibinfo {author} {\bibfnamefont {Y.}~\bibnamefont {Machida}},
  \bibinfo {author} {\bibfnamefont {K.}~\bibnamefont {Izawa}}, \bibinfo
  {author} {\bibfnamefont {Y.}~\bibnamefont {Shimada}}, \bibinfo {author}
  {\bibfnamefont {N.}~\bibnamefont {Nagasawa}}, \bibinfo {author}
  {\bibfnamefont {T.}~\bibnamefont {Onimaru}}, \bibinfo {author} {\bibfnamefont
  {T.}~\bibnamefont {Takabatake}}, \bibinfo {author} {\bibfnamefont
  {A.}~\bibnamefont {Gourgout}}, \bibinfo {author} {\bibfnamefont
  {A.}~\bibnamefont {Pourret}}, \bibinfo {author} {\bibfnamefont
  {G.}~\bibnamefont {Knebel}}, \ and\ \bibinfo {author} {\bibfnamefont {J.-P.}\
  \bibnamefont {Brison}},\ }\href {\doibase 10.7566/JPSJ.86.044711} {\bibfield
  {journal} {\bibinfo  {journal} {Journal of the Physical Society of Japan}\
  }\textbf {\bibinfo {volume} {86}},\ \bibinfo {pages} {044711} (\bibinfo
  {year} {2017})}\BibitemShut {NoStop}%
\bibitem [{\citenamefont {Morin}\ \emph {et~al.}(1982)\citenamefont {Morin},
  \citenamefont {Schmitt},\ and\ \citenamefont {du~Tremolet~de
  Lacheisserie}}]{morin82}%
  \BibitemOpen
  \bibfield  {author} {\bibinfo {author} {\bibfnamefont {P.}~\bibnamefont
  {Morin}}, \bibinfo {author} {\bibfnamefont {D.}~\bibnamefont {Schmitt}}, \
  and\ \bibinfo {author} {\bibfnamefont {E.}~\bibnamefont {du~Tremolet~de
  Lacheisserie}},\ }\href {\doibase
  https://doi.org/10.1016/0304-8853(82)90206-2} {\bibfield  {journal} {\bibinfo
   {journal} {Journal of Magnetism and Magnetic Materials}\ }\textbf {\bibinfo
  {volume} {30}},\ \bibinfo {pages} {257 } (\bibinfo {year}
  {1982})}\BibitemShut {NoStop}%
\bibitem [{\citenamefont {Onimaru}\ \emph {et~al.}(2005)\citenamefont
  {Onimaru}, \citenamefont {Sakakibara}, \citenamefont {Aso}, \citenamefont
  {Yoshizawa}, \citenamefont {Suzuki},\ and\ \citenamefont
  {Takeuchi}}]{onimaru05}%
  \BibitemOpen
  \bibfield  {author} {\bibinfo {author} {\bibfnamefont {T.}~\bibnamefont
  {Onimaru}}, \bibinfo {author} {\bibfnamefont {T.}~\bibnamefont {Sakakibara}},
  \bibinfo {author} {\bibfnamefont {N.}~\bibnamefont {Aso}}, \bibinfo {author}
  {\bibfnamefont {H.}~\bibnamefont {Yoshizawa}}, \bibinfo {author}
  {\bibfnamefont {H.~S.}\ \bibnamefont {Suzuki}}, \ and\ \bibinfo {author}
  {\bibfnamefont {T.}~\bibnamefont {Takeuchi}},\ }\href {\doibase
  10.1103/PhysRevLett.94.197201} {\bibfield  {journal} {\bibinfo  {journal}
  {Phys. Rev. Lett.}\ }\textbf {\bibinfo {volume} {94}},\ \bibinfo {pages}
  {197201} (\bibinfo {year} {2005})}\BibitemShut {NoStop}%
\bibitem [{\citenamefont {Kawae}\ \emph {et~al.}(2006)\citenamefont {Kawae},
  \citenamefont {Kinoshita}, \citenamefont {Nakaie}, \citenamefont {Tateiwa},
  \citenamefont {Takeda}, \citenamefont {Suzuki},\ and\ \citenamefont
  {Kitai}}]{kawae06}%
  \BibitemOpen
  \bibfield  {author} {\bibinfo {author} {\bibfnamefont {T.}~\bibnamefont
  {Kawae}}, \bibinfo {author} {\bibfnamefont {K.}~\bibnamefont {Kinoshita}},
  \bibinfo {author} {\bibfnamefont {Y.}~\bibnamefont {Nakaie}}, \bibinfo
  {author} {\bibfnamefont {N.}~\bibnamefont {Tateiwa}}, \bibinfo {author}
  {\bibfnamefont {K.}~\bibnamefont {Takeda}}, \bibinfo {author} {\bibfnamefont
  {H.~S.}\ \bibnamefont {Suzuki}}, \ and\ \bibinfo {author} {\bibfnamefont
  {T.}~\bibnamefont {Kitai}},\ }\href {\doibase 10.1103/PhysRevLett.96.027210}
  {\bibfield  {journal} {\bibinfo  {journal} {Phys. Rev. Lett.}\ }\textbf
  {\bibinfo {volume} {96}},\ \bibinfo {pages} {027210} (\bibinfo {year}
  {2006})}\BibitemShut {NoStop}%
\bibitem [{\citenamefont {Sato}\ \emph {et~al.}(2010)\citenamefont {Sato},
  \citenamefont {Morodomi}, \citenamefont {Ienaga}, \citenamefont {Inagaki},
  \citenamefont {Kawae}, \citenamefont {S.~Suzuki},\ and\ \citenamefont
  {Onimaru}}]{sato10}%
  \BibitemOpen
  \bibfield  {author} {\bibinfo {author} {\bibfnamefont {Y.}~\bibnamefont
  {Sato}}, \bibinfo {author} {\bibfnamefont {H.}~\bibnamefont {Morodomi}},
  \bibinfo {author} {\bibfnamefont {K.}~\bibnamefont {Ienaga}}, \bibinfo
  {author} {\bibfnamefont {Y.}~\bibnamefont {Inagaki}}, \bibinfo {author}
  {\bibfnamefont {T.}~\bibnamefont {Kawae}}, \bibinfo {author} {\bibfnamefont
  {H.}~\bibnamefont {S.~Suzuki}}, \ and\ \bibinfo {author} {\bibfnamefont
  {T.}~\bibnamefont {Onimaru}},\ }\href {\doibase 10.1143/JPSJ.79.093708}
  {\bibfield  {journal} {\bibinfo  {journal} {Journal of the Physical Society
  of Japan}\ }\textbf {\bibinfo {volume} {79}},\ \bibinfo {pages} {093708}
  (\bibinfo {year} {2010})}\BibitemShut {NoStop}%
\bibitem [{\citenamefont {Yatskar}\ \emph {et~al.}(1996)\citenamefont
  {Yatskar}, \citenamefont {Beyermann}, \citenamefont {Movshovich},\ and\
  \citenamefont {Canfield}}]{yatskar96}%
  \BibitemOpen
  \bibfield  {author} {\bibinfo {author} {\bibfnamefont {A.}~\bibnamefont
  {Yatskar}}, \bibinfo {author} {\bibfnamefont {W.~P.}\ \bibnamefont
  {Beyermann}}, \bibinfo {author} {\bibfnamefont {R.}~\bibnamefont
  {Movshovich}}, \ and\ \bibinfo {author} {\bibfnamefont {P.~C.}\ \bibnamefont
  {Canfield}},\ }\href {\doibase 10.1103/PhysRevLett.77.3637} {\bibfield
  {journal} {\bibinfo  {journal} {Phys. Rev. Lett.}\ }\textbf {\bibinfo
  {volume} {77}},\ \bibinfo {pages} {3637} (\bibinfo {year}
  {1996})}\BibitemShut {NoStop}%
\bibitem [{\citenamefont {Tanida}\ \emph {et~al.}(2006)\citenamefont {Tanida},
  \citenamefont {S.~Suzuki}, \citenamefont {Takagi}, \citenamefont {Onodera},\
  and\ \citenamefont {Tanigaki}}]{tanida06}%
  \BibitemOpen
  \bibfield  {author} {\bibinfo {author} {\bibfnamefont {H.}~\bibnamefont
  {Tanida}}, \bibinfo {author} {\bibfnamefont {H.}~\bibnamefont {S.~Suzuki}},
  \bibinfo {author} {\bibfnamefont {S.}~\bibnamefont {Takagi}}, \bibinfo
  {author} {\bibfnamefont {H.}~\bibnamefont {Onodera}}, \ and\ \bibinfo
  {author} {\bibfnamefont {K.}~\bibnamefont {Tanigaki}},\ }\href {\doibase
  10.1143/JPSJ.75.073705} {\bibfield  {journal} {\bibinfo  {journal} {Journal
  of the Physical Society of Japan}\ }\textbf {\bibinfo {volume} {75}},\
  \bibinfo {pages} {073705} (\bibinfo {year} {2006})}\BibitemShut {NoStop}%
\bibitem [{\citenamefont {Andrei}\ and\ \citenamefont
  {Coleman}(1989)}]{andrei89}%
  \BibitemOpen
  \bibfield  {author} {\bibinfo {author} {\bibfnamefont {N.}~\bibnamefont
  {Andrei}}\ and\ \bibinfo {author} {\bibfnamefont {P.}~\bibnamefont
  {Coleman}},\ }\href {\doibase 10.1103/PhysRevLett.62.595} {\bibfield
  {journal} {\bibinfo  {journal} {Physical Review Letters}\ }\textbf {\bibinfo
  {volume} {62}},\ \bibinfo {pages} {595} (\bibinfo {year} {1989})}\BibitemShut
  {NoStop}%
\bibitem [{\citenamefont {Senthil}\ \emph {et~al.}(2003)\citenamefont
  {Senthil}, \citenamefont {Sachdev},\ and\ \citenamefont {Vojta}}]{senthil03}%
  \BibitemOpen
  \bibfield  {author} {\bibinfo {author} {\bibfnamefont {T.}~\bibnamefont
  {Senthil}}, \bibinfo {author} {\bibfnamefont {S.}~\bibnamefont {Sachdev}}, \
  and\ \bibinfo {author} {\bibfnamefont {M.}~\bibnamefont {Vojta}},\ }\href
  {\doibase 10.1103/PhysRevLett.90.216403} {\bibfield  {journal} {\bibinfo
  {journal} {Physical Review Letters}\ }\textbf {\bibinfo {volume} {90}},\
  \bibinfo {pages} {216403} (\bibinfo {year} {2003})}\BibitemShut {NoStop}%
\bibitem [{\citenamefont {Tsvelik}\ and\ \citenamefont
  {Ventura}(2000)}]{tsvelik00}%
  \BibitemOpen
  \bibfield  {author} {\bibinfo {author} {\bibfnamefont {A.~M.}\ \bibnamefont
  {Tsvelik}}\ and\ \bibinfo {author} {\bibfnamefont {C.~I.}\ \bibnamefont
  {Ventura}},\ }\href {\doibase 10.1103/PhysRevB.61.15538} {\bibfield
  {journal} {\bibinfo  {journal} {Phys. Rev. B}\ }\textbf {\bibinfo {volume}
  {61}},\ \bibinfo {pages} {15538} (\bibinfo {year} {2000})}\BibitemShut
  {NoStop}%
\bibitem [{\citenamefont {Schauerte}\ \emph {et~al.}(2005)\citenamefont
  {Schauerte}, \citenamefont {Cox}, \citenamefont {Noack}, \citenamefont {van
  Dongen},\ and\ \citenamefont {Batista}}]{schauerte05}%
  \BibitemOpen
  \bibfield  {author} {\bibinfo {author} {\bibfnamefont {T.}~\bibnamefont
  {Schauerte}}, \bibinfo {author} {\bibfnamefont {D.~L.}\ \bibnamefont {Cox}},
  \bibinfo {author} {\bibfnamefont {R.~M.}\ \bibnamefont {Noack}}, \bibinfo
  {author} {\bibfnamefont {P.~G.~J.}\ \bibnamefont {van Dongen}}, \ and\
  \bibinfo {author} {\bibfnamefont {C.~D.}\ \bibnamefont {Batista}},\ }\href
  {\doibase 10.1103/PhysRevLett.94.147201} {\bibfield  {journal} {\bibinfo
  {journal} {Phys. Rev. Lett.}\ }\textbf {\bibinfo {volume} {94}},\ \bibinfo
  {pages} {147201} (\bibinfo {year} {2005})}\BibitemShut {NoStop}%
\bibitem [{\citenamefont {Coleman}(1983)}]{Coleman83}%
  \BibitemOpen
  \bibfield  {author} {\bibinfo {author} {\bibfnamefont {P.}~\bibnamefont
  {Coleman}},\ }\href {\doibase 10.1103/PhysRevB.28.5255} {\bibfield  {journal}
  {\bibinfo  {journal} {Phys. Rev. B}\ }\textbf {\bibinfo {volume} {28}},\
  \bibinfo {pages} {5255} (\bibinfo {year} {1983})}\BibitemShut {NoStop}%
\bibitem [{\citenamefont {Coqblin}\ and\ \citenamefont
  {Schrieffer}(1969)}]{Coqblin69}%
  \BibitemOpen
  \bibfield  {author} {\bibinfo {author} {\bibfnamefont {B.}~\bibnamefont
  {Coqblin}}\ and\ \bibinfo {author} {\bibfnamefont {J.~R.}\ \bibnamefont
  {Schrieffer}},\ }\href {\doibase 10.1103/PhysRev.185.847} {\bibfield
  {journal} {\bibinfo  {journal} {Phys. Rev.}\ }\textbf {\bibinfo {volume}
  {185}},\ \bibinfo {pages} {847} (\bibinfo {year} {1969})}\BibitemShut
  {NoStop}%
\bibitem [{\citenamefont {Arovas}\ and\ \citenamefont
  {Auerbach}(1988)}]{Arovas88}%
  \BibitemOpen
  \bibfield  {author} {\bibinfo {author} {\bibfnamefont {D.~P.}\ \bibnamefont
  {Arovas}}\ and\ \bibinfo {author} {\bibfnamefont {A.}~\bibnamefont
  {Auerbach}},\ }\href {\doibase 10.1103/PhysRevB.38.316} {\bibfield  {journal}
  {\bibinfo  {journal} {Phys. Rev. B}\ }\textbf {\bibinfo {volume} {38}},\
  \bibinfo {pages} {316} (\bibinfo {year} {1988})}\BibitemShut {NoStop}%
\bibitem [{\citenamefont {Abrahams}\ \emph {et~al.}(1995)\citenamefont
  {Abrahams}, \citenamefont {Balatsky}, \citenamefont {Scalapino},\ and\
  \citenamefont {Schrieffer}}]{abrahams}%
  \BibitemOpen
  \bibfield  {author} {\bibinfo {author} {\bibfnamefont {E.}~\bibnamefont
  {Abrahams}}, \bibinfo {author} {\bibfnamefont {A.}~\bibnamefont {Balatsky}},
  \bibinfo {author} {\bibfnamefont {D.~J.}\ \bibnamefont {Scalapino}}, \ and\
  \bibinfo {author} {\bibfnamefont {J.~R.}\ \bibnamefont {Schrieffer}},\ }\href
  {\doibase 10.1103/PhysRevB.52.1271} {\bibfield  {journal} {\bibinfo
  {journal} {Phys. Rev. B}\ }\textbf {\bibinfo {volume} {52}},\ \bibinfo
  {pages} {1271} (\bibinfo {year} {1995})}\BibitemShut {NoStop}%
\bibitem [{\citenamefont {Coleman}\ \emph {et~al.}(1999)\citenamefont
  {Coleman}, \citenamefont {Tsvelik}, \citenamefont {Andrei},\ and\
  \citenamefont {Kee}}]{catk}%
  \BibitemOpen
  \bibfield  {author} {\bibinfo {author} {\bibfnamefont {P.}~\bibnamefont
  {Coleman}}, \bibinfo {author} {\bibfnamefont {A.~M.}\ \bibnamefont
  {Tsvelik}}, \bibinfo {author} {\bibfnamefont {N.}~\bibnamefont {Andrei}}, \
  and\ \bibinfo {author} {\bibfnamefont {H.~Y.}\ \bibnamefont {Kee}},\ }\href
  {\doibase 10.1103/PhysRevB.60.3608} {\bibfield  {journal} {\bibinfo
  {journal} {Phys. Rev. B}\ }\textbf {\bibinfo {volume} {60}},\ \bibinfo
  {pages} {3608} (\bibinfo {year} {1999})}\BibitemShut {NoStop}%
\bibitem [{\citenamefont {Anders}(2002)}]{anders02}%
  \BibitemOpen
  \bibfield  {author} {\bibinfo {author} {\bibfnamefont {F.~B.}\ \bibnamefont
  {Anders}},\ }\href {\doibase 10.1103/PhysRevB.66.020504} {\bibfield
  {journal} {\bibinfo  {journal} {Phys. Rev. B}\ }\textbf {\bibinfo {volume}
  {66}},\ \bibinfo {pages} {020504} (\bibinfo {year} {2002})}\BibitemShut
  {NoStop}%
\bibitem [{\citenamefont {Flint}\ \emph {et~al.}(2008)\citenamefont {Flint},
  \citenamefont {Dzero},\ and\ \citenamefont {Coleman}}]{flint08}%
  \BibitemOpen
  \bibfield  {author} {\bibinfo {author} {\bibfnamefont {R.}~\bibnamefont
  {Flint}}, \bibinfo {author} {\bibfnamefont {M.}~\bibnamefont {Dzero}}, \ and\
  \bibinfo {author} {\bibfnamefont {P.}~\bibnamefont {Coleman}},\ }\href
  {\doibase 10.1038/nphys1024} {\bibfield  {journal} {\bibinfo  {journal}
  {Nature Physics}\ }\textbf {\bibinfo {volume} {4}},\ \bibinfo {pages} {643}
  (\bibinfo {year} {2008})}\BibitemShut {NoStop}%
\bibitem [{\citenamefont {Flint}\ and\ \citenamefont
  {Coleman}(2010)}]{flint10}%
  \BibitemOpen
  \bibfield  {author} {\bibinfo {author} {\bibfnamefont {R.}~\bibnamefont
  {Flint}}\ and\ \bibinfo {author} {\bibfnamefont {P.}~\bibnamefont
  {Coleman}},\ }\href {\doibase 10.1103/PhysRevLett.105.246404} {\bibfield
  {journal} {\bibinfo  {journal} {Phys. Rev. Lett.}\ }\textbf {\bibinfo
  {volume} {105}},\ \bibinfo {pages} {246404} (\bibinfo {year}
  {2010})}\BibitemShut {NoStop}%
\bibitem [{\citenamefont {Hoshino}(2014)}]{hoshino142}%
  \BibitemOpen
  \bibfield  {author} {\bibinfo {author} {\bibfnamefont {S.}~\bibnamefont
  {Hoshino}},\ }\href {\doibase 10.1103/PhysRevB.90.115154} {\bibfield
  {journal} {\bibinfo  {journal} {Phys. Rev. B}\ }\textbf {\bibinfo {volume}
  {90}},\ \bibinfo {pages} {115154} (\bibinfo {year} {2014})}\BibitemShut
  {NoStop}%
\bibitem [{\citenamefont {Mathur}\ \emph {et~al.}(1998)\citenamefont {Mathur},
  \citenamefont {Grosche}, \citenamefont {Julian}, \citenamefont {Walker},
  \citenamefont {Freye}, \citenamefont {Haselwimmer},\ and\ \citenamefont
  {Lonzarich}}]{mathur}%
  \BibitemOpen
  \bibfield  {author} {\bibinfo {author} {\bibfnamefont {N.~D.}\ \bibnamefont
  {Mathur}}, \bibinfo {author} {\bibfnamefont {F.~M.}\ \bibnamefont {Grosche}},
  \bibinfo {author} {\bibfnamefont {S.~R.}\ \bibnamefont {Julian}}, \bibinfo
  {author} {\bibfnamefont {I.~R.}\ \bibnamefont {Walker}}, \bibinfo {author}
  {\bibfnamefont {D.~M.}\ \bibnamefont {Freye}}, \bibinfo {author}
  {\bibfnamefont {R.~K.~W.}\ \bibnamefont {Haselwimmer}}, \ and\ \bibinfo
  {author} {\bibfnamefont {G.~G.}\ \bibnamefont {Lonzarich}},\ }\href
  {http://dx.doi.org/10.1038/27838} {\bibfield  {journal} {\bibinfo  {journal}
  {Nature}\ }\textbf {\bibinfo {volume} {394}},\ \bibinfo {pages} {39 EP }
  (\bibinfo {year} {1998})},\ \bibinfo {note} {article}\BibitemShut {NoStop}%
\bibitem [{\citenamefont {Miyake}\ \emph {et~al.}(1986)\citenamefont {Miyake},
  \citenamefont {Schmitt-Rink},\ and\ \citenamefont {Varma}}]{miyake}%
  \BibitemOpen
  \bibfield  {author} {\bibinfo {author} {\bibfnamefont {K.}~\bibnamefont
  {Miyake}}, \bibinfo {author} {\bibfnamefont {S.}~\bibnamefont
  {Schmitt-Rink}}, \ and\ \bibinfo {author} {\bibfnamefont {C.~M.}\
  \bibnamefont {Varma}},\ }\href {\doibase 10.1103/PhysRevB.34.6554} {\bibfield
   {journal} {\bibinfo  {journal} {Phys. Rev. B}\ }\textbf {\bibinfo {volume}
  {34}},\ \bibinfo {pages} {6554} (\bibinfo {year} {1986})}\BibitemShut
  {NoStop}%
\bibitem [{\citenamefont {Scalapino}\ \emph {et~al.}(1986)\citenamefont
  {Scalapino}, \citenamefont {Loh},\ and\ \citenamefont {Hirsch}}]{scalapino}%
  \BibitemOpen
  \bibfield  {author} {\bibinfo {author} {\bibfnamefont {D.~J.}\ \bibnamefont
  {Scalapino}}, \bibinfo {author} {\bibfnamefont {E.}~\bibnamefont {Loh}}, \
  and\ \bibinfo {author} {\bibfnamefont {J.~E.}\ \bibnamefont {Hirsch}},\
  }\href {\doibase 10.1103/PhysRevB.34.8190} {\bibfield  {journal} {\bibinfo
  {journal} {Phys. Rev. B}\ }\textbf {\bibinfo {volume} {34}},\ \bibinfo
  {pages} {8190} (\bibinfo {year} {1986})}\BibitemShut {NoStop}%
\bibitem [{\citenamefont {B\'eal-Monod}\ \emph {et~al.}(1986)\citenamefont
  {B\'eal-Monod}, \citenamefont {Bourbonnais},\ and\ \citenamefont
  {Emery}}]{bourbonnais}%
  \BibitemOpen
  \bibfield  {author} {\bibinfo {author} {\bibfnamefont {M.~T.}\ \bibnamefont
  {B\'eal-Monod}}, \bibinfo {author} {\bibfnamefont {C.}~\bibnamefont
  {Bourbonnais}}, \ and\ \bibinfo {author} {\bibfnamefont {V.~J.}\ \bibnamefont
  {Emery}},\ }\href {\doibase 10.1103/PhysRevB.34.7716} {\bibfield  {journal}
  {\bibinfo  {journal} {Phys. Rev. B}\ }\textbf {\bibinfo {volume} {34}},\
  \bibinfo {pages} {7716} (\bibinfo {year} {1986})}\BibitemShut {NoStop}%
\bibitem [{\citenamefont {Coleman}\ \emph {et~al.}(2005)\citenamefont
  {Coleman}, \citenamefont {Marston},\ and\ \citenamefont
  {Schofield}}]{Coleman2005}%
  \BibitemOpen
  \bibfield  {author} {\bibinfo {author} {\bibfnamefont {P.}~\bibnamefont
  {Coleman}}, \bibinfo {author} {\bibfnamefont {J.~B.}\ \bibnamefont
  {Marston}}, \ and\ \bibinfo {author} {\bibfnamefont {A.~J.}\ \bibnamefont
  {Schofield}},\ }\href {\doibase 10.1103/PhysRevB.72.245111} {\bibfield
  {journal} {\bibinfo  {journal} {Phys. Rev. B}\ }\textbf {\bibinfo {volume}
  {72}},\ \bibinfo {pages} {245111} (\bibinfo {year} {2005})}\BibitemShut
  {NoStop}%
\bibitem [{\citenamefont {Shiina}\ \emph {et~al.}(1997)\citenamefont {Shiina},
  \citenamefont {Shiba},\ and\ \citenamefont {Thalmeier}}]{Shiina97}%
  \BibitemOpen
  \bibfield  {author} {\bibinfo {author} {\bibfnamefont {R.}~\bibnamefont
  {Shiina}}, \bibinfo {author} {\bibfnamefont {H.}~\bibnamefont {Shiba}}, \
  and\ \bibinfo {author} {\bibfnamefont {P.}~\bibnamefont {Thalmeier}},\ }\href
  {\doibase 10.1143/JPSJ.66.1741} {\bibfield  {journal} {\bibinfo  {journal}
  {Journal of the Physical Society of Japan}\ }\textbf {\bibinfo {volume}
  {66}},\ \bibinfo {pages} {1741} (\bibinfo {year} {1997})}\BibitemShut
  {NoStop}%
\bibitem [{\citenamefont {Sigrist}\ \emph {et~al.}(1991)\citenamefont
  {Sigrist}, \citenamefont {Tsunetsuga},\ and\ \citenamefont
  {Ueda}}]{sigrist91}%
  \BibitemOpen
  \bibfield  {author} {\bibinfo {author} {\bibfnamefont {M.}~\bibnamefont
  {Sigrist}}, \bibinfo {author} {\bibfnamefont {H.}~\bibnamefont {Tsunetsuga}},
  \ and\ \bibinfo {author} {\bibfnamefont {K.}~\bibnamefont {Ueda}},\ }\href
  {\doibase 10.1103/PhysRevLett.67.2211} {\bibfield  {journal} {\bibinfo
  {journal} {Phys. Rev. Lett.}\ }\textbf {\bibinfo {volume} {67}},\ \bibinfo
  {pages} {2211} (\bibinfo {year} {1991})}\BibitemShut {NoStop}%
\bibitem [{\citenamefont {Nagaoka}(1966)}]{Nagaoka66}%
  \BibitemOpen
  \bibfield  {author} {\bibinfo {author} {\bibfnamefont {Y.}~\bibnamefont
  {Nagaoka}},\ }\href {\doibase 10.1103/PhysRev.147.392} {\bibfield  {journal}
  {\bibinfo  {journal} {Phys. Rev.}\ }\textbf {\bibinfo {volume} {147}},\
  \bibinfo {pages} {392} (\bibinfo {year} {1966})}\BibitemShut {NoStop}%
\bibitem [{\citenamefont {Wen}(2002)}]{Wen02}%
  \BibitemOpen
  \bibfield  {author} {\bibinfo {author} {\bibfnamefont {X.-G.}\ \bibnamefont
  {Wen}},\ }\href {\doibase 10.1103/PhysRevB.65.165113} {\bibfield  {journal}
  {\bibinfo  {journal} {Phys. Rev. B}\ }\textbf {\bibinfo {volume} {65}},\
  \bibinfo {pages} {165113} (\bibinfo {year} {2002})}\BibitemShut {NoStop}%
\bibitem [{\citenamefont {Altarawneh}\ \emph {et~al.}(2012)\citenamefont
  {Altarawneh}, \citenamefont {Harrison}, \citenamefont {Li}, \citenamefont
  {Balicas}, \citenamefont {Tobash}, \citenamefont {Ronning},\ and\
  \citenamefont {Bauer}}]{Altarawneh12}%
  \BibitemOpen
  \bibfield  {author} {\bibinfo {author} {\bibfnamefont {M.~M.}\ \bibnamefont
  {Altarawneh}}, \bibinfo {author} {\bibfnamefont {N.}~\bibnamefont
  {Harrison}}, \bibinfo {author} {\bibfnamefont {G.}~\bibnamefont {Li}},
  \bibinfo {author} {\bibfnamefont {L.}~\bibnamefont {Balicas}}, \bibinfo
  {author} {\bibfnamefont {P.~H.}\ \bibnamefont {Tobash}}, \bibinfo {author}
  {\bibfnamefont {F.}~\bibnamefont {Ronning}}, \ and\ \bibinfo {author}
  {\bibfnamefont {E.~D.}\ \bibnamefont {Bauer}},\ }\href {\doibase
  10.1103/PhysRevLett.108.066407} {\bibfield  {journal} {\bibinfo  {journal}
  {Phys. Rev. Lett.}\ }\textbf {\bibinfo {volume} {108}},\ \bibinfo {pages}
  {066407} (\bibinfo {year} {2012})}\BibitemShut {NoStop}%
\bibitem [{\citenamefont {De~Silva}\ \emph {et~al.}(2002)\citenamefont
  {De~Silva}, \citenamefont {Ma},\ and\ \citenamefont {Zhang}}]{desilva}%
  \BibitemOpen
  \bibfield  {author} {\bibinfo {author} {\bibfnamefont {T.~N.}\ \bibnamefont
  {De~Silva}}, \bibinfo {author} {\bibfnamefont {M.}~\bibnamefont {Ma}}, \ and\
  \bibinfo {author} {\bibfnamefont {F.~C.}\ \bibnamefont {Zhang}},\ }\href
  {\doibase 10.1103/PhysRevB.66.104417} {\bibfield  {journal} {\bibinfo
  {journal} {Phys. Rev. B}\ }\textbf {\bibinfo {volume} {66}},\ \bibinfo
  {pages} {104417} (\bibinfo {year} {2002})}\BibitemShut {NoStop}%
\bibitem [{\citenamefont {Coleman}\ and\ \citenamefont
  {Andrei}(1989)}]{coleman89}%
  \BibitemOpen
  \bibfield  {author} {\bibinfo {author} {\bibfnamefont {P.}~\bibnamefont
  {Coleman}}\ and\ \bibinfo {author} {\bibfnamefont {N.}~\bibnamefont
  {Andrei}},\ }\href {\doibase 10.1088/0953-8984/1/26/003} {\bibfield
  {journal} {\bibinfo  {journal} {Journal of Physics: Condensed Matter}\
  }\textbf {\bibinfo {volume} {1}},\ \bibinfo {pages} {4057} (\bibinfo {year}
  {1989})}\BibitemShut {NoStop}%
\bibitem [{\citenamefont {Nakamura}\ \emph {et~al.}(1994)\citenamefont
  {Nakamura}, \citenamefont {Goto}, \citenamefont {Kunii}, \citenamefont
  {Iwashita},\ and\ \citenamefont {Tamaki}}]{Nakamura94}%
  \BibitemOpen
  \bibfield  {author} {\bibinfo {author} {\bibfnamefont {S.}~\bibnamefont
  {Nakamura}}, \bibinfo {author} {\bibfnamefont {T.}~\bibnamefont {Goto}},
  \bibinfo {author} {\bibfnamefont {S.}~\bibnamefont {Kunii}}, \bibinfo
  {author} {\bibfnamefont {K.}~\bibnamefont {Iwashita}}, \ and\ \bibinfo
  {author} {\bibfnamefont {A.}~\bibnamefont {Tamaki}},\ }\href {\doibase
  10.1143/JPSJ.63.623} {\bibfield  {journal} {\bibinfo  {journal} {Journal of
  the Physical Society of Japan}\ }\textbf {\bibinfo {volume} {63}},\ \bibinfo
  {pages} {623} (\bibinfo {year} {1994})}\BibitemShut {NoStop}%
\bibitem [{\citenamefont {Hazama}\ \emph {et~al.}(2000)\citenamefont {Hazama},
  \citenamefont {Goto}, \citenamefont {Nemoto}, \citenamefont {Tomioka},
  \citenamefont {Asamitsu},\ and\ \citenamefont {Tokura}}]{Hazama00}%
  \BibitemOpen
  \bibfield  {author} {\bibinfo {author} {\bibfnamefont {H.}~\bibnamefont
  {Hazama}}, \bibinfo {author} {\bibfnamefont {T.}~\bibnamefont {Goto}},
  \bibinfo {author} {\bibfnamefont {Y.}~\bibnamefont {Nemoto}}, \bibinfo
  {author} {\bibfnamefont {Y.}~\bibnamefont {Tomioka}}, \bibinfo {author}
  {\bibfnamefont {A.}~\bibnamefont {Asamitsu}}, \ and\ \bibinfo {author}
  {\bibfnamefont {Y.}~\bibnamefont {Tokura}},\ }\href {\doibase
  10.1103/PhysRevB.62.15012} {\bibfield  {journal} {\bibinfo  {journal} {Phys.
  Rev. B}\ }\textbf {\bibinfo {volume} {62}},\ \bibinfo {pages} {15012}
  (\bibinfo {year} {2000})}\BibitemShut {NoStop}%
\bibitem [{\citenamefont {Coleman}(2016)}]{Coleman_book}%
  \BibitemOpen
  \bibfield  {author} {\bibinfo {author} {\bibfnamefont {P.}~\bibnamefont
  {Coleman}},\ }\enquote {\bibinfo {title} {Introduction to many-body
  physics},}\ \ (\bibinfo  {publisher} {Cambridge University Press},\ \bibinfo
  {year} {2016})\ pp.\ \bibinfo {pages} {709--713},\ \bibinfo {edition} {1st}\
  ed.\BibitemShut {Stop}%
\bibitem [{\citenamefont {Coleman}\ \emph {et~al.}(2001)\citenamefont
  {Coleman}, \citenamefont {Pépin}, \citenamefont {Si},\ and\ \citenamefont
  {Ramazashvili}}]{Coleman01}%
  \BibitemOpen
  \bibfield  {author} {\bibinfo {author} {\bibfnamefont {P.}~\bibnamefont
  {Coleman}}, \bibinfo {author} {\bibfnamefont {C.}~\bibnamefont {Pépin}},
  \bibinfo {author} {\bibfnamefont {Q.}~\bibnamefont {Si}}, \ and\ \bibinfo
  {author} {\bibfnamefont {R.}~\bibnamefont {Ramazashvili}},\ }\href
  {http://stacks.iop.org/0953-8984/13/i=35/a=202} {\bibfield  {journal}
  {\bibinfo  {journal} {Journal of Physics: Condensed Matter}\ }\textbf
  {\bibinfo {volume} {13}},\ \bibinfo {pages} {R723} (\bibinfo {year}
  {2001})}\BibitemShut {NoStop}%
\bibitem [{\citenamefont {Paschen}\ \emph {et~al.}(2004)\citenamefont
  {Paschen}, \citenamefont {L{\"u}hmann}, \citenamefont {Wirth}, \citenamefont
  {Gegenwart}, \citenamefont {Trovarelli}, \citenamefont {Geibel},
  \citenamefont {Steglich}, \citenamefont {Coleman},\ and\ \citenamefont
  {Si}}]{Paschen04}%
  \BibitemOpen
  \bibfield  {author} {\bibinfo {author} {\bibfnamefont {S.}~\bibnamefont
  {Paschen}}, \bibinfo {author} {\bibfnamefont {T.}~\bibnamefont
  {L{\"u}hmann}}, \bibinfo {author} {\bibfnamefont {S.}~\bibnamefont {Wirth}},
  \bibinfo {author} {\bibfnamefont {P.}~\bibnamefont {Gegenwart}}, \bibinfo
  {author} {\bibfnamefont {O.}~\bibnamefont {Trovarelli}}, \bibinfo {author}
  {\bibfnamefont {C.}~\bibnamefont {Geibel}}, \bibinfo {author} {\bibfnamefont
  {F.}~\bibnamefont {Steglich}}, \bibinfo {author} {\bibfnamefont
  {P.}~\bibnamefont {Coleman}}, \ and\ \bibinfo {author} {\bibfnamefont
  {Q.}~\bibnamefont {Si}},\ }\href {http://dx.doi.org/10.1038/nature03129}
  {\bibfield  {journal} {\bibinfo  {journal} {Nature}\ }\textbf {\bibinfo
  {volume} {432}},\ \bibinfo {pages} {881 EP } (\bibinfo {year}
  {2004})}\BibitemShut {NoStop}%
\bibitem [{\citenamefont {Nakatsuji}()}]{Nakatsuji_talk}%
  \BibitemOpen
  \bibfield  {author} {\bibinfo {author} {\bibfnamefont {S.}~\bibnamefont
  {Nakatsuji}},\ }\href@noop {} {}\bibinfo {howpublished} {private
  communication}\BibitemShut {NoStop}%
\bibitem [{\citenamefont {Onimaru}\ and\ \citenamefont
  {Kusunose}(2016)}]{onimaru16a}%
  \BibitemOpen
  \bibfield  {author} {\bibinfo {author} {\bibfnamefont {T.}~\bibnamefont
  {Onimaru}}\ and\ \bibinfo {author} {\bibfnamefont {H.}~\bibnamefont
  {Kusunose}},\ }\href {\doibase 10.7566/JPSJ.85.082002} {\bibfield  {journal}
  {\bibinfo  {journal} {Journal of the Physical Society of Japan}\ }\textbf
  {\bibinfo {volume} {85}},\ \bibinfo {pages} {082002} (\bibinfo {year}
  {2016})}\BibitemShut {NoStop}%
\bibitem [{\citenamefont {Endo}\ \emph {et~al.}(2002)\citenamefont {Endo},
  \citenamefont {Kimura}, \citenamefont {Aoki}, \citenamefont {Terashima},
  \citenamefont {Uji}, \citenamefont {Terakura},\ and\ \citenamefont
  {Matsumoto}}]{endo02}%
  \BibitemOpen
  \bibfield  {author} {\bibinfo {author} {\bibfnamefont {M.}~\bibnamefont
  {Endo}}, \bibinfo {author} {\bibfnamefont {N.}~\bibnamefont {Kimura}},
  \bibinfo {author} {\bibfnamefont {H.}~\bibnamefont {Aoki}}, \bibinfo {author}
  {\bibfnamefont {T.}~\bibnamefont {Terashima}}, \bibinfo {author}
  {\bibfnamefont {S.}~\bibnamefont {Uji}}, \bibinfo {author} {\bibfnamefont
  {C.}~\bibnamefont {Terakura}}, \ and\ \bibinfo {author} {\bibfnamefont
  {T.}~\bibnamefont {Matsumoto}},\ }\href {\doibase 10.1143/JPSJS.71S.127}
  {\bibfield  {journal} {\bibinfo  {journal} {Journal of the Physical Society
  of Japan}\ }\textbf {\bibinfo {volume} {71}},\ \bibinfo {pages} {127}
  (\bibinfo {year} {2002})}\BibitemShut {NoStop}%
\bibitem [{\citenamefont {Ohkuni}\ \emph {et~al.}(1999)\citenamefont {Ohkuni},
  \citenamefont {Inada}, \citenamefont {Tokiwa}, \citenamefont {Sakurai},
  \citenamefont {Settai}, \citenamefont {Honma}, \citenamefont {Haga},
  \citenamefont {Yamamoto}, \citenamefont {Ōnuki}, \citenamefont {Yamagami},
  \citenamefont {Takahashi},\ and\ \citenamefont {Yanagisawa}}]{Ohkuni99}%
  \BibitemOpen
  \bibfield  {author} {\bibinfo {author} {\bibfnamefont {H.}~\bibnamefont
  {Ohkuni}}, \bibinfo {author} {\bibfnamefont {Y.}~\bibnamefont {Inada}},
  \bibinfo {author} {\bibfnamefont {Y.}~\bibnamefont {Tokiwa}}, \bibinfo
  {author} {\bibfnamefont {K.}~\bibnamefont {Sakurai}}, \bibinfo {author}
  {\bibfnamefont {R.}~\bibnamefont {Settai}}, \bibinfo {author} {\bibfnamefont
  {T.}~\bibnamefont {Honma}}, \bibinfo {author} {\bibfnamefont
  {Y.}~\bibnamefont {Haga}}, \bibinfo {author} {\bibfnamefont {E.}~\bibnamefont
  {Yamamoto}}, \bibinfo {author} {\bibfnamefont {Y.}~\bibnamefont {Ōnuki}},
  \bibinfo {author} {\bibfnamefont {H.}~\bibnamefont {Yamagami}}, \bibinfo
  {author} {\bibfnamefont {S.}~\bibnamefont {Takahashi}}, \ and\ \bibinfo
  {author} {\bibfnamefont {T.}~\bibnamefont {Yanagisawa}},\ }\href {\doibase
  10.1080/13642819908214859} {\bibfield  {journal} {\bibinfo  {journal}
  {Philosophical Magazine B}\ }\textbf {\bibinfo {volume} {79}},\ \bibinfo
  {pages} {1045} (\bibinfo {year} {1999})}\BibitemShut {NoStop}%
\bibitem [{\citenamefont {Abrahams}\ and\ \citenamefont
  {W\"olfle}(2008)}]{Abrahams08}%
  \BibitemOpen
  \bibfield  {author} {\bibinfo {author} {\bibfnamefont {E.}~\bibnamefont
  {Abrahams}}\ and\ \bibinfo {author} {\bibfnamefont {P.}~\bibnamefont
  {W\"olfle}},\ }\href {\doibase 10.1103/PhysRevB.78.104423} {\bibfield
  {journal} {\bibinfo  {journal} {Phys. Rev. B}\ }\textbf {\bibinfo {volume}
  {78}},\ \bibinfo {pages} {104423} (\bibinfo {year} {2008})}\BibitemShut
  {NoStop}%
\bibitem [{\citenamefont {Dyke}\ \emph {et~al.}()\citenamefont {Dyke},
  \citenamefont {Zhang},\ and\ \citenamefont {Flint}}]{vandyke18}%
  \BibitemOpen
  \bibfield  {author} {\bibinfo {author} {\bibfnamefont {J.~S.~V.}\
  \bibnamefont {Dyke}}, \bibinfo {author} {\bibfnamefont {G.}~\bibnamefont
  {Zhang}}, \ and\ \bibinfo {author} {\bibfnamefont {R.}~\bibnamefont
  {Flint}},\ }\href@noop {} {\enquote {\bibinfo {title} {A realistic model of
  cubic ferrohastatic order},}\ }\Eprint
  {http://arxiv.org/abs/arXiv:1802.05752} {arXiv:1802.05752} \BibitemShut
  {NoStop}%
\bibitem [{\citenamefont {Moreno}\ \emph {et~al.}(2001)\citenamefont {Moreno},
  \citenamefont {Qin}, \citenamefont {Coleman},\ and\ \citenamefont
  {Yu}}]{moreno01}%
  \BibitemOpen
  \bibfield  {author} {\bibinfo {author} {\bibfnamefont {J.}~\bibnamefont
  {Moreno}}, \bibinfo {author} {\bibfnamefont {S.}~\bibnamefont {Qin}},
  \bibinfo {author} {\bibfnamefont {P.}~\bibnamefont {Coleman}}, \ and\
  \bibinfo {author} {\bibfnamefont {L.}~\bibnamefont {Yu}},\ }\href {\doibase
  10.1103/PhysRevB.64.085116} {\bibfield  {journal} {\bibinfo  {journal} {Phys.
  Rev. B}\ }\textbf {\bibinfo {volume} {64}},\ \bibinfo {pages} {085116}
  (\bibinfo {year} {2001})}\BibitemShut {NoStop}%
\bibitem [{\citenamefont {Anders}(1999)}]{anders99}%
  \BibitemOpen
  \bibfield  {author} {\bibinfo {author} {\bibfnamefont {F.~B.}\ \bibnamefont
  {Anders}},\ }\href {\doibase 10.1103/PhysRevLett.83.4638} {\bibfield
  {journal} {\bibinfo  {journal} {Phys. Rev. Lett.}\ }\textbf {\bibinfo
  {volume} {83}},\ \bibinfo {pages} {4638} (\bibinfo {year}
  {1999})}\BibitemShut {NoStop}%
\bibitem [{\citenamefont {Kuramoto}\ and\ \citenamefont
  {Hoshino}(2014)}]{kuramoto14}%
  \BibitemOpen
  \bibfield  {author} {\bibinfo {author} {\bibfnamefont {Y.}~\bibnamefont
  {Kuramoto}}\ and\ \bibinfo {author} {\bibfnamefont {S.}~\bibnamefont
  {Hoshino}},\ }\href {\doibase 10.7566/JPSJ.83.061007} {\bibfield  {journal}
  {\bibinfo  {journal} {Journal of the Physical Society of Japan}\ }\textbf
  {\bibinfo {volume} {83}},\ \bibinfo {pages} {061007} (\bibinfo {year}
  {2014})}\BibitemShut {NoStop}%
\bibitem [{\citenamefont {Kusunose}(2016)}]{kusunose16}%
  \BibitemOpen
  \bibfield  {author} {\bibinfo {author} {\bibfnamefont {H.}~\bibnamefont
  {Kusunose}},\ }\href {\doibase 10.7566/JPSJ.85.113701} {\bibfield  {journal}
  {\bibinfo  {journal} {Journal of the Physical Society of Japan}\ }\textbf
  {\bibinfo {volume} {85}},\ \bibinfo {pages} {113701} (\bibinfo {year}
  {2016})}\BibitemShut {NoStop}%
\bibitem [{\citenamefont {Hoshino}\ \emph {et~al.}(2012)\citenamefont
  {Hoshino}, \citenamefont {Otsuki},\ and\ \citenamefont
  {Kuramoto}}]{hoshino12}%
  \BibitemOpen
  \bibfield  {author} {\bibinfo {author} {\bibfnamefont {S.}~\bibnamefont
  {Hoshino}}, \bibinfo {author} {\bibfnamefont {J.}~\bibnamefont {Otsuki}}, \
  and\ \bibinfo {author} {\bibfnamefont {Y.}~\bibnamefont {Kuramoto}},\ }\href
  {http://stacks.iop.org/1742-6596/391/i=1/a=012155} {\bibfield  {journal}
  {\bibinfo  {journal} {Journal of Physics: Conference Series}\ }\textbf
  {\bibinfo {volume} {391}},\ \bibinfo {pages} {012155} (\bibinfo {year}
  {2012})}\BibitemShut {NoStop}%
\bibitem [{\citenamefont {Senthil}\ \emph {et~al.}(2004)\citenamefont
  {Senthil}, \citenamefont {Vishwanath}, \citenamefont {Balents}, \citenamefont
  {Sachdev},\ and\ \citenamefont {Fisher}}]{Senthil04science}%
  \BibitemOpen
  \bibfield  {author} {\bibinfo {author} {\bibfnamefont {T.}~\bibnamefont
  {Senthil}}, \bibinfo {author} {\bibfnamefont {A.}~\bibnamefont {Vishwanath}},
  \bibinfo {author} {\bibfnamefont {L.}~\bibnamefont {Balents}}, \bibinfo
  {author} {\bibfnamefont {S.}~\bibnamefont {Sachdev}}, \ and\ \bibinfo
  {author} {\bibfnamefont {M.~P.~A.}\ \bibnamefont {Fisher}},\ }\href {\doibase
  10.1126/science.1091806} {\ \textbf {\bibinfo {volume} {303}},\ \bibinfo
  {pages} {1490} (\bibinfo {year} {2004})}\BibitemShut {NoStop}%
\bibitem [{\citenamefont {Si}(2006)}]{si06}%
  \BibitemOpen
  \bibfield  {author} {\bibinfo {author} {\bibfnamefont {Q.}~\bibnamefont
  {Si}},\ }\href {\doibase https://doi.org/10.1016/j.physb.2006.01.156}
  {\bibfield  {journal} {\bibinfo  {journal} {Physica B: Condensed Matter}\
  }\textbf {\bibinfo {volume} {378-380}},\ \bibinfo {pages} {23 } (\bibinfo
  {year} {2006})},\ \bibinfo {note} {proceedings of the International
  Conference on Strongly Correlated Electron Systems}\BibitemShut {NoStop}%
\bibitem [{\citenamefont {Yamamoto}\ and\ \citenamefont
  {Si}(2007)}]{yamamoto07}%
  \BibitemOpen
  \bibfield  {author} {\bibinfo {author} {\bibfnamefont {S.~J.}\ \bibnamefont
  {Yamamoto}}\ and\ \bibinfo {author} {\bibfnamefont {Q.}~\bibnamefont {Si}},\
  }\href {\doibase 10.1103/PhysRevLett.99.016401} {\bibfield  {journal}
  {\bibinfo  {journal} {Phys. Rev. Lett.}\ }\textbf {\bibinfo {volume} {99}},\
  \bibinfo {pages} {016401} (\bibinfo {year} {2007})}\BibitemShut {NoStop}%
\bibitem [{\citenamefont {Hertz}(1976)}]{hertz}%
  \BibitemOpen
  \bibfield  {author} {\bibinfo {author} {\bibfnamefont {J.~A.}\ \bibnamefont
  {Hertz}},\ }\href {\doibase 10.1103/PhysRevB.14.1165} {\bibfield  {journal}
  {\bibinfo  {journal} {Phys. Rev. B}\ }\textbf {\bibinfo {volume} {14}},\
  \bibinfo {pages} {1165} (\bibinfo {year} {1976})}\BibitemShut {NoStop}%
\bibitem [{\citenamefont {Millis}(1993)}]{millis}%
  \BibitemOpen
  \bibfield  {author} {\bibinfo {author} {\bibfnamefont {A.~J.}\ \bibnamefont
  {Millis}},\ }\href {\doibase 10.1103/PhysRevB.48.7183} {\bibfield  {journal}
  {\bibinfo  {journal} {Phys. Rev. B}\ }\textbf {\bibinfo {volume} {48}},\
  \bibinfo {pages} {7183} (\bibinfo {year} {1993})}\BibitemShut {NoStop}%
\bibitem [{\citenamefont {Moriya}(1985)}]{moriya}%
  \BibitemOpen
  \bibfield  {author} {\bibinfo {author} {\bibfnamefont {T.}~\bibnamefont
  {Moriya}},\ }\href@noop {} {\emph {\bibinfo {title} {Spin Fluctuations in
  Itinerant Electron Magnetism}}}\ (\bibinfo  {publisher} {Springer-Verlag},\
  \bibinfo {address} {Berlin, New York},\ \bibinfo {year} {1985})\BibitemShut
  {NoStop}%
\bibitem [{\citenamefont {Si}\ \emph {et~al.}(2001)\citenamefont {Si},
  \citenamefont {Rabello}, \citenamefont {Ingersent},\ and\ \citenamefont
  {Smith}}]{si01}%
  \BibitemOpen
  \bibfield  {author} {\bibinfo {author} {\bibfnamefont {Q.}~\bibnamefont
  {Si}}, \bibinfo {author} {\bibfnamefont {S.}~\bibnamefont {Rabello}},
  \bibinfo {author} {\bibfnamefont {K.}~\bibnamefont {Ingersent}}, \ and\
  \bibinfo {author} {\bibfnamefont {J.~L.}\ \bibnamefont {Smith}},\ }\href
  {http://dx.doi.org/10.1038/35101507} {\bibfield  {journal} {\bibinfo
  {journal} {Nature}\ }\textbf {\bibinfo {volume} {413}},\ \bibinfo {pages}
  {804 EP } (\bibinfo {year} {2001})},\ \bibinfo {note} {article}\BibitemShut
  {NoStop}%
\bibitem [{\citenamefont {Hoshino}\ and\ \citenamefont
  {Kuramoto}(2014)}]{hoshino143}%
  \BibitemOpen
  \bibfield  {author} {\bibinfo {author} {\bibfnamefont {S.}~\bibnamefont
  {Hoshino}}\ and\ \bibinfo {author} {\bibfnamefont {Y.}~\bibnamefont
  {Kuramoto}},\ }\href {\doibase 10.1103/PhysRevLett.112.167204} {\bibfield
  {journal} {\bibinfo  {journal} {Phys. Rev. Lett.}\ }\textbf {\bibinfo
  {volume} {112}},\ \bibinfo {pages} {167204} (\bibinfo {year}
  {2014})}\BibitemShut {NoStop}%
\end{thebibliography}%

\end{document}